\documentclass[12pt,eqno]{article}
\usepackage{amsmath,amssymb,epsfig,slashbox}
\numberwithin{equation}{section}

\def\ignore#1{{}}
 
\newcommand{\alp}{\alpha}
\newcommand{\bt}{\beta}

\newcommand{\dlt}{\delta}
\newcommand{\Dlt}{\Delta}
\newcommand{\ep}{\epsilon}

\newcommand{\tht}{\theta}

\newcommand{\kp}{\kappa}
\newcommand{\lmd}{\lambda}
\newcommand{\Lmd}{\Lambda}
\newcommand{\sgm}{\sigma}
\newcommand{\Sgm}{\Sigma}
\newcommand{\vph}{\varphi}

\newcommand{\Omg}{\Omega}

\newcommand{\be}{\begin{equation}}
\newcommand{\ee}{\end{equation}}
\newcommand{\bea}{\begin{eqnarray}}
\newcommand{\eea}{\end{eqnarray}}
\newcommand{\eql}{\!\!\!&=\!\!\!&}

\newcommand{\defa}{\!\!\!&\equiv\!\!\!&}

\newcommand{\exch}{\leftrightarrow}
\newcommand{\simgt}{\stackrel{>}{{}_\sim}}

\newcommand{\sma}{\!\!\!&\simeq\!\!\!&}

\newcommand{\tl}[1]{\tilde{#1}}
\newcommand{\bdm}[1]{{\mbox{\boldmath $#1$}}}
\newcommand{\tr}{{\rm tr}\,}

\newcommand{\diag}{{\rm diag}\,}
\newcommand{\der}{\partial}
\newcommand{\dr}{\!\!d}
\newcommand{\hc}{{\rm h.c.}}
\newcommand{\ie}{{\it i.e.}}
\newcommand{\id}{\mbox{\boldmath $1$}}

\newcommand{\vev}[1]{\langle #1 \rangle}
\newcommand{\Lvev}[1]{\left\langle #1 \right\rangle}
\newcommand{\brkt}[1]{\left( #1 \right)}
\newcommand{\brc}[1]{\left\{ #1 \right\}}
\newcommand{\sbk}[1]{\left[ #1 \right]}
\newcommand{\abs}[1]{| #1 |}

\renewcommand{\Re}{{\rm Re}\,}

\newcommand{\cD}{{\cal D}}
\newcommand{\cE}{{\cal E}}
\newcommand{\cF}{{\cal F}}
\newcommand{\cG}{{\cal G}}
\newcommand{\cH}{{\cal H}}

\newcommand{\cL}{{\cal L}}

\newcommand{\cN}{{\cal N}}
\newcommand{\cO}{{\cal O}}
\newcommand{\cP}{{\cal P}}
\newcommand{\cQ}{{\cal Q}}
\newcommand{\cR}{{\cal R}}

\newcommand{\cT}{{\cal T}}
\newcommand{\cU}{{\cal U}}
\newcommand{\cV}{{\cal V}}
\newcommand{\cW}{{\cal W}}
\newcommand{\cX}{{\cal X}}
\newcommand{\cY}{{\cal Y}}

\newcommand{\NP}[1]{{\it Nucl.~Phys.}~{\bf #1}}
\newcommand{\PL}[1]{{\it Phys.~Lett.}~{\bf #1}}

\newcommand{\PR}[1]{{\it Phys.~Rev.}~{\bf #1}}
\newcommand{\PRL}[1]{{\it Phys.~Rev.~Lett.}~{\bf #1}}
\newcommand{\PTP}[1]{{\it Prog.~Theor.~Phys.}~{\bf #1}}

\newcommand{\JH}[1]{{\it JHEP}~{\bf #1}}

\begin{document}

\begin{titlepage}
\null
\begin{flushright}
WU-HEP-11-03
\\
KEK-TH-1491
\\
December, 2011
\end{flushright}

\vskip 1.5cm
\begin{center}
\baselineskip 0.8cm
{\LARGE \bf 
SUSY flavor structure of generic 5D supergravity models
}

\lineskip .75em
\vskip 1.5cm

\normalsize

{\large\bf Hiroyuki Abe,}${}^1\!${\def\thefootnote{\fnsymbol{footnote}}
\footnote[1]{E-mail address: abe@waseda.jp}}
{\large\bf Hajime Otsuka,}${}^1\!${\def\thefootnote{\fnsymbol{footnote}}
\footnote[2]{E-mail address: hajime.13.gologo@akane.waseda.jp}} 
{\large\bf Yutaka Sakamura,}${}^2\!${\def\thefootnote{\fnsymbol{footnote}}
\footnote[3]{E-mail address: sakamura@post.kek.jp}} \\ 
{\large\bf and Yusuke Yamada}${}^1\!${\def\thefootnote{\fnsymbol{footnote}}
\footnote[4]{E-mail address: yuusuke-yamada@asagi.waseda.jp}}

\vskip 1.0em

${}^1${\small\it Department of Physics, Waseda University, \\ 
Tokyo 169-8555, Japan}

\vskip 1.0em

${}^2${\small\it KEK Theory Center, Institute of Particle and Nuclear Studies, 
KEK, \\ Tsukuba, Ibaraki 305-0801, Japan} \\ \vspace{1mm}
{\small\it Department of Particles and Nuclear Physics, \\
The Graduate University for Advanced Studies (Sokendai), \\
Tsukuba, Ibaraki 305-0801, Japan} 

\vspace{12mm}

{\bf Abstract}\\[5mm]
{\parbox{13cm}{\hspace{5mm} \small
We perform a comprehensive and systematic analysis of the SUSY flavor 
structure of generic 5D supergravity models on $S^1/Z_2$ 
with multiple $Z_2$-odd vector multiplets 
that generate multiple moduli. 
The SUSY flavor problem can be avoided due to contact terms 
in the 4D effective K\"ahler potential peculiar to the multi-moduli case. 
A detailed phenomenological analysis is provided based on 
an illustrative model. 
}}

\end{center}

\end{titlepage}


\section{Introduction}

The standard model (SM) of elementary particles is a successful 
theory without any contradiction to the observations up to now. 
However, it contains many free parameters, most of 
which come from Yukawa couplings. 
The eigenvalues and eigenvectors 
of the Yukawa coupling matrices determine the mass ratios and mixings 
between generations, respectively, and their observed values are quite 
hierarchical. 
Models beyond the SM should explain such hierarchical structures 
of quarks and leptons. 


Extra dimensions provide a simple way to realize 
the hierarchical flavor structures, \ie,  
a wave function localization of matter fields 
in extra dimensions~\cite{Arkani-Schmaltz}. 
Actually, the most promising 
candidate for the unified theory of the SM and the gravity, 
\ie, the superstring theory, predicts the existence of extra dimensions. 
They can also explain other problems of the SM, such as 
the gauge hierarchy problem~\cite{RS}, 
a candidate for dark matter~\cite{Servant:2002aq}, and so on. 

Supersymmetry (SUSY) is an interesting extension of the SM. 
It softens the divergences in quantum field theories 
and then protects the electroweak scale against large radiative 
corrections. 
The three SM gauge couplings are unified in the minimal supersymmetric 
standard model (MSSM) at $M_{\rm GUT}\equiv 2\times 10^{16}$~GeV, 
which suggests the grand unified theory. 
It also has a candidate for dark matter if the R-parity forbids decays 
of the lightest SUSY particle. 
Besides, the existence of SUSY is predicted by the superstring theory. 

\ignore{
Among various extensions of the SM, supersymmetric extensions are quite 
interesting because, in addition to the fact that superstring theory 
predicts supersymmetry (SUSY), it softens the divergences in quantum 
field theories and then protects the electroweak (EW) scale against 
the radiative corrections. Furthermore, the superpartners of SM particles 
would contain dark matter which are required by cosmological observations, 
and the three gauge coupling constants of the minimal supersymmetric 
standard model (MSSM) are unified at the so-called grand unification 
theory (GUT) scale. 
}

From the above reasons, we consider an extension of the SM by 
introducing SUSY and extra dimensions. 
The minimal setup for such an extension is a five-dimensional (5D) 
supergravity (SUGRA) compactified on an orbifold~$S^1/Z_2$. 
The chiral structure of the SM can be realized 
by the orbifold $Z_2$ projection, which preserves $N=1$ SUSY 
in a four-dimensional (4D) sense. 
The local SUSY (\ie, SUGRA) is required in order to 
avoid the existence of a massless Goldstino which contradicts to many 
observations, and to discuss the moduli stabilization. 

The off-shell formulation of 5D SUGRA~\cite{Kugo-Ohashi,Kugo:2002js} 
provides the most general way to construct 5D SUGRA models. 
There is, moreover, a systematic way to obtain 4D effective theories of 
such models, which we call the off-shell dimensional 
reduction~\cite{Abe:2007} based on the $N=1$ superfield description 
of 5D SUGRA~\cite{Paccetti Correia:2004ri,Abe:2004}. 
This method can be applied to general 5D SUGRA models. 
For example, we analyzed some class of models 
by this method~\cite{Abe:2007zv} 
that include SUSY extension of the Randall-Sundrum model~\cite{SUSY_RS} 
and the 5D heterotic M theory~\cite{5D_Mtheory} as special limits 
of parameters. 

The wave function localization of matter fields (hypermultiplets) 
is realized by controlling bulk mass parameters for the matter fields 
in 5D models. 
In 5D SUGRA on $S^1/Z_2$, the bulk mass parameters are obtained 
as $U(1)$ charges (times 5D Planck mass) under $Z_2$-odd vector 
multiplets including the graviphoton multiplet. 
One of the minimal models with the realistic flavor structure 
was constructed with a single $Z_2$-odd vector multiplet. 
However, it suffers from the SUSY flavor problem as well as 
tachyonic squarks and/or sleptons~\cite{Abe:2008ka}. 
In our previous paper~\cite{Abe:2008an}, 
we constructed models with two $Z_2$-odd vector multiplets, which 
induce a modulus chiral multiplet other than the radion multiplet 
in the 4D effective theory, and showed that 
there is an important contribution of the multiple moduli multiplets 
to the effective K\"{a}hler potential that may solve problems 
mentioned above. 
In this paper, we extend the previous work to more generic set-up, 
which has an arbitrary number of $Z_2$-odd vector multiplets 
(\ie, moduli multiplets) and a nontrivial warping along 
the extra dimension. 
We perform a comprehensive and systematic analysis to understand 
the SUSY flavor structure of such generic 5D SUGRA models, and 
provide a phenomenological analysis based on an illustrative model.


This paper is organized as follows. 
In Sec.~\ref{review_5DSUGRA}, we set up our model 
with a brief review of the off-shell formulation of 5D SUGRA. 
In Sec.~\ref{4Deffective}, we derive 4D effective theory 
of our 5D model and study its properties. 
In Sec.~\ref{specific_model}, 
we perform a phenomenological analysis based on an illustrative model. 
Sec.~\ref{summary} is devoted to a summary. 
In Appendix~\ref{derive:L_eff}, some details of the derivation 
of the effective action are shown. 
In Appendix~\ref{comment:4-fermi}, we provide a comment on some peculiar 
structure of the 4D effective theory to 5D SUGRA. 
In Appendix~\ref{explicit_forms}, explicit expressions of some quantities 
in the model in Sec.~\ref{specific_model} are collected.

\section{Set-up and brief review of 5D SUGRA}  \label{review_5DSUGRA} 
\subsection{Brief picture of set-up} \label{setup}
The set-up we consider in this article is as follows. 
\begin{itemize}
\item The fifth dimension is compactified on the orbifold~$S^1/Z_2$, 
 and the background 5D metric is the warped metric. 
 In contrast to the original warped model~\cite{RS}, 
 the warp factor is supposed to be $\cO(10^2)$ and explains 
 the small hierarchy between the Planck scale~$M_{\rm Pl}=2.4\times 10^{18}$~GeV 
 and the GUT scale~$M_{\rm GUT}=2\times 10^{16}$~GeV. 
 
\item The compactification scale is around $M_{\rm GUT}$, 
 and below this scale, 4D effective theory becomes MSSM. 
 
\item All the standard model fields are identified with 
 zero-modes of the 5D bulk fields. 
 
\item The hierarchical flavor structure of the SM fermions 
 is realized by the quasi-localization of the wave functions 
 for the zero-modes~\cite{Arkani-Schmaltz}. 
 
\item SUSY is broken at some scale below $M_{\rm GUT}$, and 
 the main source of SUSY breaking is the $F$-term of 
 a single chiral superfield~$X$ in 4D effective theory. 
\end{itemize}
When we study a model with extra dimensions, 
it is indispensable to stabilize the size of the compactified internal space 
to a finite value. 
Especially in SUSY models, such stabilization mechanisms affect 
the sfermion mass spectrum in the MSSM sector. 
In order to take into account the stabilization of the extra dimension, 
we have to work in the context of SUGRA. 

Although the above type of set-up has been studied in many papers, 
most of them do not consider the full SUGRA effects 
or only consider a {\it limited} case from the viewpoint of SUGRA. 
In our previous work~\cite{Abe:2008an}, we pointed out 
a possibility to solve the SUSY flavor problem 
thanks to some peculiar terms 
in the K\"{a}hler potential of 4D effective theory, 
in a case that the theory has two moduli 
multiplets (we will provide the definition of the moduli multiplet 
in the next subsection).

In this article, we extend our previous work  
to the case of an arbitrary number of the moduli multiplets 
and a non-trivial warp factor, and discuss some phenomenological 
aspects. 
Before specifying the model, we start with {\it general} 5D SUGRA, 
derive 4D effective theory in a systematic way, 
and analyze the flavor structure of the soft SUSY breaking terms 
in the effective theory. 
After that, we will construct an illustrative model 
in Sec~\ref{specific_model} to realize the above set-up. 
For our purpose, the superconformal gravity formulation, 
which is an off-shell description of 5D SUGRA, is useful. 
So we briefly review this formulation 
and explain the structure of 5D SUGRA 
in the rest of this section.

\subsection{Superconformal multiplets in 5D SUGRA}
The background 5D metric with 4D Poincar\'{e} invariance is parametrized as 
\be
 ds^2 = e^{2\sgm(y)}\eta_{\mu\nu}dx^\mu dx^\nu-dy^2, 
\ee
where $\mu,\nu=0,1,2,3$, $\eta_{\mu\nu}=\diag(1,-1,-1,-1)$, 
and $e^{\sgm(y)}$ is a warp factor, which is a function of only $y$ 
and determined by the dynamics. 
We take the fundamental region of the orbifold as $0\leq y\leq L$. 
Since we choose the coordinate~$y$ such that $\vev{e_y^{\;\;4}}=1$, 
the constant~$L$ denotes the size of the extra dimension. 

Our formalism is based on the superconformal formulation
developed in Ref.~\cite{Kugo-Ohashi,Kugo:2002js}. 
5D superconformal multiplets relevant to our study are summarized 
in Table~\ref{sc_multiplets}. 
\begin{table}[t]
\begin{center}
\begin{tabular}{l|c|c} 
\rule[-2mm]{0mm}{7mm} 5D superconformal multiplet & 
$N=1$ decomposition & $Z_2$-parity  \\ \hline\hline
\rule[-2mm]{0mm}{7mm} Weyl multiplet (gravity) & 
$\bdm{E_W}=(E_W,L^\alp_E,V_E)$ & $(+,-,+)$  \\ \hline
\rule[-2mm]{0mm}{7mm} Vector multiplet (moduli) & 
$\bdm{V}^{I'}=(V^{I'},\Sgm^{I'})$ & $(-,+)$  
\\ \hline
\rule[-2mm]{0mm}{7mm} Vector multiplet (gauge)& 
$\bdm{V}^{I''}=(V^{I''},\Sgm^{I''})$ & $(+,-)$ \\ \hline
\rule[-2mm]{0mm}{7mm} Hypermultiplet (compensator) & 
$\bdm{H}^{a=1}=(\Phi^1,\Phi^2)$ & $(-,+)$ \\ \hline
\rule[-2mm]{0mm}{7mm} Hypermultiplet (matter) & 
$\bdm{H}^{a\geq 2}=(\Phi^{2a-1},\Phi^{2a})$ & $(-,+)$ \\ \hline 
\end{tabular}
\end{center}
\caption{Relevant 5D superconformal multiplets. Each multiplet is 
decomposed into $N=1$ multiplets. }
\label{sc_multiplets}
\end{table}
Each multiplet can be decomposed into $N=1$ multiplets~\cite{Kugo:2002js}. 
The signs in the last column of Table~\ref{sc_multiplets} 
denote the orbifold $Z_2$ parities of the $N=1$ multiplets 
in the next column. 
We assume that each $N=1$ multiplet has the same $Z_2$ parity 
at both boundaries~$y=0,L$, for simplicity.\footnote{ 
In general, vector and hypermultiplets can have different 
$Z_2$ parities at different boundaries, 
but only $Z_2$-even fields at both boundaries have zero-modes 
which are relevant to 4D effective theory. 
All the other fields are decoupled and should be integrated out. 
Thus the treatment of multiplets with different $Z_2$ parities 
is the same as that of multiplets which are $Z_2$-odd at both boundaries. }

\begin{description}
\item[Weyl multiplet~$\bdm{E_W}$]
 This corresponds to the gravitational multiplet, 
 and is decomposed into the $N=1$ Weyl multiplet~$E_W$, 
 a complex general multiplet~$L_E^\alp$ ($\alp$: spinor index), 
 and a real general multiplet~$V_E$. 
 Among them, $E_W$ and $V_E$ are $Z_2$-even, 
 and includes the 4D parts of the vierbein~$e_\mu^{\;\;\underline{\nu}}$ 
 and the extra-dimensional component~$e_y^{\;\;4}$, respectively.  
 The $Z_2$-odd multiplet~$L_E^\alp$ includes 
 the ``off-diagonal'' parts~$e_\mu^{\;\;4}$ and $e_y^{\;\;\underline{\mu}}$. 
 The latter is irrelevant to the following discussion, and is neglected. 
 When the loop corrections are taken into account, however, 
 the contribution from $L_E^\alp$ has to be included.   

\item[Vector multiplet~$\bdm{V}^I$] 
 This is decomposed into 
 $N=1$ vector and chiral multiplets~$V^I$ and $\Sgm^I$, 
 which have opposite $Z_2$-parities. 
 The vector multiplets are divided into two classes 
 according to their $Z_2$ parities. 
 One is a class of the gauge multiplets, which are denoted as $\bdm{V}^{I''}$.  
 In this class, $V^{I''}$ are $Z_2$-even and have zero-modes 
 that are identified with the gauge multiplets in 4D effective theory. 
 In the other classes ($\bdm{V}^{I'}$), 
 the 4D vector components have no zero-modes. 
 Instead, the chiral multiplets~$\Sgm^{I'}$ have zero-modes. 
 They include the scalar fields and their potential is flat 
 at the classical level. 
 Thus we refer to $\Sgm^{I'}$ (or $\bdm{V}^{I'}$) as the moduli multiplets 
 in this article.\footnote{
 These moduli fields are actually identified with the shape moduli 
 of the compactified space for a 5D effective 
 theory of the heterotic M-theory compactified 
 on the Calabi-Yau manifold~\cite{5D_Mtheory}, for example. }
 
 At least one vector multiplet belongs to the latter category. 
 In the pure SUGRA, the vector component of such a multiplet 
 is identified with the graviphoton.\footnote{
 In this article, the terminology~``graviphoton'' represents 
 the vector field in the gravitational multiplet of 
 the {\it on-shell} formulation. 
 It should be distinguished from the off-diagonal components of 
 the 5D metric, which are included in $L_E^\alp$ in the current formulation. }
 All the other components are auxiliary fields that are eliminated 
 by the superconformal gauge-fixing. 
 Thus, when there are $n_V$ moduli multiplets in the off-shell action, 
 only $(n_V-1)$ degrees of freedom are physical. 
 (See Sec.~\ref{sc_GF}.)

\item[Hypermultiplet~$\bdm{H}^a$]
 This is decomposed into 
 two chiral multiplets~$\Phi^{2a-1}$ and $\Phi^{2a}$, 
 which have opposite $Z_2$-parities. 
 We can always choose their $Z_2$-parities 
 as listed in Table~\ref{sc_multiplets} 
 by using $SU(2)_U$, which is an automorphism of the superconformal algebra. 
 The hypermultiplets are also divided into two classes. 
 One is the compensator multiplets~$a=1,2,\cdots,n_C$ and the other is 
 the physical matter multiplets~$a=n_C+1,\cdots,n_C+n_H$. 
 The former is an auxiliary degree of freedom and eliminated 
 by the superconformal gauge-fixing. 
 In contrast to 4D SUGRA, both types of hypermultiplets have the same 
 quantum numbers of the superconformal symmetries in 5D SUGRA. 
 They are discriminated only by signs of their kinetic terms in the action. 
 Thus, in principle, it is possible to introduce 
 an arbitrary number of the compensator multiplets in the theory. 
 In this article, we consider the single compensator case ($n_C=1$) 
 for simplicity.\footnote{
 The number of the compensator multiplets determines the target 
 manifold of the hyperscalars. 
 For instance, it is 
 $USp(2,2n_H)/USp(2)\times USp(2n_H)$ for $n_C=1$, 
 and $SU(2,n_H)/SU(2)\times U(n_H)$ for $n_C=2$. }
\end{description}

\subsection{$\bdm{N=1}$ description of 5D action}
For our purpose, it is convenient to describe the 5D action 
in terms of the $N=1$ multiplets~\cite{Paccetti Correia:2004ri,Abe:2004}. 
This corresponds to the extension of the result in Ref.~\cite{ArkaniHamed:2001tb} 
to the local SUSY case. 
We can see that $V_E$ has no kinetic term in this description.\footnote{
This does not mean that $e_y^{\;\;4}$ is an auxiliary field. 
It is also contained in $\Sgm^I$, which have their own kinetic terms. }
After integrating it out, the 5D Lagrangian is expressed 
as~\cite{Correia:2006pj}  
\bea
 \cL \eql -3e^{2\sgm}\int\dr^4\tht\;\cN^{1/3}(\cV)
 \brc{d_{\hat{a}}^{\;\;\hat{b}}\bar{\Phi}_{\hat{b}}
 \brkt{e^{-2g_It_IV^I}}^{\hat{a}}_{\;\;\hat{c}}\Phi^{\hat{c}}}^{2/3} 
 \nonumber\\
 &&-e^{3\sgm}\sbk{\int\dr^2\tht\;\Phi^{\hat{a}}d_{\hat{a}}^{\;\;\hat{b}}
 \rho_{\hat{b}\hat{c}}\brkt{\der_y-2ig_It_I\Sgm^I}^{\hat{c}}_{\;\;\hat{d}}
 \Phi^{\hat{d}}+\hc} \nonumber\\
 &&+\cL_{\rm vec}+2\sum_{y_*=0,L}\cL^{(y_*)}\dlt(y-y_*),  \label{5D_action}
\eea
where $d_{\hat{a}}^{\;\;\hat{b}}=\diag(\id_{2n_C},-\id_{2n_H})$, 
$\rho_{\hat{a}\hat{b}}=i\sgm_2\otimes\id_{n_C+n_H}$, 
$\bar{\Phi}_{\hat{b}}\equiv (\Phi^{\hat{b}})^\dagger$, 
and $\cL_{\rm vec}$ is defined as 
\be
 \cL_{\rm vec} \equiv \int\dr^2\tht\brc{
 -\frac{\cN_{IJ}(\Sgm)}{4}\cW^I\cW^J
 +\frac{\cN_{IJK}}{48}\bar{D}^2\brkt{
 V^ID^\alp\der_y V^J-D^\alp V^I\der_y V^J}\cW^K_\alp}+\hc.  \label{def:L_CS}
\ee
This contains the kinetic terms 
for the vector multiplets~$V^I$ and the Chern-Simons terms. 
The indices~$\hat{a},\hat{b}$ run over the whole $2(n_C+n_H)$ chiral 
multiplets coming from the hypermultiplets. 
As mentioned in the previous subsection, 
we consider the case of $n_C=1$ in the following. 
Here $\sgm_2$ in $\rho_{\hat{a}\hat{b}}$ acts 
on each hypermultiplet~$(\Phi^{2a-1},\Phi^{2a})$. 
$\cN$ is a cubic polynomial called the norm function,\footnote{
This corresponds to the prepotential in the $N=2$ global SUSY case. } 
which is defined by 
\be
 \cN(X) \equiv C_{IJK}X^IX^JX^K. 
\ee
A real constant tensor~$C_{IJK}$ is completely symmetric for the indices, 
and $\cN_I(X)\equiv \der\cN/\der X^I$, 
$\cN_{IJ}(X)\equiv\der^2\cN/\der X^I\der X^J$, and so on. 
The superfield strength~$\cW^I_\alp\equiv -\frac{1}{4}\bar{D}^2D_\alp V^I$ 
and $\cV^I\equiv -\der_y V^I+\Sgm^I+\bar{\Sgm}^I$ 
are gauge-invariant quantities. 
The generators~$t_I$ are anti-hermitian. 
For a gauge multiplet of a non-abelian gauge group~$G$, 
the indices~$I$, $J$ run over $\dim G$ values and $\cN_{IJ}$ 
are common for them. 
The index~$a$ for the hypermultiplets are divided into irreducible 
representations of $G$. 
The fractional powers in the first line of (\ref{5D_action}) 
appear after integrating $V_E$ out. 
The boundary Lagrangian~$\cL^{(y_*)} (y_*=0,L)$ can be introduced 
independently of the bulk Lagrangian. 

Note that (\ref{5D_action}) is a shorthand expression of the full SUGRA action. 
We can always restore the full action by the promotion, 
\be
 \int\dr^4\tht\brc{\cdots} \to \frac{1}{2}\sbk{\cdots}_D, \;\;\;\;\;
 \int\dr^2\tht\brc{\cdots}+\hc \to \sbk{\cdots}_F,  \label{FDpromote}
\ee 
where $\sbk{\cdots}_D$ and $\sbk{\cdots}_F$ denote the $D$- and $F$-term action 
formulae of the $N=1$ superconformal formulation~\cite{4Doffshell}, 
which are compactly listed in Appendix~C of Ref.~\cite{Kugo:2002js}.\footnote{ 
The superfield descriptions of the $D$- and $F$-term formulae 
on the ordinary superspace are provided 
at the linear order in the fields belonging to $E_W$ 
in Ref.~\cite{Sakamura:2011df}. }
Here we omitted the spacetime integral. 
This promotion restores 
the dependence of the action on the components 
of the $N=1$ Weyl multiplet~$E_W$, 
such as the Einstein-Hilbert term and the gravitino-dependent terms. 
The Weyl multiplet also contains some auxiliary fields. 
After integrating them out, some terms in (\ref{5D_action}) are modified. 
Practically, the kinetic terms for $V^I$ are the only such terms  
that are relevant to the phenomenological discussions. 
Their kinetic functions are read off as 
$-\cN_{IJ}(\Sgm)/4$ from the first term in (\ref{def:L_CS}). 
This will be modified after integrating out the above-mentioned 
auxiliary fields, and 
the correct kinetic function is obtained as $\brc{\cN a_{IJ}}(\Sgm)/2$, 
where~\cite{Kugo-Ohashi} 
\be
 a_{IJ} \equiv -\frac{1}{2\cN}\brkt{\cN_{IJ}-\frac{\cN_I\cN_J}{\cN}}. 
 \label{def:a_IJ}
\ee

\subsection{Superconformal gauge-fixing in 5D} \label{sc_GF}
Here we provide some comments on the superconformal gauge-fixing in 5D SUGRA. 
Since we will not impose the gauge-fixing conditions 
at the 5D stage,\footnote{
The superconformal gauge-fixing will be imposed after 
4D effective action is derived because it breaks the $N=1$ off-shell 
structure of the action. } 
the readers can skip this subsection. 
Nevertheless, the comments presented here may help the readers 
to understand the structure of 5D SUGRA set-up. 

The Lagrangian~(\ref{5D_action}) with the promotion~(\ref{FDpromote}) 
is invariant (up to total derivatives) 
under the superconformal symmetries. 
In order to obtain the usual Poincar\'{e} SUGRA, 
we have to eliminate the extra symmetries by imposing 
the gauge-fixing conditions. 
A conventional choice of such conditions is expressed 
in our $N=1$ superfield notation as 
\bea
 \left.\cN\brkt{\frac{\cV}{V_E}}\right|_0 
 \eql \left. 
 d_{\hat{a}}^{\;\;\hat{b}}\bar{\Phi}_{\hat{b}}\Phi^{\hat{a}}\right|_0 = M_5^3, 
 \nonumber\\
 \left.\cN\brkt{\frac{\cV}{V_E}}\right|_\tht \eql 
 \left.\cN\brkt{\frac{\cV}{V_E}}\right|_{\bar{\tht}^2\tht} = 
 \left.d_{\hat{a}}^{\;\;\hat{b}}\bar{\Phi}_{\hat{b}}\Phi^{\hat{a}}
 \right|_\tht = 
 \left.\Phi^{\hat{a}}d_{\hat{a}}^{\;\;\hat{b}}\rho_{\hat{b}\hat{c}}\Phi^{\hat{c}}
 \right|_\tht = 0, \;\;\; \cdots,  \label{GF_cond}
\eea
where $M_5$ is the 5D Planck mass, 
and the symbols~$|_0$, $|_\tht$ and $|_{\bar{\tht}^2\tht}$ 
denote the lowest, $\tht$- and $\bar{\tht}^2\tht$-components 
in the superfields, respectively. 
$V_E=e_y^{\;\;4}+\tht\psi_y^++\bar{\tht}\bar{\psi}_y^++\cdots$ 
is the real general multiplet 
coming from the 5D Weyl multiplet (see Table~\ref{sc_multiplets}), 
where $\psi_y^+$ is the $Z_2$-even 5th-component of the gravitino.  
The conditions in the first line fix the dilatation, 
and those in the second line fix the conformal SUSY. 
They reproduce the Einstein-Hilbert 
term~$\cL=-\frac{M_5^3}{2}e^{(5)}\cR^{(5)}+\cdots$, 
where $e^{(5)}$ is the determinant of the f\"unfbein, 
$\cR^{(5)}$ is the 5D Ricci scalar, from the $D$-term action formula. 

The conditions in (\ref{GF_cond}) indicate that 
there is one multiplet whose components are not physical 
in each of the vector and hypermultiplet sectors. 
Such a multiplet is the graviphoton multiplet in the vector multiplet sector, 
and the compensator multiplet in the hypermultiplet sector. 
However, the graviphoton~$B_M$ ($M=\mu,y$) itself is exceptional. 
Since (\ref{GF_cond}) does not involve the vector components,  
the graviphoton is always physical. 
The first condition in (\ref{GF_cond}) suggests that 
$\Sgm^{I'}|_0$ generically have nonvanishing VEVs. 
(Since $\Sgm^{I''}$ are $Z_2$-odd, they do not have zero-modes nor VEVs.)
Specifically, the lowest components of $V^I$ and $\Sgm^I$ are 
\be
 V^I = \tht\sgm^\mu\bar{\tht}W_\mu^I+\cdots, \;\;\;\;\;
 \Sgm^I = \frac{1}{2}\brkt{e_y^{\;\;4}M^I-iW_y^I}+\cdots, 
\ee
where $M^I$ and $W^I_M$ are the real scalar and vector components 
of the 5D vector multiplet~$\bdm{V}^I$, 
and the 4D vector part of the graviphoton multiplet is identified as 
\be 
 V_B \equiv \frac{\cN_I}{3\cN}(2\Re\vev{\Sgm})V^I,  \label{def:V_B}
\ee 
whose fermionic component vanishes by (\ref{GF_cond}). 
The corresponding chiral multiplet part is defined as 
\be
 \cT \equiv \frac{\cN_I}{3\cN}(2\Re\vev{\Sgm})\Sgm^I. \label{def:cT}
\ee
In contrast to $V_B$, this remains physical under the gauge-fixing conditions. 
In fact, (\ref{GF_cond}) and (\ref{def:V_B}) suggest that
\be
 \cT = \frac{1}{2}\brkt{e_y^{\;\;4}-iB_y}+\tht\psi_y^++\cdots, 
 \label{comp:cT}
\ee
where $B_M=\frac{\cN_I}{3\cN}(2\Re\vev{\Sgm})W^I_M$ is the graviphoton.  
We have chosen the coordinate~$y$ such that $\vev{e_y^{\;\;4}}=1$. 
\ignore{
\be
 \cN(2\Re\Sgm)|_0 = \brkt{M_5 e_y^{\;\;4}}^3, \;\;\;\;\;
 \cN_I(2\Re\Sgm)|_0 \Sgm^I|_\tht = 3M_5^3\brkt{e_y^{\;\;4}}^2\psi_y^+. 
\ee}
Eq.(\ref{comp:cT}) shows that $\cT$ no longer belong 
to the vector multiplet sector 
after the gauge-fixing~(\ref{GF_cond}), 
but it is the ``5D radion multiplet'' belonging to the 5D gravitational sector. 


Therefore, among $\Sgm^I$, one combination~$\cT$ has a different origin 
from the others. 
We can also see this fact explicitly from the action. 
The condition~(\ref{GF_cond}) suggests that $\Phi^2$ must have 
a nonzero VEV, and it plays a similar role to the chiral compensator 
multiplet in 4D SUGRA.  
($\Phi^1$ does not have a VEV because it is $Z_2$-odd.)
To emphasize this point, we rewrite the hypermultiplets as 
$\phi\equiv\brkt{\Phi^2}^{2/3}$ and 
$\hat{\Phi}^{\hat{a}}\equiv\Phi^{\hat{a}}/\Phi^2$ ($\hat{a}\neq 2$) 
so that their Weyl weights are one and zero, respectively.  
Then the first line of (\ref{5D_action}) is expanded around $\vev{\Sgm^I}$ as 
\bea
 \cL_D \eql -3\hat{\cN}^{1/3}(\vev{\cV})
 \int\dr^4\tht\;\abs{\phi}^2\brc{
 1+\frac{\cN_I}{\cN}(\vev{\cV})\tl{\cV}^I
 +\frac{3\cN\cN_{IJ}-2\cN_I\cN_J}{18\cN^2}(\vev{\cV})
 \tl{\cV}^I\tl{\cV}^J}+\cdots \nonumber\\
 \eql -\hat{\cN}^{1/3}(2\Re\vev{\Sgm})\int\dr^4\tht\;\abs{\phi}^2
 \brc{3\brkt{\cT+\bar{\cT}}-\brkt{a\cdot\cP}_{IJ}\tl{\cV}^I\tl{\cV}^J
 +\cO(\hat{\Phi}^2,\tl{\cV}^3)},  \label{5DL_expand}
\eea
where $\tl{\cV}^I=-\der_y V^I+\tl{\Sgm}^I+\bar{\tl{\Sgm}}$ 
denotes the fluctuation part of $\cV^I$, 
and $\cP^I_{\;\;J}$ is a projection operator defined as 
\be
 \cP^I_{\;\;J}(\cX) \equiv \dlt^I_{\;\:J}-\frac{\cX^I\cN_J}{3\cN}(\cX),  
 \label{def:cP}
\ee
which has a property, 
\be
 \brc{\cN_I\cP^I_{\;\;J}}(\cX) = \cP^I_{\;\;J}(\cX)\cX^J = 0. 
\ee
The argument of $\brkt{a\cdot\cP}_{IJ}\equiv a_{IK}\cP^K_{\;\;J}$ is 
$(2\Re\vev{\Sgm})$. 
In the second line of (\ref{5DL_expand}), we can see that 
the first term has the no-scale structure peculiar 
to the 5D radion multiplet~\cite{Marti:2001iw}, 
and the second term does not include $\cT$ due to the projection 
operator~$\cP^I_{\;\;J}$. 
Hence $\Sgm^I$ are divided into two categories according to their origins.

\ignore{
In order to reproduce this result from (\ref{5D_action}), 
we need to consider the replacement~(\ref{FDpromote}). 
Then the gauge field~$A_\mu$ for $U(1)_A$,\footnote{
$U(1)_A$ is an automorphism of the $N=1$ superconformal 
algebra~\cite{4Doffshell}. } 
which is a vector component of $E_W$, is restored in the action 
by promoting $\der_\mu$ to the covariant derivative~$\cD_\mu$. 
This gauge field does not have a kinetic term in the action, 
and should be treated as an auxiliary field. 
After integrating it out, the correct coefficients~(\ref{def:a_IJ}) 
are reproduced. 
}

\subsection{Gauging and mass scales}
In SUGRA, an introduction of any mass scales into the action requires 
gauging some of the isometries on the hyperscalar manifold 
by the moduli multiplets~$\bdm{V}^{I'}$, \ie, 
we have to deal with the gauged SUGRA.  
For example, the bulk cosmological constant is induced 
when the compensator hypermultiplet~$(\Phi^1,\Phi^2)$ is charged, 
and a bulk mass parameter for a physical hypermultiplet 
is generated when it is charged for $\bdm{V}^{I'}$. 
The gauging by $\bdm{V}^{I''}$ leads to the usual gauging 
by a 4D massless gauge multiplet in 4D effective theory. 
We omit the latter type of gauging in the following expressions 
because it does not play a significant role in the derivation 
of the effective theory and can be easily 
restored in the 4D effective action. 
We assume that all the gaugings by $\bdm{V}^{I'}$ 
are abelian, and are chosen to $\sgm_3$-direction 
in the $(\Phi^{2a-1},\Phi^{2a})$-space, for simplicity.  
Thus, the generators and the gauge couplings are chosen as 
\be
 \brkt{ig_{I'}t_{I'}}^{\hat{a}}_{\;\;\hat{b}} 
 = \sgm_3\otimes\diag\brkt{\frac{3}{2}k_{I'},c_{2I'},c_{3I'},
 \cdots,c_{(n_H+1)I'}}, 
\ee
where $\sgm_3$ acts on each hypermultiplet~$(\Phi^{2a-1},\Phi^{2a})$. 
Note that these coupling constants are $Z_2$-odd. 
Such kink-type couplings can be realized in SUGRA context 
by the mechanism proposed in Ref.~\cite{Bergshoeff:2000zn}. 
In contrast to Ref.~\cite{Abe:2008an}, the compensator multiplet 
also has non-vanishing charges, 
which lead to the warping of the 5D spacetime. 

Then, after rescaling chiral multiplets by a factor~$e^{3\sgm/2}$, 
we obtain
\bea
 \cL \eql -3\int\dr^4\tht\;\cN^{1/3}(\cV)\left\{
 e^{-3k\cdot V}\abs{\Phi^1}^2+e^{3k\cdot V}\abs{\Phi^2}^2 \right. \nonumber\\
 &&\hspace{30mm} \left. 
 -\sum_{a=2}^{n_H+1}\brkt{e^{-2c_a\cdot V}\abs{\Phi^{2a-1}}^2
 +e^{2c_a\cdot V}\abs{\Phi^{2a}}^2}\right\}^{2/3} \nonumber\\
 &&-2\sbk{\int\dr^2\tht\;\brc{
 \Phi^1\brkt{\der_y+3k\cdot\Sgm}\Phi^2
 -\sum_{a=2}^{n_H+1}\Phi^{2a-1}\brkt{\der_y+2c_a\cdot\Sgm}\Phi^{2a}}+\hc} 
 \nonumber\\
 &&+\cL_{\rm vec}
 +2\sum_{y_*=0,L}\sbk{\int\dr^2\tht\;\brkt{\Phi^2}^2
 W^{(y_*)}\brkt{\hat{\Phi}^{2a}}+\hc}\dlt(y-y_*),  \label{5D_action2}
\eea
where $k\cdot V\equiv\sum_{I'}k_{I'}V^{I'}$, 
$c_a\cdot V\equiv\sum_{I'}c_{aI'}V^{I'}$, 
and $\hat{\Phi}^{2a}=\Phi^{2a}/\Phi^2$. 
For simplicity, we have introduced only superpotentials~$W^{(y_*)}$ 
in the boundary Lagrangians. 
The hypermultiplets appear in $W^{(y_*)}$ only through $\hat{\Phi}^{2a}$ 
because physical chiral multiplets must have zero Weyl weights 
in $N=1$ superconformal formulation~\cite{4Doffshell} and 
$\Phi^{2a-1}$ vanish at the boundaries due to their orbifold parities. 
The boundary Lagrangians can also depend on 
boundary-localized 4D superfields, which are not considered 
in this article. 
The power of $\Phi^2$ in front of $W^{(y_*)}$ is determined by
the requirement that the argument of the $F$-term action formula 
must have the Weyl weight 3. 
(The Weyl weight of $\Phi^{\hat{a}}$ is $3/2$.)

\section{4D effective theory and its properties} \label{4Deffective}
\subsection{4D effective Lagrangian}
Following the off-shell dimensional reduction developed in Ref.~\cite{Abe:2007}, 
we can derive the 4D effective action, keeping the $N=1$ off-shell structure. 
A detailed derivation is summarized in Appendix~\ref{derive:L_eff}. 
The result is 
\bea
 \cL_{\rm eff} \eql -\frac{1}{4}\sbk{\int\dr^2\tht\;
 \sum_r f_{\rm eff}^r(T)\tr\brkt{\cW^r\cW^r}+\hc} \nonumber\\
 &&+\int\dr^4\tht\;\abs{\phi}^2\Omg_{\rm eff}\brkt{\abs{Q}^2,\Re T} 
 +\sbk{\int\dr^2\tht\;\phi^3 W_{\rm eff}(Q,T)+\hc}, 
 \label{L_eff}
\eea
where the gauge multiplets are summarized in the matrix forms 
for the non-abelian gauge groups, 
the index~$r$ indicates the gauge sectors, and  
$\cW^r$ is the field strength supermultiplet for 
a massless 4D vector multiplet~$V^r$. 
$Q_a$ ($a\geq 2$) and $T^{I'}$ are the zero-modes 
for $\hat{\Phi}^{2a}$ and $\Sgm^{I'}$, respectively. 
We have used the same symbols for the zero-modes~$V^r$ as 
the corresponding 5D multiplets. 
Each function in (\ref{L_eff}) is expressed as  
\bea
 f_{\rm eff}^r(T) \eql \sum_{I'}\xi^r_{I'}T^{I'}, \nonumber\\
 \Omg_{\rm eff}\brkt{\abs{Q}^2,\Re T}
 \defa -3e^{-K_{\rm eff}/3} \nonumber\\ 
 \eql \hat{\cN}^{1/3}(\Re T)\left[
 -3Y(k\cdot T)+2\sum_a Y((k+d_a)\cdot T)\abs{Q_a}^2 \right. \nonumber\\
 &&\hspace{12mm}\left.
 +\sum_{a,b}\tl{\Omg}^{(4)}_{a,b}(\Re T)\abs{Q_a}^2\abs{Q_b}^2
 +\cO\brkt{(k\cdot\cP)^2}+\cO\brkt{\abs{Q}^6} \right], \nonumber\\
 W_{\rm eff}(Q,T) \eql W^{(0)}(Q)+e^{-3k\cdot T} 
 W^{(L)}\brkt{e^{-d_a\cdot T}Q_a}, 
 \label{defining_fcns}
\eea
where $\xi_{I'}^r$ are real constants determined from $C_{I'J''K''}$, 
$d_{aI'}\equiv c_{aI'}-\frac{3}{2}k_{I'}$, and  
\be
 Y(z) \equiv \frac{1-e^{-2\Re z}}{2\Re z}. \label{def:Y}
\ee
The functions~$\tl{\Omg}^{(4)}_{a,b}$ are defined as 
\be
 \tl{\Omg}^{(4)}_{a,b} \equiv -\frac{\brkt{d_a\cdot\cP a^{-1}\cdot d_b}
 \brc{Y((k+d_a+d_b)\cdot T)-\frac{Y(d_a\cdot T)Y(d_b\cdot T)}{Y(-k\cdot T)}}}
 {\brc{(k+d_a)\cdot\Re T}\brc{(k+d_b)\cdot\Re T}}
 +\frac{Y((k+d_a+d_b)\cdot T)}{3}.  \label{tlOmg4}
\ee
In the derivation of $\Omg_{\rm eff}$ summarized 
in Appendix~(\ref{derive:K_eff}), we have assumed that 
\be
 k_{I'}\cP^{I'}_{\;\;J'}(\Re T) = 0,  \label{moduli_align}
\ee
in order to obtain an analytic expression. 
Thus we focus on a case that the moduli VEVs are (at least approximately) 
aligned to satisfy (\ref{moduli_align}) by some mechanism. 
When the number of the moduli is two, the above effective Lagrangian 
reduces to that in Ref.~\cite{Abe:2008an} 
in the limit of $k_{I'}\to 0$.\footnote{
There are typos in Ref.~\cite{Abe:2008an}. 
The indices of the derivatives of the norm functions should be replaced as 
$\hat{\cN}_1\exch\hat{\cN}_2$ and $\hat{\cN}_{11}\exch\hat{\cN}_{22}$ 
in (2.15) and Sec.3.1 of Ref.~\cite{Abe:2008an}. }
The constraint~(\ref{moduli_align}) disappears in this limit.

\subsection{Superconformal gauge-fixing and mass dimension}
Here we mention the superconformal gauge-fixing in 4D SUGRA. 
Since the effective action~(\ref{L_eff}) 
(with the promotion~(\ref{FDpromote})) 
has the $N=1$ superconformal symmetries, 
we have to impose the gauge-fixing conditions 
in order to obtain the usual Poincar\'{e} SUGRA. 
The extra symmetries to eliminate are 
the dilatation~$\bdm{D}$, the $U(1)_A$ automorphism, 
the conformal SUSY~$\bdm{S}$, and the conformal boost~$\bdm{K}$. 
For a real general multiplet~$U=C+\tht\zeta+\bar{\tht}\bar{\zeta}+\cdots
+\frac{1}{2}\tht^2\bar{\tht}^2D$, 
the $D$-term action formula is written as~\cite{Kugo:2002js,4Doffshell} 
\be
 \sbk{U}_D = e^{(4)}\brc{D+\frac{1}{3}C\brkt{
 \cR^{(4)}-4\ep^{\mu\nu\rho\tau}\bar{\psi}_\mu\bar{\sgm}_\nu
 \der_\rho\psi_\tau}
 +\frac{4i}{3}\zeta\sgm^{\mu\nu}\der_\mu\psi_\nu+\cdots}, 
\ee
where $e^{(4)}$ is the determinant of the vierbein, 
$\cR^{(4)}$ is the 4D Ricci scalar, 
and $\sgm^{\mu\nu}\equiv\frac{1}{4}\brkt{\sgm^\mu\bar{\sgm}^\nu
-\sgm^\nu\bar{\sgm}^\mu}$.\footnote{
We follow the spinor notation of Ref.~\cite{Wess:1992cp}. }
We omitted the spacetime integral. 
A conventional choice of the gauge-fixing is as follows. 
The $\bdm{D}$-gauge is fixed so that the Einstein-Hilbert term is realized, 
and the $\bdm{S}$-gauge is fixed so that the kinetic mixing between 
the matter fermions and the gravitino is absent. 
The corresponding gauge-fixing conditions are given by  
\be
 \left.\brkt{\abs{\phi}^2\Omg_{\rm eff}}\right|_0 = -3M_{\rm Pl}^2, \;\;\;\;\;
 \left.\brkt{\abs{\phi}^2\Omg_{\rm eff}}\right|_\tht = 0.  \label{GF_cond:4D}
\ee
The $U(1)_A$-symmetry is eliminated by fixing the phase of $\phi|_0$ to zero. 
The $\bdm{K}$-gauge fixing condition is irrelevant to the current discussion. 
\ignore{
Namely, 
\bea
 \phi|_0 \eql M_{\rm Pl}\brc{\hat{\cN}^{1/3}(\Re T)Y(k\cdot T)}^{-1/2}
 +\cO(\abs{Q_a}^2), 
 \nonumber\\
 \phi|_\tht \eql 
\eea
}
The resultant Lagrangian in the gravitational sector is obtained as   
\be
 \cL_{\rm gravi} = -\frac{M_{\rm Pl}^2}{2}e^{(4)}\brkt{\cR^{(4)}
 -4\ep^{\mu\nu\rho\tau}\bar{\psi}_\mu\bar{\sgm}_\tau\der_\nu\psi_\rho}
 +\cdots.  \label{L_gravi}
\ee 

We comment on the relations to the 5D gauge-fixing we chose 
in (\ref{GF_cond}). 
From the definitions of $T^{I'}$ and $\phi$ in Appendix~\ref{derive:fW}, 
the $\bdm{D}$-gauge fixing in (\ref{GF_cond}) corresponds to 
the following conditions in 4D effective theory. 
\bea
 \hat{\cN}^{1/3}(\Re\vev{T}) \eql \int_0^L\dr y\;\cN^{1/3}(2\Re\vev{\Sgm}) 
 = M_5L, \nonumber\\
 \vev{\phi} \eql \brkt{\vev{\Phi^2}}^{2/3} \simeq M_5. \label{rel:GF_cond}
\eea
We have assumed that VEVs of all the physical hypermultiplets 
are much smaller than $M_{\rm Pl}^{3/2}$, 
and used the facts that $\vev{\Sgm^{I''}}=0$ and 
all $\vev{\Sgm^{I'}}$ have the same $y$-dependence 
under the condition~(\ref{moduli_align}). 
Namely we can read off the 5D scales~$M_5$ and $L^{-1}$ 
from (\ref{rel:GF_cond}). 

Before the gauge fixing, all quantities in the action do not have 
the mass dimension. 
It can be defined after the $\bdm{D}$-gauge fixing 
that introduces the mass scale into the theory. 
Since $\bdm{D}$ corresponds to the scale transformation, 
the mass dimension seems to be identified with 
the $\bdm{D}$ charge, \ie, the Weyl weight. 
However, they are completely different as shown in Table~\ref{mass_dimension}. 
\begin{table}[t]
\begin{center}
\begin{tabular}{c||c|c|c|c|c|c|c} 
\rule[-2mm]{0mm}{7mm}  & 
$x^\mu$ & $\tht^\alp$ & $e_\mu^{\;\;\underline{\nu}}$ 
& $\psi_\mu^\alp$ & $Q_a$ & $T^{I'}$ & $V^{I''}$  \\ \hline
\rule[-2mm]{0mm}{7mm} Weyl weight & 
0 & $(-1/2)$ & $-1$ & $-1/2$ & 0 & 0 & 0  \\ \hline
\rule[-2mm]{0mm}{7mm} mass dimension & 
$-1$ & $-1/2$ & 0 & 1/2 & 1 & $-1$ & 0 
\\ \hline
\end{tabular}
\end{center}
\caption{The Weyl weights and the mass dimensions of 
the coordinates and fields in 4D effective theory.  }
\label{mass_dimension}
\end{table}
For instance, the former assigns a nonzero value to the coordinate~$x^\mu$ 
while the latter does not. 
The mass dimensions of the gravitational fields are determined from 
(\ref{L_gravi}). 
As for the chiral and vector multiplets, the numbers in the table 
denote those for the lowest components. 
They increase for higher components by $1/2$. 
The mass dimension of the moduli~$T^{I'}$ is determined 
so that their VEVs have a dimension of length. 
Note that the gauge-fixing conditions break the $N=1$ off-shell structure 
and the theory cannot be expressed in terms of superfields any longer. 
In order to express the action in terms of the component fields 
that have the mass dimensions listed in Table~\ref{mass_dimension}, 
we rescale each quantity as 
\bea
 x^\mu & \to &  M_{\rm Pl}x^\mu, \nonumber\\
 \brkt{e_\mu^{\;\;\underline{\nu}}, \; \psi_\mu^\alp} & \to & 
 \frac{1}{M_{\rm Pl}}\brkt{e_\mu^{\;\;\underline{\nu}}, \; 
 \psi_\mu^\alp}, \nonumber\\
 \brkt{Q_a, \; T^{I'}, \; V^{I''}} & \to & 
 \brkt{\frac{Q_a}{M_{\rm Pl}}, \; M_{\rm Pl}T^{I'}, \; V^{I''}}. \nonumber\\
\eea
Then the coupling constants~$k_{I'}$ and $c_{I'}$ are accompanied 
with $M_{\rm Pl}$ in the rescaled action while $g_{I''}$ are not. 
So we also rescale these constants as 
\be
 M_{\rm Pl}k_{I'}, \;\; M_{\rm Pl}c_{I'}, \;\; g_{I''} \;\;
 \to \;\; k_{I'}, \;\; c_{I'}, \;\; g_{I''}. 
\ee
Hence the moduli couplings~$k_{I'}$ and $c_{I'}$ are regarded as 
mass parameters while the gauge couplings~$g_{I''}$ are dimensionless constants. 
After this procedure, 
$M_{\rm Pl}$ appears only in (\ref{L_gravi}) and the gravitational interactions. 

\ignore{
\begin{enumerate}
\item  Rescale the coordinates as 
 \be
  \brkt{x^\mu, \;\; \tht^\alp} \to 
  \brkt{M_{\rm Pl}x^\mu, \;\; M_{\rm Pl}^{1/2}\tht^\alp}. 
  \label{rescale:cd}
 \ee
\item Rescale the superfields as 
 \be
  \brkt{\phi, \;\; Q_a, \;\; T^{I'}, \;\; V^{I''}} \to 
  \brkt{M_{\rm Pl}\phi, \;\; \frac{Q_a}{M_{\rm Pl}}, 
  \;\; M_{\rm Pl}T^{I'}, \;\; V^{I''}},  
  \label{rescale:fd}
 \ee
 so that the mass dimensions of $\phi$, $Q_a$, $T^{I'}$ and $V^{I''}$ 
 become 0, 1, $-1$, 0, respectively.
\item Rescale the coupling constants as 
 \be
  \brkt{k_{I'}, \;\; c_{I'}, \;\; g_{I''}}  
  \to \brkt{\frac{k_{I'}}{M_{\rm Pl}}, \;\; \frac{c_{I'}}{M_{\rm Pl}}, 
  \;\; g_{I''}}.    \label{rescale:cp}
 \ee
 Then, the coupling constants for $\bdm{V}^{I'}$, $k_{I'}$ and $c_{I'}$, 
 are regarded as the mass parameters 
 while those for $\bdm{V}^{I''}$, $g_{I''}$, 
 are the usual dimensionless gauge coupling constants. 
\end{enumerate}
Each component in a superfield is also rescaled in order to 
absorb $M_{\rm Pl}$ induced by (\ref{rescale:cd}). 
After this procedure, 
$M_{\rm Pl}$ appears only in the gravitational interactions. 
}

\subsection{Moduli kinetic terms}
Since the moduli VEVs satisfy (\ref{moduli_align}), 
the combination~$k\cdot T$ is expressed as 
\be
 k\cdot T = \kp T_{\rm rad}, 
\ee
where 
\be
 \kp \equiv \frac{1}{L}\brkt{k\cdot\Re\vev{T}}, \;\;\;\;\;
 T_{\rm rad} \equiv 
 L\frac{\hat{\cN}_{I'}}{3\hat{\cN}}(\Re\vev{T})T^{I'}.  
 \label{def:Trad}
\ee
We have determined the coefficient so that $\Re\vev{T_{\rm rad}}=L$. 
Note that $T_{\rm rad}$ is identified with the zero-mode 
of the 5D radion multiplet~$\cT$ defined in (\ref{def:cT}). 
Since $\hat{\cN}^{1/3}(\Re T)$ is expanded around $T=\vev{T}$ as 
\bea
 \hat{\cN}^{1/3}(\Re(\vev{T}+\dlt T)) 
 \eql \hat{\cN}^{1/3}(\Re\vev{T})\brc{
 1+\frac{\hat{\cN}_{I'}}{3\hat{\cN}}(\Re\vev{T})
 \Re\dlt T^{I'}+\cO(\dlt T^2)} 
 \nonumber\\
 \eql \frac{1}{L}\hat{\cN}^{1/3}(\Re\vev{T})\Re T_{\rm rad}+\cO(\dlt T^2), 
\eea
$\Omg_{\rm eff}$ in (\ref{defining_fcns}) becomes 
\bea
 \Omg_{\rm eff} \eql -\frac{3}{L}\hat{\cN}^{1/3}(\Re\vev{T})
 \frac{1-e^{-2\kp\Re T_{\rm rad}}}{2\kp}+\cO(\dlt T^2)+\cdots,   
\eea
where the ellipsis denotes terms involving other multiplets 
than $T_{\rm rad}$. 
This is the kinetic term of the radion multiplet 
in the Randall-Sundrum spacetime~\cite{Luty:2000ec}. 
Thus, the alignment of moduli VEVs in (\ref{moduli_align}) 
is interpreted as the condition 
for the spacetime to be the Randall-Sundrum spacetime, 
and $\kp$ defined in (\ref{def:Trad}) is identified with 
the AdS curvature scale that is related to the bulk cosmological constant. 
After the gauge-fixing~(\ref{GF_cond:4D}), we obtain 
\be
 M_{\rm Pl}^2 = -\frac{1}{3}\vev{\abs{\phi}^2\Omg_{\rm eff}} 
 \simeq -\frac{1}{3}M_5^2\cdot
 \brkt{-\frac{3}{L}}\cdot M_5L \cdot \frac{1-e^{-2\kp L}}{2\kp}
 = \frac{M_5^3\brkt{1-e^{-2\kp L}}}{2\kp}.  \label{rel:Mpl}
\ee
We have used (\ref{rel:GF_cond}). 
Eq.(\ref{rel:Mpl}) is the well-known relation 
in the Randall-Sundrum spacetime~\cite{RS}. 

The other moduli are orthogonal to $T_{\rm rad}$ in the moduli space, 
When there exist $n_V$ moduli, they are parametrized 
by the coordinate system~$\brc{\vph^i}$ 
($i=1,\cdots,n_V-1$) on the $(n_V-1)$-dimensional submanifold determined 
by~\footnote{
One choice of $\brc{\vph^i}$ is $\vph^i\equiv T^{I'=i}-\vev{T^{I'=i}}$.  
In this case, $T^{n_V}$ becomes a function of $\vph^i$ through (\ref{VSmanifold}). 
} 
\be
 \hat{\cN}(\Re T(\vph))=\hat{\cN}(\Re\vev{T}). 
 \label{VSmanifold}
\ee
Since $\hat{\cN}_{I'}(\Re T)\frac{\der T^{I'}}{\der\vph^i}=0$, 
$(k\cdot\cP)_{I'}$ is expressed in terms of $\vph^i$ as 
\be
 \brkt{k\cdot\cP}_{I'} = -\brc{\frac{(k\cdot\Re T)\hat{\cN}_{I'J'}}
 {3\hat{\cN}}}_{\vph=0}
 \Re\brkt{\left.\frac{\der T^{J'}}{\der\vph^i}\right|_{\vph=0}\vph^i}
 +\cO(\vph^2). 
\ee
The kinetic terms for $\vph^i$ are contained in the $\cO((k\cdot\cP)^2)$-terms 
in (\ref{defining_fcns}), which start from~\footnote{
This can be calculated by the perturbative expansion of 
the first equation in (\ref{dg/dU})
in terms of $\abs{k}\equiv(\sum_{I'}k_{I'}^2)^{1/2}$. 
}
\be
 \dlt\Omg_{\rm eff} = -\frac{1}{4}\brc{\hat{\cN}^{1/3}
 \brkt{k\cdot\cP}_{I'} a^{I'J'}\brkt{k\cdot\cP}_{J'}}(\Re T)
 +\cO\brkt{k^4,\abs{Q}^2}. 
\ee

\subsection{Quadratic terms in $\bdm{\Omg_{\rm eff}}$ and Yukawa hierarchy} 
\label{Yukawa_hierarchy}
The coefficients of $\abs{Q_a}^2$ in $\Omg_{\rm eff}$ are important 
for generating the fermion mass hierarchy. 
The Yukawa couplings can be introduced only in the boundary actions 
due to the $N=2$ SUSY in the bulk. 
We assume that they are contained in $W^{(0)}$ at $y=0$, 
\be
 W^{(0)} = \sum_{a,b,c}\lmd_{abc}\hat{\Phi}^{2a}\hat{\Phi}^{2b}\hat{\Phi}^{2c}
 +\cdots, 
\ee
where $\lmd_{abc}$ are the holomorphic Yukawa coupling constants 
and are supposed to be of $\cO(1)$.  
Then the effective theory has the Yukawa couplings,  
\be
 W_{\rm eff} = \sum_{a,b,c}\lmd_{abc}Q_a Q_b Q_c+\cdots. 
\ee
The physical Yukawa couplings~$y_{abc}$ are obtained 
by the canonical normalization of the chiral superfields~$Q_a$, and we have
\be
 y_{abc} = \frac{\lmd_{abc}}{\sqrt{\vev{Y_aY_bY_c}}}, 
 \label{yabc}
\ee
where 
\be
 Y_a \equiv 2\hat{\cN}^{1/3}(\Re T)
 \brc{Y((k+d_a)\cdot T)+\tl{\Omg}_{a,X}^{(4)}(\Re T)\abs{X}^2+\cO(\abs{X}^4)}. 
 \label{def:Y_a}
\ee
The function~$Y(z)$ is always positive, 
and approximated as 
\be
 Y(z) \simeq \begin{cases}  \frac{1}{2\Re z},  & \Re z>0 \\ 
 \frac{1}{2\abs{\Re z}}\exp\brc{2\abs{\Re z}}. & \Re z<0 \end{cases}
\ee
From the 5D viewpoint, the wave function of $Q_a$ is localized 
toward $y=0$ ($y=L$) 
in the case that $(k+d_a)\cdot\vev{\Re T}$ is positive (negative). 
As we can see from (\ref{yabc}), $y_{abc}$ is of $\cO(1)$  
when all the relevant fields are localized toward $y=0$, 
while it is exponentially small when there is a field localized toward $y=L$ 
among them. 
This is the well-known split fermion mechanism~\cite{Arkani-Schmaltz}.

\subsection{Quartic couplings in $\bdm{\Omg_{\rm eff}}$ 
and soft SUSY-breaking masses}   \label{quartic_cp}
The coefficients of $\abs{Q_a}^2\abs{Q_b}^2$ have a peculiar form 
to 5D SUGRA. 
This type of terms are important because they lead to 
the soft SUSY-breaking masses 
for the sfermions when $Q_a$ and $Q_b$ are identified with 
the quark or lepton superfield and the SUSY-breaking superfield~$X$, 
respectively. 
Notice that the first term in (\ref{tlOmg4}) is absent 
in the single modulus case due to the projection operator~$\cP^{I'}_{\;\;J'}$. 
It is induced by integrating out the vector multiplets~$V^{I'}$, 
which are the $N=2$ partners of the moduli 
multiplets~$\Sgm^{I'}$~\cite{Abe:2008an}. 
\begin{figure}[t,b]
\centering \leavevmode
\includegraphics[width=70mm]{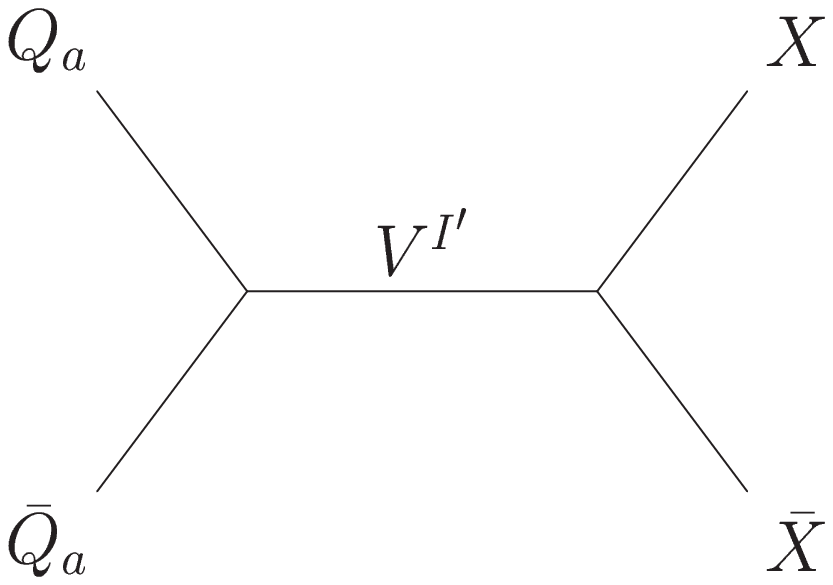}
\caption{Feynmann diagrams contributing to $\tl{\Omg}_{a,X}^{(4)}$ 
in the multi moduli case. 
}
\label{diagram}
\end{figure}
The relevant Feynmann diagrams are depicted in Fig.~\ref{diagram}.  
This can be seen from the fact that the coefficient 
of the first term in (\ref{tlOmg4}) involves the inverse matrix of $a_{I'J'}$, 
which comes from the propagator for $V^{I'}$. 
The existence of the projection operator~$\cP^{I'}_{\;\;J'}$ indicates that 
the graviphoton multiplet~$V_B$ defined in (\ref{def:V_B}) does not 
contribute to $\tl{\Omg}_{a,b}^{(4)}$. 
This can be understood from the fact that 
most of the components of $V_B$ are auxiliary fields, 
as mentioned in Sec.~\ref{sc_GF}. 

In the next section, we will consider a case that
the $F$-term of one chiral multiplet~$X$ in the effective theory 
provides the main source of SUSY breaking and $\vev{X}\simeq 0$.   
In such a case, the $\abs{Q_a}^2\abs{X}^2$-term contributes 
to the soft SUSY-breaking mass for the scalar component of $Q_a$, 
and it is expressed as 
(see (\ref{sp}) in Sec.~\ref{visible})
\be
 m_a^2 \simeq -\abs{F^X}^2\frac{\tl{\Omg}^{(4)}_{a,X}(\Re\vev{T})}
 {Y((k+d_a)\cdot\vev{T})}.  \label{ap_m_soft}
\ee

Let us first consider the single modulus case. 
In this case, $\tl{\Omg}^{(4)}_{a,X}$ is always positive 
because the first term in (\ref{tlOmg4}) is absent. 
Thus the soft scalar masses in (\ref{ap_m_soft}) become 
\be
 m_a^2 \simeq -\abs{F^X}^2\frac{Y((\kp+d_a+d_X)L)}
 {3Y((\kp+d_a)L)},  \label{m_soft:single}
\ee
and are found to be tachyonic. 
These tachyonic masses can be saved by quantum effects 
in some cases. 
The soft masses in (\ref{m_soft:single}) are exponentially suppressed 
when $\kp+d_a<0$ and $\kp+d_X>0$. 
This corresponds to a case that the matter~$Q_a$ is localized 
around $y=L$ while $X$ is around $y=0$. 
In such a case, quantum effects to the soft scalar masses become dominant 
and may lead to non-tachyonic masses. 
However the large top quark mass cannot be realized 
because the top Yukawa coupling is suppressed in that case. 
(Recall that the Yukawa couplings are localized at the $y=0$ boundary.)

This problem can be evaded in the multi moduli case. 
Let us consider a case that 
\be
 d_a\cdot\Re\vev{T} < -\kp L < 0 < d_X\cdot\Re\vev{T}. 
 \label{rel:dTs}
\ee
In this case, the $y=0$ boundary is identified with the UV brane, and 
$Q_a$ and $X$ are localized around the IR and UV branes respectively  
since $(k+d_a)\cdot\Re\vev{T}<0$ and $(k+d_X)\cdot\Re\vev{T}>0$. 
VEV of $\tl{\Omg}_{a,b}^{(4)}$ is approximately expressed as 
\bea
 \tl{\Omg}_{a,X}^{(4)}(\Re\vev{T}) \sma 
 \frac{d_a\cdot\cP a^{-1}\cdot d_X}
 {\brc{(k+d_a)\cdot\Re\vev{T}}\brc{(k+d_X)\cdot\Re\vev{T}}}
 \frac{Y(d_a\cdot\vev{T})Y(d_X\cdot\vev{T})}{Y(-\kp L)} \nonumber\\
 \sma \frac{d_a\cdot\cP a^{-1}\cdot d_X}
 {\brc{(k+d_a)\cdot\Re\vev{T}}\brc{(k+d_X)\cdot\Re\vev{T}}}\cdot
 \frac{-\kp L e^{-2(k+d_a)\cdot\Re\vev{T}}}
 {2\brkt{d_a\cdot\Re\vev{T}}\brkt{d_X\cdot\Re\vev{T}}}. \nonumber\\
 \label{ap_tlOmg4}
\eea
Therefore, (\ref{ap_m_soft}) becomes 
\be
 m_a^2 \simeq -\abs{F^X}^2\frac{(d_a\cdot\cP a^{-1}\cdot d_X)\kp L}
 {\brkt{d_a\cdot\Re\vev{T}}\brkt{d_X\cdot\Re\vev{T}}
 \brc{(k+d_X)\cdot\Re\vev{T}}}.  \label{ap:soft_mass}
\ee
The sign of $m_a^2$ now depends on 
the (truncated) norm function~$\hat{\cN}$ and the directions of the gauging 
for $V^{I'}$. 
In fact, we can always realize non-tachyonic soft masses 
for any choices of $\hat{\cN}$ 
by choosing the directions of the gauging such that 
\be
 \frac{d_a\cdot\cP a^{-1}\cdot d_X}
 {\brkt{d_a\cdot\Re\vev{T}}\brkt{d_X\cdot\Re\vev{T}}} < 0. 
 \label{cond:non-tachyonic}
\ee
Furthermore, if $n_V$-dimensional vectors~$\vec{d}_{a}$ 
point to the same direction, 
\be
 \vec{d}_a\propto\vec{n},  \label{cond:d_a}
\ee
the soft masses~$m_a^2$ become independent of the ``flavor index''~$a$. 
(The direction~$\vec{n}$ must not be parallel to $\vec{k}$, 
otherwise all $m_a^2$ vanish.)
This opens up the possibility to solve the SUSY flavor problem. 
We will discuss this issue in the next section. 

\ignore{
Finally we provide an interpretation of the above flavor universality 
from the 5D viewpoint. 
Recall that the dominant part in $\tl{\Omg}_{aX}^{(4)}$ 
under the condition~(\ref{rel:dts}) originates from the diagram 
in Fig.~\ref{diagram}. 
The internal line represents the $n$-th KK modes of $V^{I'}$, 
which are integrated out. 
The effective couplings~$g_a^{(n)}$ and $g_X^{(n)}$ are defined as 
}
A similar result is also obtained in a case that 
\be
 d_a\cdot\Re\vev{T} < 0 < -\kp L < d_X\cdot\Re\vev{T}. 
\ee
Conditions for obtaining non-tachyonic (and flavor-universal) soft masses 
are the same as (\ref{cond:non-tachyonic}) (and (\ref{cond:d_a})). 
In this case, however, the $y=0$ boundary becomes the IR brane, 
and the approximate expressions of the soft masses are suppressed 
from (\ref{ap:soft_mass}) by a factor~$e^{-2\kp L}\gg 1$.\footnote{ 
The typical SUSY-breaking mass scale~$M_{\rm SB}$ is also suppressed 
by the same factor in this case. 
So a ratio~$m_a/M_{\rm SB}$ is not suppressed. 
(See Sec.~\ref{visible}.)}

\section{Illustrative model} \label{specific_model}
Now we specify the model that realizes the set-up mentioned in Sec.~\ref{setup}, 
and show some phenomenological analysis. 

\subsection{Hidden (mediation) sector contents} \label{hidden}
In this paper we assume that a single chiral multiplet~$X$ originating 
from a 5D hypermultiplet is responsible for the spontaneous SUSY breaking. 
We do not specify the potential of the hidden sector $X$ and moduli $T^{I'}$ 
chiral multiplets. There are various ways to stabilize the moduli, 
including the size of the extra dimension. 
In general, the mechanism for the moduli stabilization determines 
the $F$-terms of $T^{I'}$, and thus affects the mediation of SUSY breaking 
to the MSSM sector. 
Here we do not specify the moduli stabilization and SUSY breaking mechanism, 
and treat (VEVs of) $F^X$ and $F^{T^{I'}}$ as free parameters,\footnote{
We do not consider the $D$-term SUSY breaking in this article.} 
while VEV of (the lowest component of) $X$ is assumed to be almost 
vanishing $\langle X \rangle \ll \langle T^{I'} \rangle \simeq {\cal O}(1)$ 
in the 4D Planck mass unit.\footnote{
Concrete moduli stabilization and SUSY breaking mechanisms were studied 
in our previous work~\cite{Abe:2008an} based on Ref.~\cite{Abe:2006xp}, 
which are also applicable here.} 

Instead of the non-vanishing $F$-terms, $F^{T^{I'}}$ ($I'=1,2,3$) and $F^X$, 
we use ratios of them~$\alp_{I'}$ and a typical scale of SUSY breaking~$M_{\rm SB}$ 
in order to parametrize the soft SUSY breaking parameters. 
They are defined as 
\bea
 \alp_{I'} \defa \frac{\cF^{T^{I'}}}{\cF^X}, \;\;\;\;\;
 M_{\rm SB} \equiv \left|\cF^X\right|,  \nonumber\\
 \cF^A \defa E^A_{\;\;B}F^B,  \label{def:M_SB}
\eea
where $E^A_{\;\;B}$ is the vielbein\footnote{
We use the same index for the flat and curved coordinates 
on the K\"ahler manifold to save characters. } 
for the K\"{a}hler metric~$K_{A\bar{B}}\equiv\der_A\der_{\bar{B}}K_{\rm eff}$.
\ignore{
\be
 \alp_{I'} \equiv \frac{\sqrt{K_{I'\bar{I'}}}
 \left|F^{T^{I'}}\right|}
 {\sqrt{K_{X\bar{X}}}\left|F^X\right|}, \;\;\;\;\;
 M_{\rm SB} \equiv \sqrt{K_{X\bar{X}}}\left|F^X\right|, 
 \label{def:M_SB}
\ee
where $K_{I'\bar{I}'}$ and $K_{X\bar{X}}$ are the diagonal components 
of the K\"ahler metric shown in (\ref{K-metric}). 
}
Then the vacuum value of the scalar potential is written as 
\bea
 V \eql \sum_A\left|\cF^A\right|^2-3e^{K} \left| W \right|^2 \nonumber\\
 \eql \left( 1+\sum_{I'=1}^3\abs{\alpha_{I'}}^2 \right) M_{\rm SB}^2 -3m_{3/2}^2, 
\eea
where $m_{3/2}= \langle e^{K/2}|W| \rangle$ is the gravitino mass. 
The moduli stabilization at a SUSY breaking Minkowski minimum 
$\langle V \rangle \simeq 0$ required by the observation 
leads to a relation 
\begin{eqnarray}
m_{3/2} &\simeq& \sqrt{ \frac{1}{3}
\left( 1+\sum_{I'=1}^3\abs{\alpha_{I'}}^2 \right)}\,M_{\rm SB}. 
\end{eqnarray}

The ratios~$\alp_{I'}$ parameterize contributions 
of ``the moduli mediation'' induced by $F^{T^{I'}}$ 
compared to that of ``the direct mediation'' induced by $F^X$. 
In the following analysis we mainly consider cases that 
\begin{eqnarray}
 \abs{\alpha_{I'}} \ = \ {\cal O}(1) 
 \ &\textrm{or}& \ {\cal O} \left( 1/(4 \pi^2) \right). 
\end{eqnarray}
The latter values can be realized by the numerical value of 
$1/\ln (M_{\rm Pl}/{\rm TeV})$ if the moduli $T^{I'}$ 
are stabilized by nonperturbative effects like gaugino 
condensations at a SUSY anti-de Sitter vacuum, which is 
uplifted to the almost Minkowski minimum by the vacuum energy of 
the (TeV scale) SUSY breaking sector~\cite{Choi:2004sx,Choi:2005ge}. 
We can identify $F^X$ with a source of the uplifting vacuum energy 
based on a scenario of F-term uplifting~\cite{Abe:2006xp,Dudas:2006gr}. 

For the phenomenological analysis, 
we consider a case with three $Z_2$-odd $U(1)_{I'}$ vector multiplets 
$\bdm{V}^{I'}$ in 5D, where $I'=1,2,3$. 
They generate three moduli chiral multiplets 
\be
T^{I'}=(T^1, T^2, T^3), 
\ee
in the 4D effective theory 
and we choose the (truncated) norm function as 
\be
 \hat{\cN}(\Re T) = (\Re T^1)(\Re T^2)(\Re T^3). 
\label{nft1t2t3}
\ee
Then the matrix~$a_{I'J'}$ defined in (\ref{def:a_IJ}) becomes diagonal. 
Explicit forms of the K\"ahler metric~$K_{A\bar{B}}$ and $a_{I'J'}$ are 
shown in Appendix~\ref{explicit_forms}. 
In the flat case ($\kp L=0$), the K\"ahler metric also becomes diagonal. 

For concreteness, we further assume the moduli VEVs as 
\be
 \brkt{\Re\vev{T^1},\Re\vev{T^2},\Re\vev{T^3}} = (4,4,1). \;\;\;\;\;
 \mbox{(in the $M_{\rm Pl}$ unit)}  \label{VEV:assumption}
\ee
Then the condition to realize the Randall-Sundrum spacetime~(\ref{moduli_align})
fixes the gauging direction of the compensator multiplets as  
\be
 (k_1,k_2,k_3) = \brkt{\frac{1}{4},\frac{1}{4},1}k_3. 
 \label{cpst_gauging}
\ee 
The value of $k_3$ is determined by the warp factor 
through $\kp L=k\cdot\Re\vev{T}=3k_3$.

\subsection{Visible sector contents and soft SUSY-breaking parameters} 
\label{visible}
We assume that the visible sector consists of 
the following MSSM matter contents: 
\bea
 (V_1,V_2,V_3) &:& \mbox{gauge vector multiplets}, \nonumber\\
 (\cQ_i,\cU_i,\cD_i) &:& \mbox{quark chiral multiplets}, \nonumber\\
 (\cL_i,\cE_i) &:& \mbox{lepton chiral multiplets}, \nonumber\\
 (\cH_u,\cH_d) &:& \mbox{Higgs chiral multiplets},
\eea
where $V_1$, $V_2$, $V_3$ denote 
the gauge multiplets for $U(1)_Y,SU(2)_L,SU(3)_C$ 
originating from $\bdm{V}^{I''}$, 
and the others are chiral multiplets from 5D hypermultiplets. 
The index~$i=1,2,3$ denotes the generation. 

We also assume an approximate global~$U(1)_R$-symmetry that is responsible 
for the dynamical SUSY breaking~\cite{Nelson:1993nf}. 
We assign the R-charge as shown in Table~\ref{Rcharge}. 
\begin{table}[t]
\begin{center}
\begin{tabular}{c||c|c|c} \hline
\rule[-2mm]{0mm}{7mm} 
Multiplet & $\cQ_i$, $\cU_i$, $\cD_i$, $\cL_i$, $\cE_i$ & 
$\cH_u$, $\cH_d$ & $X$ \\ \hline
\rule[-2mm]{0mm}{7mm} R-charge & 1/2 & 1 & 2 \\ \hline
\end{tabular}
\end{center}
\caption{$U(1)_R$-charges of the chiral multiplets. }
\label{Rcharge}
\end{table}
It is supposed to be broken by the nonperturbative effects. 
Then, the holomorphic Yukawa couplings and the $\mu$-term 
in the boundary superpotentials as well as the gauge kinetic functions 
are independent of $X$. 
We further assume that these terms exist only 
at the $y=0$ boundary.
The resulting gauge kinetic functions and the superpotential 
for the visible sector in the 4D effective theory are parametrized as 
\bea
 f^r_{\rm eff}(T) \eql \sum_{I'}\xi^r_{I'}T^{I'}, \nonumber\\
 W_{\rm MSSM} \eql \zeta_0\cH_u\cH_d+\lmd_{ij}^u\cH_u\cQ_i\cU_j
 +\lmd_{ij}^d\cH_d\cQ_i\cD_j+\lmd_{ij}^e\cH_d\cL_i\cE_j, 
\label{wmssm}
\eea
where $r=1,2,3$ for $U(1)_Y$, $SU(2)_L$, $SU(3)_C$, 
and $\xi^r_{I'}$ are real constants determined by 
the coefficients $C_{I'J''K''}$ in the norm function ${\cal N}$, 
while  $\zeta_0$ and $\lmd_{ij}^{u,d,e}$ are in general complex constants. 
After the canonical normalization, the $\mu$ parameter is expressed as 
\be
 \mu = \Lvev{\frac{\zeta_0}{\sqrt{\hat{\cN}^{1/3}(\Re T) Y(k\cdot T)
 Y_{\cH_u}Y_{\cH_d}}}} 
 = \Lvev{\frac{\zeta_0}{\sqrt{2Y(\kp L)Y_{\cH_u}Y_{\cH_d}}}}, 
 \label{expr:mu}
\ee
where $Y_a$ is defined in (\ref{def:Y_a}), 
and we have used (\ref{VEV:assumption}) at the second equality. 
The physical Yukawa couplings are expressed as 
\be
 y_{ij}^u = \frac{\lmd_{ij}^u}
 {\sqrt{\vev{Y_{\cH_u}Y_{\cQ_i}Y_{\cU_j}}}}, \;\;\;\;\;
 y_{ij}^d = \frac{\lmd_{ij}^d}
 {\sqrt{\vev{Y_{\cH_d}Y_{\cQ_i}Y_{\cD_j}}}}, \;\;\;\;\;
 y_{ij}^e = \frac{\lmd_{ij}^e}
 {\sqrt{\vev{Y_{\cH_d}Y_{\cL_i}Y_{\cE_j}}}}. 
\ee
The holomorphic Yukawa couplings~$\lmd_{ij}^x$ ($x=u,d,e$) 
are assumed to be of $\cO(1)$. 
The hierarchical structure of the Yukawa couplings are obtained 
by choosing the moduli couplings~$c_{a I'}$ 
as explained in Sec.~\ref{Yukawa_hierarchy}. 

The soft SUSY-breaking parameters in the MSSM are defined by 
\bea
 \cL_{\rm soft} \eql -\sum_{Q_a}m_{Q_a}^2\abs{Q_a}^2 
 -\frac{1}{2}\sum_r M_r\tr\brkt{\lmd^r\lmd^r} \\
 &&-\sum_{i,j}\brc{B\mu H_u H_d+y_{i,j}^u A_{i,j}^u H_u \tl{q}_i \tl{u}_j
 +y_{i,j}^d A_{i,j}^d H_d\tl{q}_i\tl{d}_j
 +y_{i,j}^e A_{i,j}^e H_d\tl{l}_i\tl{e}_j+\hc}, \nonumber
\eea
where $\lmd^r$ ($r=1,2,3$) and $Q_a=H_{u,d}$, $\tl{q}_i$, $\tl{u}_i$, $\tl{d}_i$, 
$\tl{l}_i$, $\tl{e}_i$ are the gauginos and 
the scalar components of $\cH_{u,d}$, $\cQ_i$, $\cU_i$, $\cD_i$, 
$\cL_i$, $\cE_i$, respectively. 
The fields are all {\it canonically normalized}. 
These parameters are induced 
by the formulae~\cite{Choi:2005ge,Kaplunovsky:1993rd}, 
\bea
 M_r \eql \vev{F^A\der_A\ln\brkt{\Re f^r_{\rm eff}}}, \;\;\;\;\;
 m_{Q_a}^2 = 
 -\vev{F^A\bar{F}^{\bar{B}}\der_A\der_{\bar{B}}\ln Y_{Q_a}}, \nonumber\\
 B \eql -\Lvev{F^A\der_A\ln\brkt{\frac{\zeta_0}{\hat{\cN}^{1/3}(\Re T)
 Y(k\cdot T)Y_{\cH_u}Y_{\cH_d}}}}-m_{3/2}e^{i\vph}, \nonumber\\
 A_{ij}^u \eql \vev{F^A\der_A\ln\brkt{Y_{\cH_u}Y_{\cQ_i}Y_{\cU_j}}}, \;\;\;\;\;
 A_{ij}^d = \vev{F^A\der_A\ln\brkt{Y_{\cH_d}Y_{\cQ_i}Y_{\cD_j}}}, \nonumber\\
 A_{ij}^e \eql \vev{F^A\der_A\ln\brkt{Y_{\cH_d}Y_{\cL_i}Y_{\cE_j}}}, 
\label{sp}
\eea
where $A,B=X,T^1,T^2,T^3$, and $\vph\equiv\arg\vev{W}$.  

\begin{figure}[t]
\centering \leavevmode
\begin{minipage}{0.3\linewidth}
\begin{center}
\includegraphics[width=\linewidth]{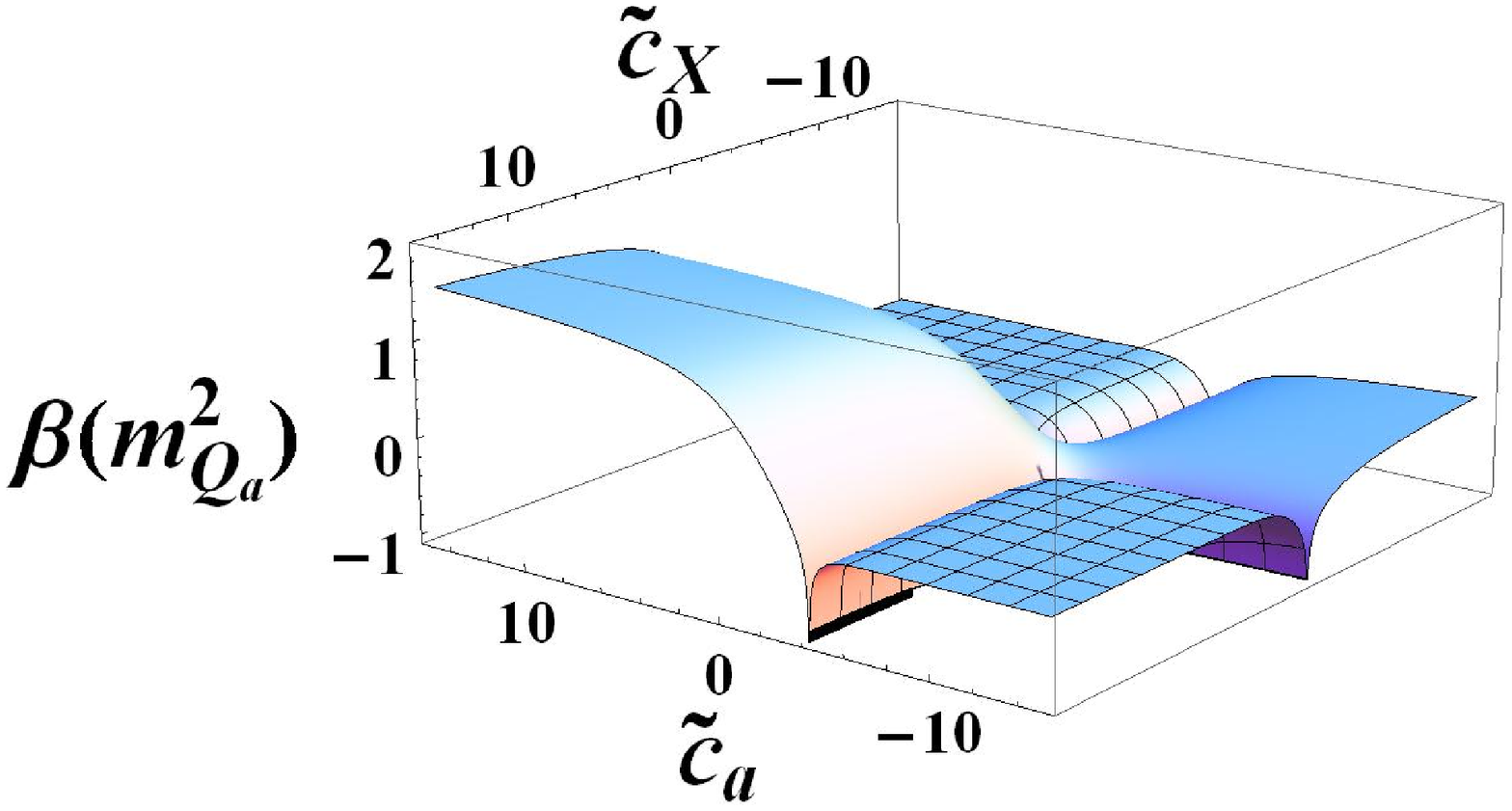}
(a) $\kp L = -3.6$
\end{center}
\end{minipage}
\hfill 
\begin{minipage}{0.3\linewidth}
\begin{center}
\includegraphics[width=\linewidth]{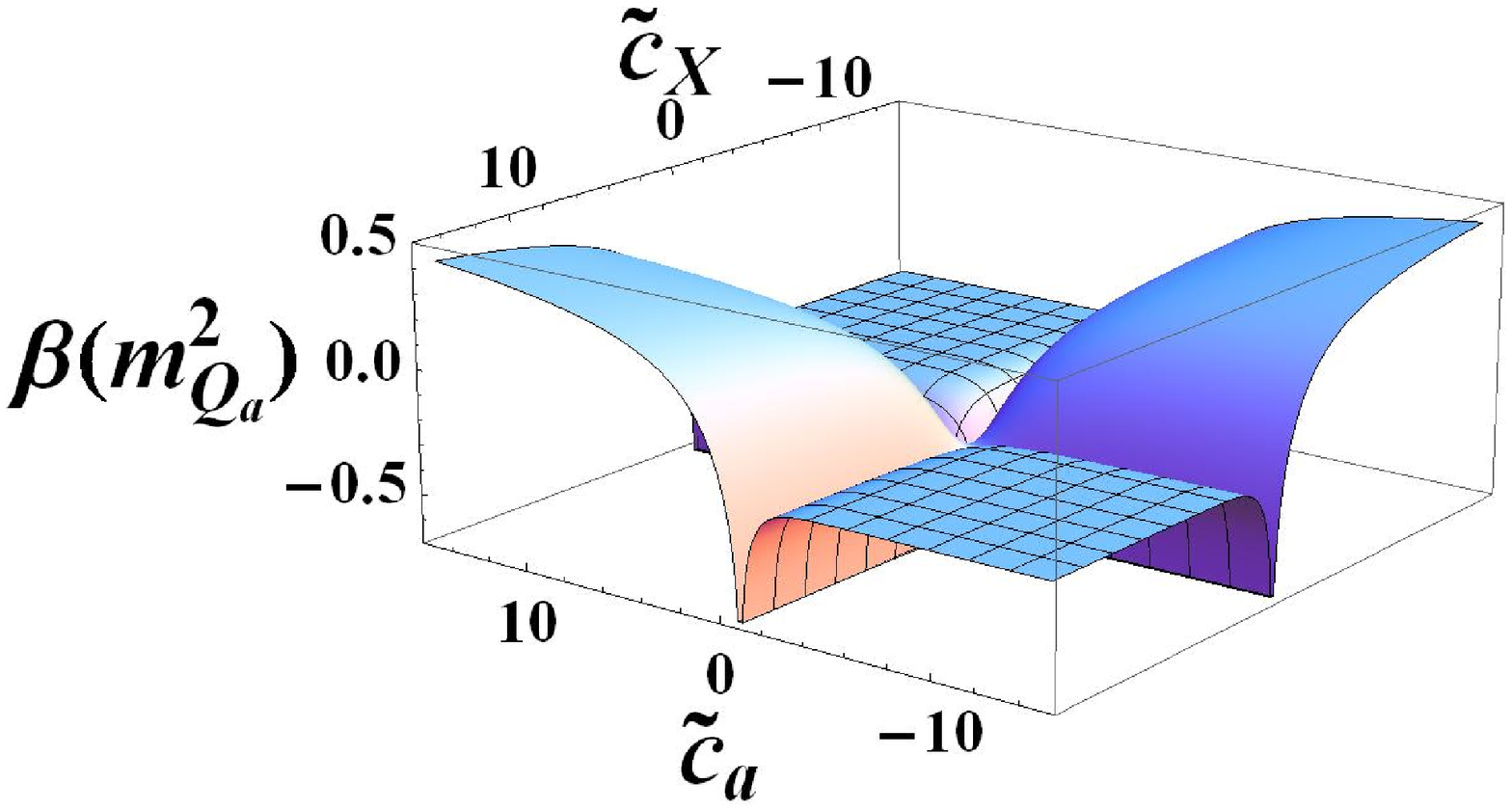}
(b) $\kp L = 0$
\end{center}
\end{minipage}
\hfill 
\begin{minipage}{0.3\linewidth}
\begin{center}
\includegraphics[width=\linewidth]{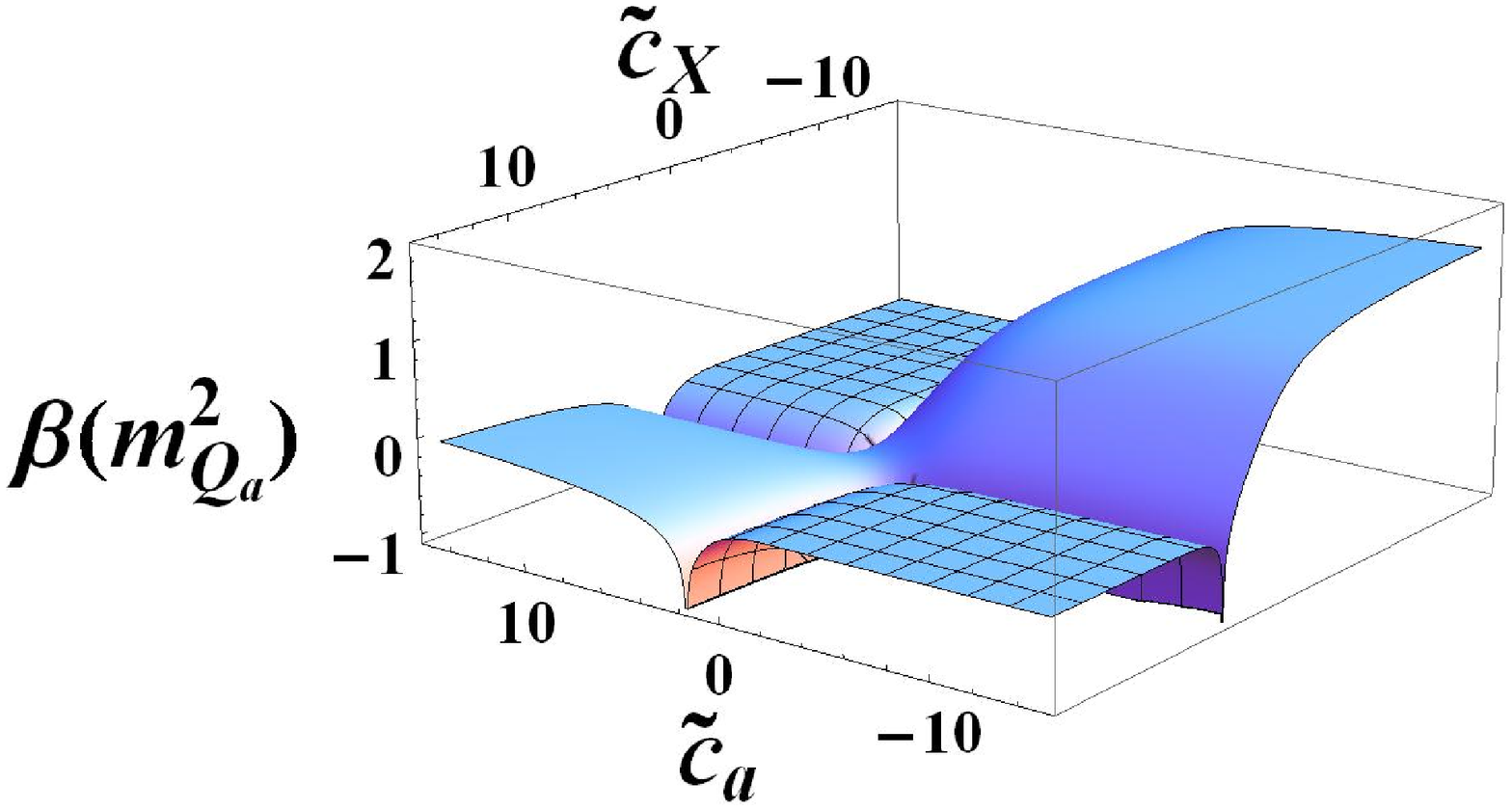}
(c) $\kp L = 3.6$
\end{center}
\end{minipage}
\caption{
The charge dependences of $\bt(m_{Q_a}^2)$ defined in 
Eq.(\ref{hatmqa}) with the norm function (\ref{nft1t2t3}) 
and $\alpha_1=\alp_3=1/(4\pi^2)$, $\alp_2=2/(4\pi^2)$. 
The charge assignment for $Q_a$ and $X$ is chosen as 
$(\tl{c}_{Q_a}^1,\tl{c}_{Q_a}^2,\tl{c}_{Q_a}^3)
=(\frac{\kp L}{3},\tl{c}_a-\frac{2\kp L}{3},\frac{\kp L}{3})$ 
and $(\tl{c}_X^1,\tl{c}_X^2,\tl{c}_X^3)
=(\tl{c}_X-\frac{2\kp L}{3},\frac{\kp L}{3},\frac{\kp L}{3})$. 
The surface with (without) a mesh 
describes the region $m_{Q_a}^2>0$ ($m_{Q_a}^2<0$). 
}
\label{cdsmsq1}
\end{figure}

\begin{figure}[t]
\centering \leavevmode
\begin{minipage}{0.3\linewidth}
\begin{center}
\includegraphics[width=\linewidth]{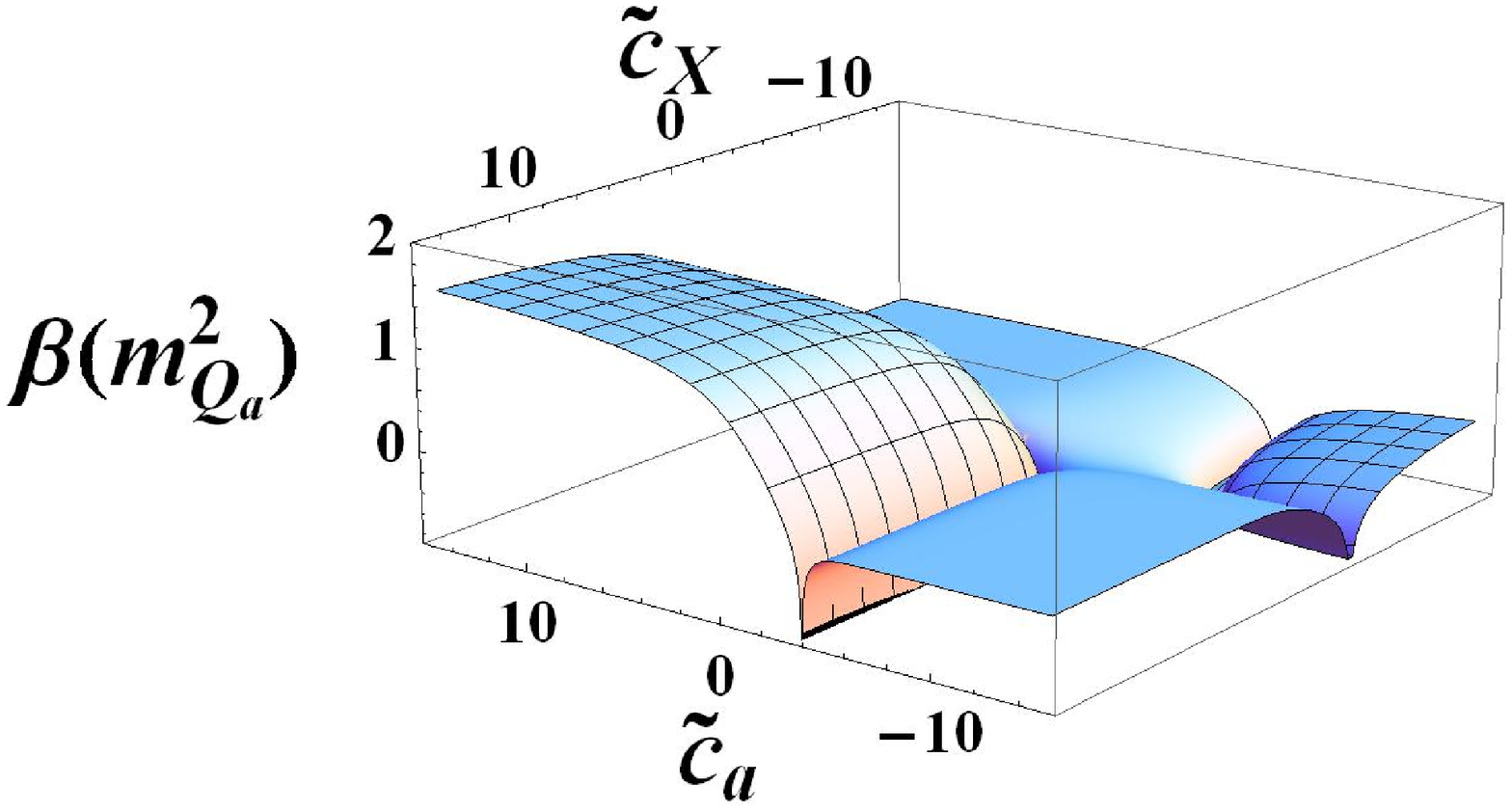}
(a) $\kp L = -3.6$
\end{center}
\end{minipage}
\hfill 
\begin{minipage}{0.3\linewidth}
\begin{center}
\includegraphics[width=\linewidth]{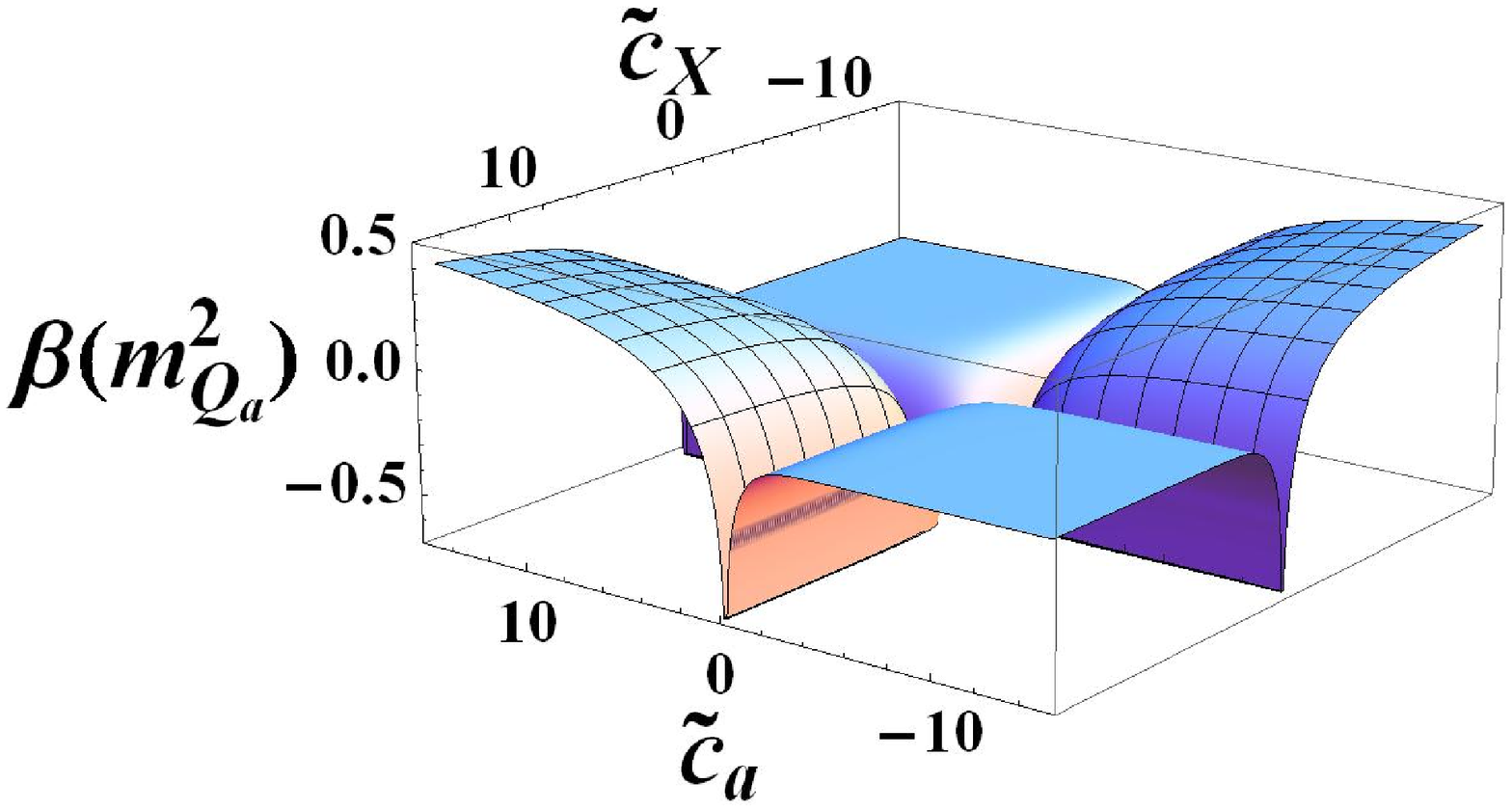}
(b) $\kp L = 0$
\end{center}
\end{minipage}
\hfill 
\begin{minipage}{0.3\linewidth}
\begin{center}
\includegraphics[width=\linewidth]{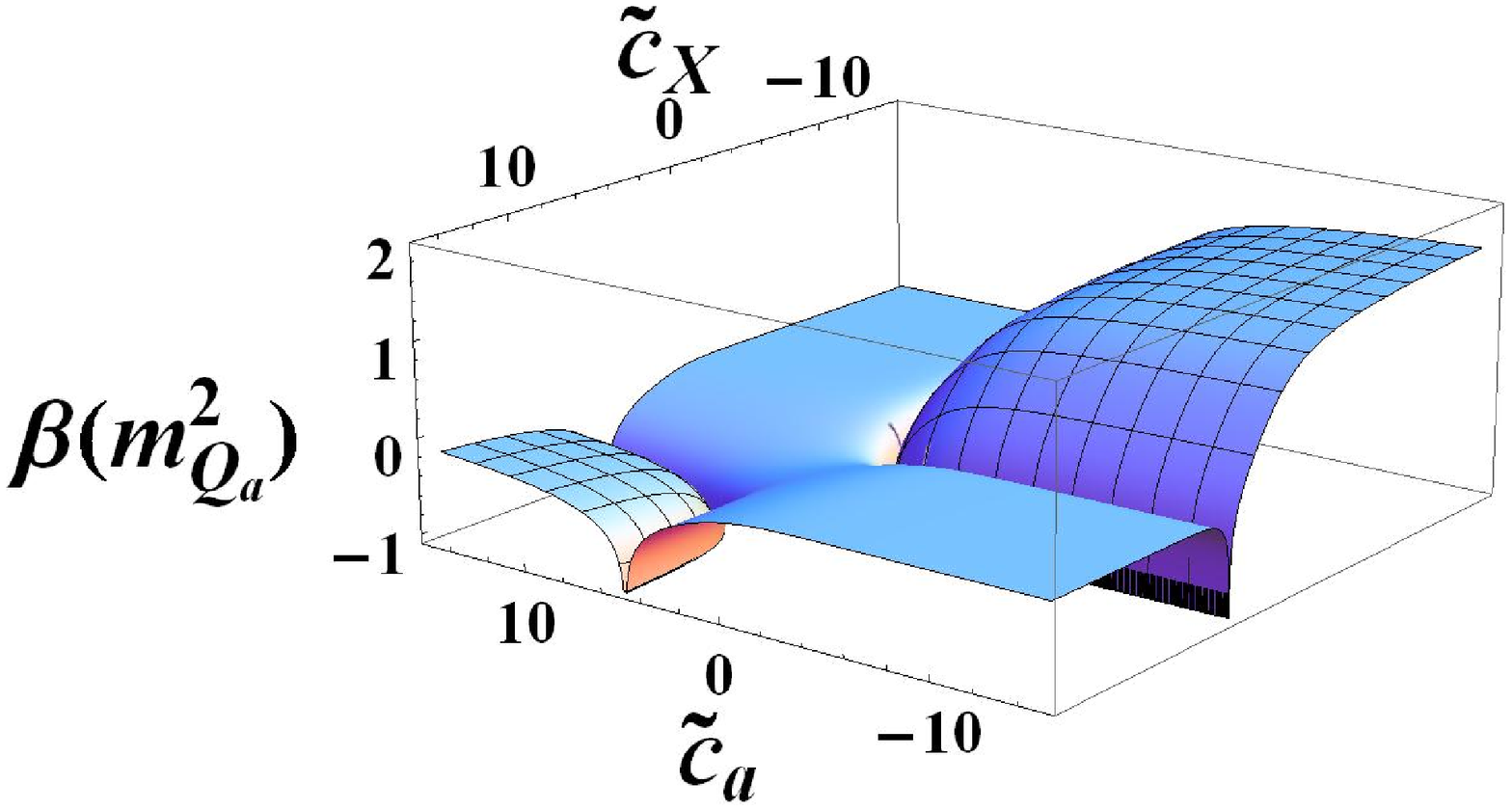}
(c) $\kp L = 3.6$
\end{center}
\end{minipage}
\caption{
The charge dependences of $\bt(m_{Q_a}^2)$ defined in 
Eq.(\ref{hatmqa}) with the norm function (\ref{nft1t2t3}) 
and $\alpha_1=\alp_3=1/(4\pi^2)$, $\alp_2=2/(4\pi^2)$. 
The charge assignment for $Q_a$ and $X$ is chosen as 
$(\tl{c}_{Q_a}^1,\tl{c}_{Q_a}^2,\tl{c}_{Q_a}^3)
=(\tl{c}_a-\frac{2\kp L}{3},\frac{\kp L}{3},\frac{\kp L}{3})$ 
and $(\tl{c}_X^1,\tl{c}_X^2,\tl{c}_X^3)
=(\tl{c}_X-\frac{2\kp L}{3},\frac{\kp L}{3},\frac{\kp L}{3})$. 
The surface with (without) a mesh 
describes the region $m_{Q_a}^2>0$ ($m_{Q_a}^2<0$). 
}
\label{cdsmsq2}
\end{figure}

\begin{figure}[t]
\centering \leavevmode
\begin{minipage}{0.3\linewidth}
\begin{center}
\includegraphics[width=\linewidth]{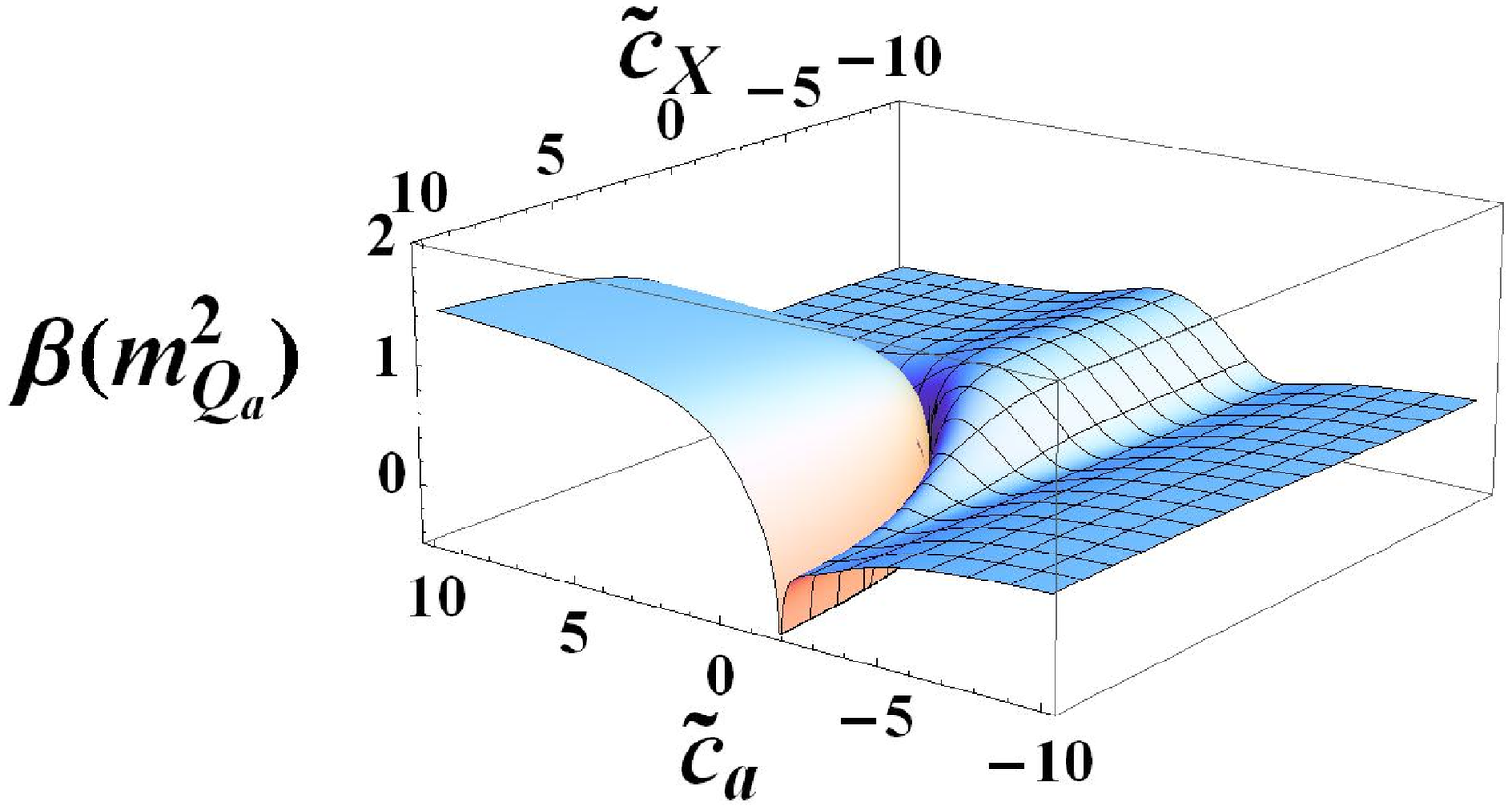}
(a) $\kp L = -3.6$
\end{center}
\end{minipage}
\hfill 
\begin{minipage}{0.3\linewidth}
\begin{center}
\includegraphics[width=\linewidth]{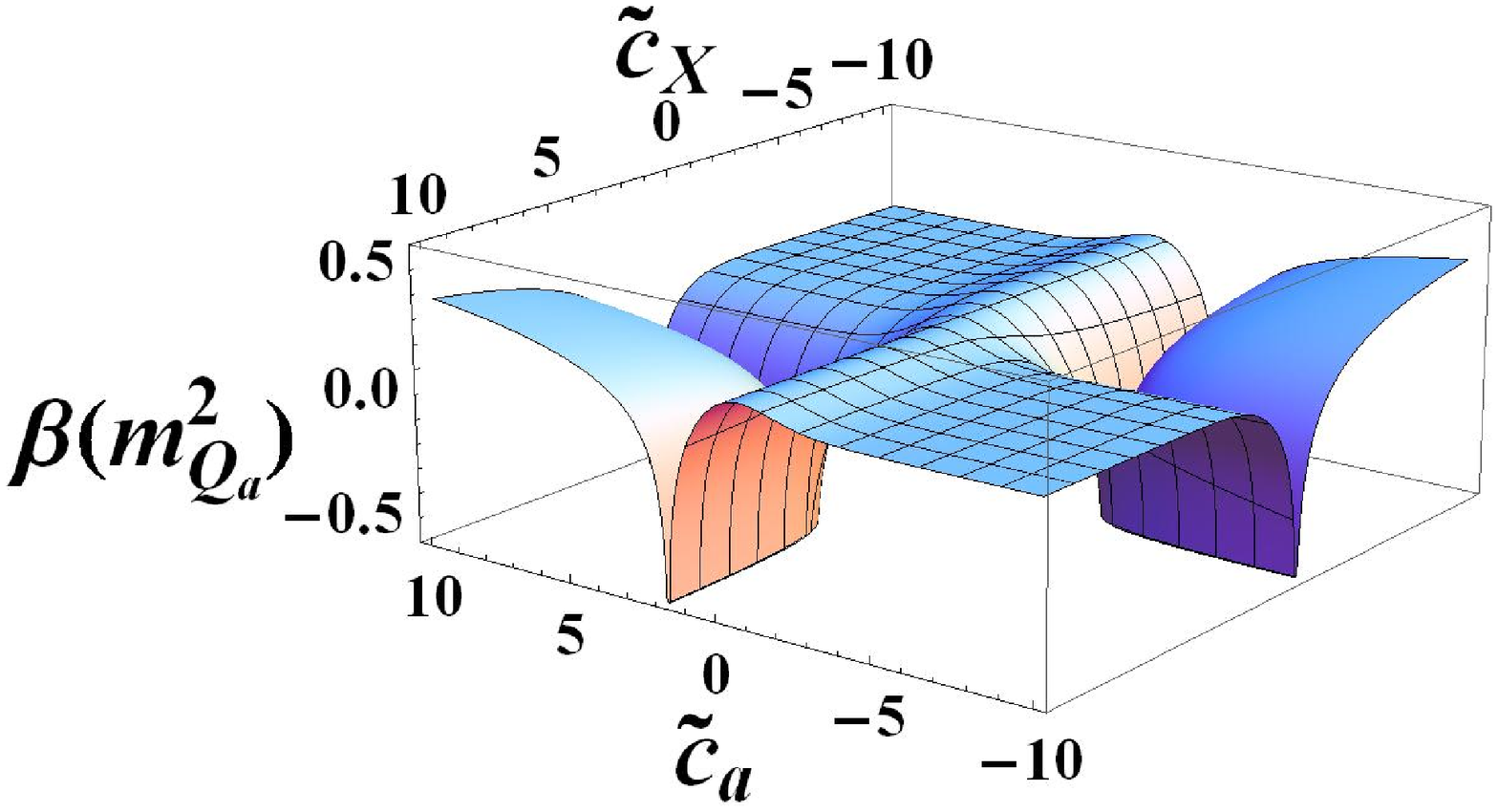}
(b) $\kp L = 0$
\end{center}
\end{minipage}
\hfill 
\begin{minipage}{0.3\linewidth}
\begin{center}
\includegraphics[width=\linewidth]{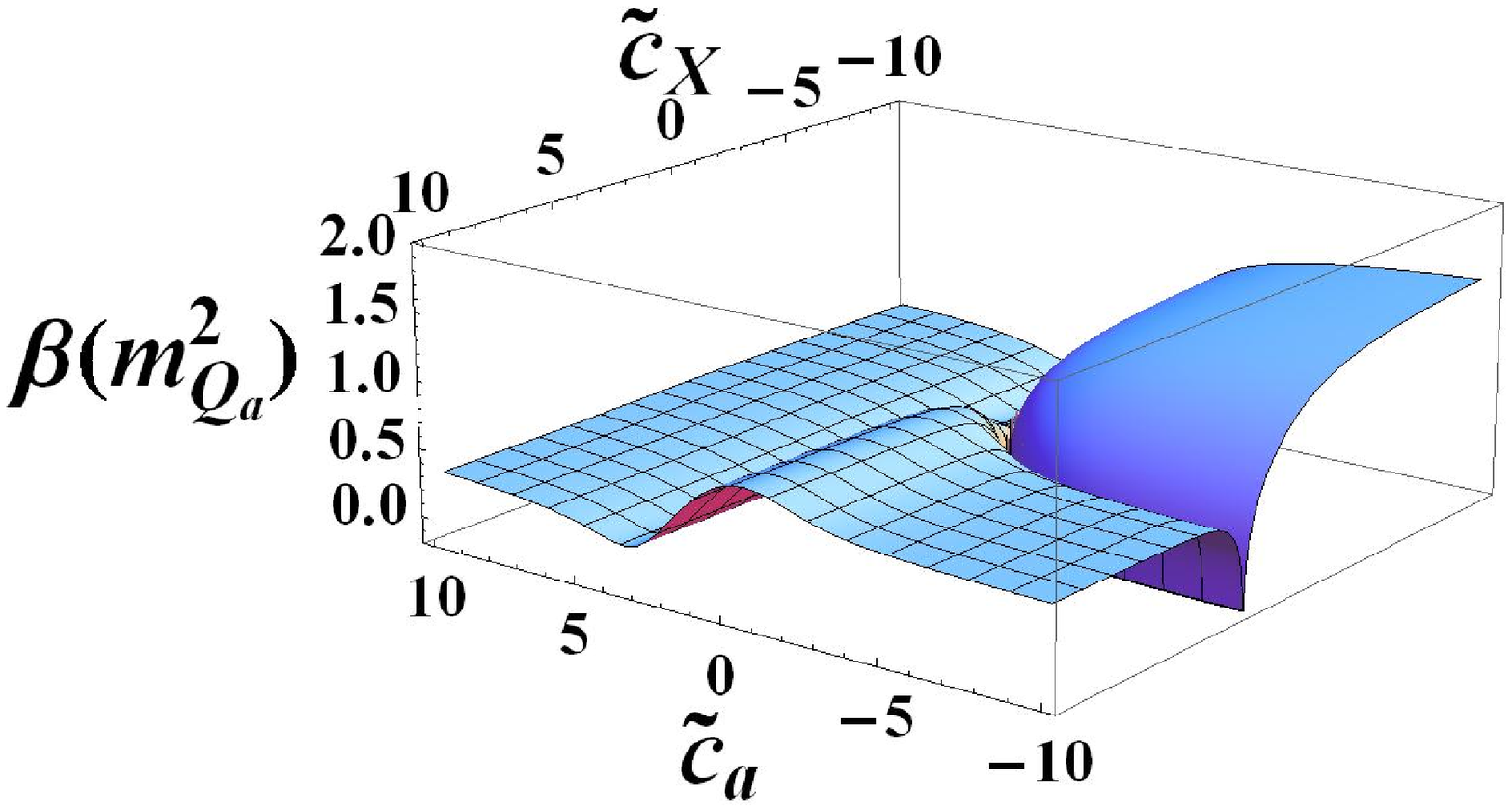}
(c) $\kp L = 3.6$
\end{center}
\end{minipage}
\caption{
The charge dependences of $\bt(m_{Q_a}^2)$ defined in 
Eq.(\ref{hatmqa}) with the norm function (\ref{nft1t2t3}) 
and $\alp_1=\alp_2=1$, $\alp_3=2$. 
The charge assignment for $Q_a$ and $X$ is chosen as 
$(\tl{c}_{Q_a}^1,\tl{c}_{Q_a}^2,\tl{c}_{Q_a}^3)
=(\frac{\kp L}{3},\tl{c}_a-\frac{2\kp L}{3},\frac{\kp L}{3})$ 
and $(\tl{c}_X^1,\tl{c}_X^2,\tl{c}_X^3)
=(\tl{c}_X-\frac{2\kp L}{3},\frac{\kp L}{3},\frac{\kp L}{3})$. 
The surface with (without) a mesh 
describes the region $m_{Q_a}^2>0$ ($m_{Q_a}^2<0$). 
}
\label{cdsmsq3}
\end{figure}

Now we show the (squared) soft scalar masses~$m_{Q_a}^2$ as functions 
of the charges for the $Z_2$-odd vector multiplets~$V^{I'}$ ($I'=1,2,3$). 
Since $\vev{Y_a}$ in the above formulae are functions of  
$(k+d_a)\cdot\Re\vev{T}$, 
we normalize each charge as 
\begin{eqnarray}
\tilde{c}^{I'}_a \defa (k_{I'}+d_{aI'})\Re\vev{T^{I'}}
 = (c_{aI'}-k_{I'}/2) \Re\vev{T^{I'}},  \label{def:tlc}
\end{eqnarray}
without summations for the index~$I'$. 
The soft scalar mass~$m_{Q_a}^2$ varies exponentially over $\cO(1)$ ranges of $\tl{c}_a^{I'}$. 
Thus we plot a quantity defined as 
\begin{eqnarray}
 \bt(m_{Q_a}^2) &\equiv& 
 \log_{10} \frac{\sqrt{|m_{Q_a}^2|}}{M_{\rm SB}}.  
 \label{hatmqa}
\end{eqnarray}
In Figs.~\ref{cdsmsq1} and \ref{cdsmsq2}, we assume that 
$\alp_1=\alp_3=1/(4\pi^2)$, $\alp_2=2/(4\pi^2)$. 
Namely, contributions of the moduli mediation are tiny. 
In Fig.~\ref{cdsmsq3}, we assume that $\alp_1=\alp_2=1$, $\alp_3=2$, 
\ie, contributions of the moduli mediation are comparable to 
that of the direct mediation. 
The charge assignment of $X$ is chosen as  
$(\tl{c}_X^1,\tl{c}_X^2,\tl{c}_X^3)=(\tl{c}_X-\frac{2\kp L}{3},
\frac{\kp L}{3},\frac{\kp L}{3})$ in all figures, 
while that of $Q_a$ is chosen as 
$(\tl{c}^1_{Q_a},\tl{c}^2_{Q_a},\tl{c}^3_{Q_a})
=(\frac{\kp L}{3},\tl{c}_a-\frac{2\kp L}{3},\frac{\kp L}{3})$ 
in Figs.~\ref{cdsmsq1} and \ref{cdsmsq3}, and 
$(\tl{c}^1_{Q_a},\tl{c}^2_{Q_a},\tl{c}^3_{Q_a})
=(\tl{c}_a-\frac{2\kp L}{3},\frac{\kp L}{3},\frac{\kp L}{3})$ in Fig.~\ref{cdsmsq2}. 
The surface with (without) a mesh describes a region~$m_{Q_a}^2>0$ 
($m_{Q_a}^2<0$). 
Note that the above charge assignments satisfy (\ref{cond:d_a}) because 
they are rewritten as 
\bea
 \brkt{d_X^1,d_X^2,d_X^3} \eql \brkt{d_X,0,0}, \nonumber\\
 \brkt{d_{Q_a}^1,d_{Q_a}^2,d_{Q_a}^3} \eql \begin{cases} \displaystyle
 \brkt{0,d_a,0}, & (\mbox{in Figs.~\ref{cdsmsq1} and 
 \ref{cdsmsq3}}) \\ \displaystyle 
 \brkt{d_a,0,0}. & (\mbox{in Fig.~\ref{cdsmsq2}}) \end{cases}
\eea
where $d_X\equiv(\tl{c}_X-\kp L)/4$ and $d_a\equiv(\tl{c}_a-\kp L)/4$. 

Let us first consider a case that $\alp_{I'}\ll 1$, \ie, 
the direct mediation dominates. 
The soft scalar mass is expressed from (\ref{ap_m_soft}) as 
\be
 \frac{m_{Q_a}^2}{M_{\rm SB}^2} \simeq 
 -\frac{Y(\kp L)\tl{\Omg}^{(4)}_{Q_a,X}(\Re\vev{T})}{2Y(\tl{c}_a)Y(\tl{c}_X)}. 
 \label{ratio:m/M}
\ee
We can see that $\abs{m_{Q_a}^2}>M_{\rm SB}^2$ for $\tl{c}_a\tl{c}_X\gg 1$ 
in Figs.~\ref{cdsmsq1} and \ref{cdsmsq2}. 
This behavior can be understood from the fact that 
$Q_a$ and $X$ localize toward the same boundary in such a region. 
Especially in the warped case, the soft scalar mass is enhanced 
when they localize toward the IR boundary. 
On the other hand, they localize toward the opposite boundaries 
in a region~$\tl{c}_a\tl{c}_X<-1$. 
Notice that $m_{Q_a}^2/M_{\rm SB}^2$ is not exponentially suppressed 
and remains to be of $\cO(1)$ in this region, 
because they are not sequestered 
due to the existence of the contact terms~$\abs{Q_a}^2\abs{X}^2$ 
in $\Omg_{\rm eff}$ induced by integrating out the heavy $Z_2$-odd vector 
multiplets~$V^{I'}$.  
This is in sharp contrast to the single modulus case 
where the sequestering occurs. 
Especially, in a region that 
$\tl{c}_a\tl{c}_X<-1$ and $(\tl{c}_a-\kp L)(\tl{c}_X-\kp L)< -1$, 
such induced contact terms dominate, 
and (\ref{ratio:m/M}) can be approximated as 
\be
 \frac{m_{Q_a}^2}{M_{\rm SB}^2} \simeq 
 -\frac{e^{2\kp L}Y(\kp L)}{2Y(-\kp L)}
 \frac{d_{Q_a}\cdot\cP a^{-1}\cdot d_X}
 {(d_{Q_a}\cdot\Re\vev{T})(d_X\cdot\Re\vev{T})} 
 \simeq -\frac{d_{Q_a}\cdot\cP a^{-1}\cdot d_X}
 {2(d_{Q_a}\cdot\Re\vev{T})(d_X\cdot\Re\vev{T})}. 
 \label{ap:m/M}
\ee
Since the charge assignments of $Q_a$ satisfy 
the condition~(\ref{cond:d_a}), the flavor dependence of (\ref{ap:m/M}) cancels. 
With our choice of the norm function and the assumption~(\ref{VEV:assumption}), 
its approximate value is $16/3$ ($-32/3$) 
in the case of Fig.~\ref{cdsmsq1} (Fig.~\ref{cdsmsq2}), 
irrespective of the value of $\kp L$. 
The non-tachyonic condition in such a flavor-universal 
region~(\ref{cond:non-tachyonic}) is satisfied 
for the charge assignment of Fig.~\ref{cdsmsq1}, 
while not for that of Fig.~\ref{cdsmsq2}, 
as we can see the figures. 

In a case that contributions from the moduli mediation are not negligible, 
\ie, $\abs{\alp_{I'}}=\cO(1)$, the above behaviors are disturbed. 
The expression of the soft scalar mass in such a case 
is given in (\ref{explicit:ma/MSB}). 
In the flat case ($\kp L=0$), for example, it is written as 
\be
 \frac{m_{Q_a}^2}{M_{\rm SB}^2} \simeq 
 -\frac{\tl{\Omg}^{(4)}_{Q_a,X}(\Re\vev{T})}{2Y(\tl{c}_a)Y(\tl{c}_X)}
 +\sum_{I'}\frac{\abs{\alp_{I'}}^2}{3}
 -\sum_{I',J'}\alp_{I'}\bar{\alp}_{J'}
 \tl{c}_{Q_a}^{I'}\tl{c}_{Q_a}^{J'}\cY(\tl{c}_a), 
\ee
where $\cY(x)$ is a function defined in (\ref{def:cY}). 
We have used the specific form of the norm function~(\ref{nft1t2t3}). 
In spite of the nontrivial $\tl{c}_a$-dependence of the third term,  
there is still a region in which $m_{Q_a}^2$ is almost flavor-universal. 
This is due to the property of the function~$\cY(x)$ 
that $x^2\cY(x)\simeq 1$ for $\abs{x}\simgt 3$. 

\ignore{
we find 
$|m_{Q_a}^2| \gtrsim M_{\rm SB}^2$ 
($|m_{Q_a}^2| \sim M_{\rm SB}^2$) 
for 
$\tilde{c}^a \tilde{c}^X \gtrsim 1$ 
($\tilde{c}^a \tilde{c}^X \lesssim 1$) 
because in this case $Q_a$ and $X$ localize toward 
the same (opposite) fixed point to each other. 
This behavior can be understood from the geometrical point of view. 
For the warped case $\kp L > 0$ ($\kp L < 0$) 
in Figs.~\ref{cdsmsq1} and \ref{cdsmsq2}, 
the absolute value of scalar mass squared 
$|m_{Q_a}^2|$ is enhanced (suppressed) for the region 
$\tilde{c}_a$, $\tilde{c}_X < 0$. 
This is the effect of the warp factor $e^{\kp L}$. 
The sign of the scalar mass squared $m_{Q_a}^2$ 
(with the same warp factor) is flipped between 
Fig.~\ref{cdsmsq1} and Fig.~\ref{cdsmsq2}
in the region $|\tilde{c}_a \tilde{c}_X| \gtrsim 1$. 
Therefore, the tachyonic soft scalar mass can be avoided for 
a suitable charge assignment even with certain fixed values of 
$\tilde{c}_a$ and $\tilde{c}_X$. This is one of the most important 
property with multiple moduli in order to realize realistic Yukawa 
hierarchies ({\it i.e.} observed quark/lepton masses and mixings) 
without tachyonic squarks/sleptons. 
}

By making use of the above properties, 
we construct a realistic model in the next subsection, 
and analyze the flavor structure of fermions and 
sfermions as well as the other phenomenological features. 
We comment that the boundary induced K\"{a}hler potentials~$K^{(y_*)}$ 
($y_*=0,L$) are neglected in this paper.  
They may disturb the flavor structure 
if they dominate the contributions from the bulk. 
Here we just assume that such boundary contributions are small enough 
compared to those from the bulk.

\subsection{Phenomenological analysis}
In the following phenomenological analysis, 
the warp factor is chosen as 
\be
 \kp L = 3.6,  \label{kL}
\ee
so that $e^{\kp L}=\cO(M_{\rm Pl}/M_{\rm GUT})$. 
This determines the compensator charges in (\ref{cpst_gauging}) as $k_3=1.2$. 
The typical KK mass scale is set to the GUT scale, 
\be
 M_{\rm KK} \equiv \frac{\kp\pi}{e^{\kp L}-1} 
 = M_{\rm GUT}.  \label{mKK}
\ee
This determines $\kp$ (and $L$) from (\ref{kL}). 
The 4D effective theory is valid below $M_{\rm KK}$. 

Here we comment on a consistency condition of our 5D setup. 
In order for the 5D description of the theory to be valid, 
the 5D curvature~$\cR^{(5)}$ must satisfy 
the condition~$\abs{\cR^{(5)}}<M_5^2$~\cite{Davoudiasl:1999tf}. 
For the Randall-Sundrum spacetime, $\cR^{(5)}=-20\kappa^2$. 
Thus, together with the relation~(\ref{rel:Mpl}) 
and the definition of $M_{\rm KK}$, the consistency condition is rewritten as 
\be
 e^{\kp L} < \frac{\sqrt{2}\pi}{20^{3/4}}\frac{M_{\rm Pl}}{M_{\rm KK}} 
 \simeq 0.47\frac{M_{\rm Pl}}{M_{\rm KK}}. 
\ee
For our choice of $M_{\rm KK}$ in (\ref{mKK}), this indicates 
that $\kp L<4.0$, which is satisfied by (\ref{kL}). 

In order to realize phenomenologically viable fermion and 
sfermion flavor structures, we assign the following 
$U(1)_{I'}$ charges for the $Z_2$-odd vector multiplets~$V^{I'}$ 
($I'=1,2,3$) to the MSSM matter multiplets 
$Q_a=({\cQ_i,\cU_i,\cD_i,\cL_i,\cE_i,\cH_u,\cH_d})$.  
For the values of 
${\rm Re}\,\langle T^{I'} \rangle$ and $k_{I'}$ 
given by (\ref{VEV:assumption}) and (\ref{cpst_gauging}), 
the $U(1)_1$ charges are chosen as 
\begin{eqnarray}
&\tl{c}^{I'=1}_{\cQ_i} = (1.2, 1.2, 0.5), \quad 
\tl{c}^{I'=1}_{\cU_i} = (1.2, 1.2, 0.5), \quad 
\tl{c}^{I'=1}_{\cD_i} = (1.2, 1.2, 1.2), & \nonumber \\
&\tl{c}^{I'=1}_{\cL_i} = (1.2, 1.2, 1.2), \quad 
\tl{c}^{I'=1}_{\cE_i} = (1.2, 1.2, 1.2), & \nonumber \\
&\tl{c}^{I'=1}_{\cH_u} = 1.0, \quad
\tl{c}^{I'=1}_{\cH_d} = 1.2, \quad 
\tl{c}^{I'=1}_{X} = 8.7, & 
\label{fc1}
\end{eqnarray}
the $U(1)_2$ charges are assigned as 
\begin{eqnarray}
&\tl{c}^{I'=2}_{\cQ_i} = (-7.9, -5.9,0), \quad 
\tl{c}^{I'=2}_{\cU_i} = (-10.4, -5.9,0), \quad 
\tl{c}^{I'=2}_{\cD_i} = (-6.4, -6.9, -4.9),& \nonumber \\
&\tl{c}^{I'=2}_{\cL_i} = (-6.9, -6.9, -4.9), \quad 
\tl{c}^{I'=2}_{\cE_i} = (-9.4, -3.9, -3.9),& \nonumber \\
&\tl{c}^{I'=2}_{\cH_u} = 0, \quad
\tl{c}^{I'=2}_{\cH_d} = -3.4, \quad 
\tl{c}^{I'=2}_{X} = 1.2, & 
\label{fc2}
\end{eqnarray}
and  the $U(1)_3$ charges are assigned as 
\begin{eqnarray}
&\tl{c}^{I'=3}_{\cQ_i} = (1.2, 1.2, 0), \quad 
\tl{c}^{I'=3}_{\cU_i} = (1.2, 1.2, 0), \quad 
\tl{c}^{I'=3}_{\cD_i} = (1.2, 1.2, 1.2),& \nonumber \\
&\tl{c}^{I'=3}_{\cL_i} = (1.2, 1.2, 1.2), \quad 
\tl{c}^{I'=3}_{\cE_i} = (1.2, 1.2, 1.2),& \nonumber \\
&\tl{c}^{I'=3}_{\cH_u} = 0, \quad
\tl{c}^{I'=3}_{\cH_d} = 1.2, \quad 
\tl{c}^{I'=3}_{X} = 1.2, & 
\label{fc3}
\end{eqnarray} 
\ignore{
These charges are almost determined 
uniquely by the requirement that the observed 
quark and charged lepton masses and the absolute values 
of CKM mixings are realized\footnote{Note that the completely 
precise matching of the predicted and observed values shown in 
Table~\ref{pqlmm} is meaningless because the former values 
are calculated from the tree-level effective theory at 
$M_{\rm GUT}$ with certain reference values of ${\cal O}(1)$ 
holomorphic Yukawa couplings $\lambda^{u,d,e}_{ij}$ and 
some one-loop renormalization effects below $M_{\rm GUT}$.} 
as shown in Table~\ref{pqlmm} with ${\cal O}(1)$ values of 
the holomorphic Yukawa couplings $\lambda^{u,d,e}_{ij}$ 
in the superpotential (\ref{wmssm}). 
}
These charges satisfy (\ref{cond:d_a}) 
for the first two generations of quark and lepton multiplets. 
With this charge assignment, 
the observed quark and charged lepton masses and the absolute values 
of CKM mixings are realized, as shown in Table~\ref{pqlmm}, 
with ${\cal O}(1)$ values of 
the holomorphic Yukawa couplings $\lambda^{u,d,e}_{ij}$ 
in the superpotential (\ref{wmssm}). 
\begin{table}[t]
\begin{center}
\begin{tabular}{|c||c|c|} \hline
 & Predicted & Observed \\ \hline
$(m_u, m_c, m_t)/m_t$ & 
$(1.4 \times 10^{-5}, 7.38 \times 10^{-3}, 1.0)$ & 
$(1.5 \times 10^{-5}, 7.37 \times 10^{-3}, 1.0)$ \\ \hline
$(m_d, m_s, m_b)/m_b$ & 
$(1.2 \times 10^{-3}, 2.41 \times 10^{-2}, 1.0)$ & 
$(1.2 \times 10^{-3}, 2.54 \times 10^{-2}, 1.0)$ \\ \hline
$(m_e, m_\mu, m_\tau)/m_\tau$ & 
$(2.871 \times 10^{-4}, 5.955 \times 10^{-2}, 1.0)$ & 
$(2.871 \times 10^{-4}, 5.959 \times 10^{-2}, 1.0)$ \\ \hline
$|V_{CKM}|$ & 
\begin{minipage}{0.38\linewidth}
\begin{eqnarray} 
\left( 
\begin{array}{ccc}
0.97324 & 0.2298 & 0.00337 \\
0.2297 & 0.97235 & 0.042 \\
0.00637 & 0.0417 & 0.999112 
\end{array}
\right) 
\nonumber
\end{eqnarray} \\*[-20pt]
\end{minipage}
& 
\begin{minipage}{0.38\linewidth}
\begin{eqnarray} 
\left( 
\begin{array}{ccc}
0.97428 & 0.2253 & 0.00347 \\
0.2252 & 0.97345 & 0.041 \\
0.00862 & 0.0403 & 0.999152 
\end{array}
\right) 
\nonumber
\end{eqnarray} \\*[-20pt]
\end{minipage}
\\ \hline
\end{tabular}
\end{center}
\caption{The predicted quark and charged lepton masses 
as well as the absolute values of CKM mixings compared 
with the experimental data~\cite{Nakamura:2010zzi}. 
The flavor charges are chosen as shown in 
Eqs.(\ref{fc1}) and (\ref{fc2}).}
\label{pqlmm}
\end{table}

After fixing all the $U(1)_{I'}$ charges, the remaining parameters 
are the coefficients $\xi^r_{I'}$ in the effective 
gauge kinetic functions $f^r_{\rm eff}(T)$ in (\ref{wmssm}). 
One of $\xi^r_{I'}$ for each $r$ is determined by matching $f^r_{\rm eff}(\vev{T})$ 
with the observed values of the SM gauge couplings, \ie, 
by the condition for the gauge coupling unification at $M_{\rm GUT}$. 
The remaining $\xi^r_{I'}$ control the gaugino masses~$M_r$ at $M_{\rm GUT}$. 
In the following analysis, all the gauge couplings, Yukawa couplings, 
the $\mu$-term and soft SUSY breaking parameters in the visible sector 
are evaluated at the EW scale~$M_{\rm EW}$ by using the 1-loop renormalization 
group (RG) equations of MSSM, where we neglect effects of all 
the Yukawa couplings except for the top Yukawa coupling. 

In order to estimate the SUSY flavor violations, 
we rotate the soft scalar mass matrices 
$(m^2_{\tl{f}_{L,R}})_{ij}=\diag\brkt{m^2_{\tl{f}_{L,R}1},m^2_{\tl{f}_{L,R}2},
m^2_{\tl{f}_{L,R}3}}$ 
and the scalar trilinear coupling matrices 
$(\tilde{A}^f)_{ij}=(y^f)_{ij}(A^f)_{ij}$ 
into the super-CKM basis and describe them as 
\be
\hat{m}^2_{\tilde{f}_L} = (V^f_L)^\dagger m^2_{\tilde{f}_L} V^f_L, \quad 
\hat{m}^2_{\tilde{f}_R} = (V^f_R)^\dagger m^2_{\tilde{f}_R} V^f_R, \quad 
\hat{A}^f = (V^f_L)^\dagger \tilde{A}^f V^f_R, 
\ee
where $f=u,d,e$ and $V^u_L=V^d_L\equiv V^q_L$. 
The unitary matrices $V^f_{L,R}$ are defined by 
\be
 (V^f_L)^\dagger y_f V^f_R = \frac{1}{v_f}\diag\brkt{m_{f1},m_{f2},m_{f3}}, 
\ee
where $v_f=(\sin \beta, \cos \beta, \cos \beta)v$ and $v \simeq 174$ GeV. 
Here we consider a case of $\tan\bt=4$ as an example.\footnote{
In fact, $\tan\bt$ is not a free parameter in our original setup. 
One way to treat $\tan\bt$ as a free parameter 
is to introduce another $\mu$-term, $\zeta_L\cH_u\cH_d$, on the $y=L$ boundary. 
Then the constant~$\zeta_0$ in (\ref{expr:mu}) and (\ref{sp}) is replaced by 
$\zeta\equiv\zeta_0+e^{-(c_{\cH_u}+c_{\cH_d})\cdot T}\zeta_L$, and 
we can control the value of $\tan\bt$ by varying $\zeta_L$ 
keeping $\vev{\zeta}$ unchanged. 
} 
Then we define mass insertion parameters as~\cite{Misiak:1997ei} 
\begin{eqnarray}
&\displaystyle 
(\delta_{LL}^f)_{ij} = 
\frac{(\hat{m}^2_{\hat{f}_L})_{ij}
+\left( \left( m_f \right)_i - \rho_{LL}^f \right) 
\delta_{ij} 
}{
\sqrt{
(\hat{m}^2_{\tilde{f}_L})_{ii} 
(\hat{m}^2_{\tilde{f}_L})_{jj}}}, \quad 
(\delta_{RR}^f)_{ij} = 
\frac{(\hat{m}^2_{\hat{f}_R})_{ij}
+\left( \left( m_f \right)_i - \rho_{RR}^f \right) 
\delta_{ij} 
}{
\sqrt{
(\hat{m}^2_{\tilde{f}_R})_{ii} 
(\hat{m}^2_{\tilde{f}_R})_{jj}}},& 
\nonumber \\
&\displaystyle 
(\delta_{LR}^f)_{ij} = 
\frac{v_f (\hat{A}^f)_{ij}
-\mu_f \left( m_f \right)_i \delta_{ij} 
}{
\sqrt{
(\hat{m}^2_{\tilde{f}_L})_{ii}
(\hat{m}^2_{\tilde{f}_R})_{jj}}}
\ = \ (\delta_{RL}^f)_{ji}^\ast,& 
\end{eqnarray}
where 
\begin{eqnarray}
\mu_f &=& \left( 
\cot \beta, \tan \beta, \tan \beta \right) \mu,
\nonumber \\
\rho_{LL}^f &=& 
\frac{\cos 2 \beta}{6} \left( 
M_Z^2-4M_W^2,\, M_Z^2+2M_W^2,\, -3M_Z^2+6M_W^2 \right), 
\nonumber \\
\rho_{RR}^f &=& 
\frac{\cos 2 \beta}{3} \sin^2 \theta_W \left( 
-2M_Z^2,\, M_Z^2,\, 3M_Z^2 \right), 
\nonumber \\
M_Z &\simeq& 91.2 \ {\rm GeV}, \quad 
M_W \ \simeq \ 80.1 \ {\rm GeV}, \quad 
\sin^2 \theta_W \ \simeq \ 0.23, 
\end{eqnarray}
and $\mu$ is determined 
by the minimization condition of the Higgs potential. 

\begin{figure}[t]
\centering \leavevmode
\hfill
\includegraphics[width=0.35\linewidth]{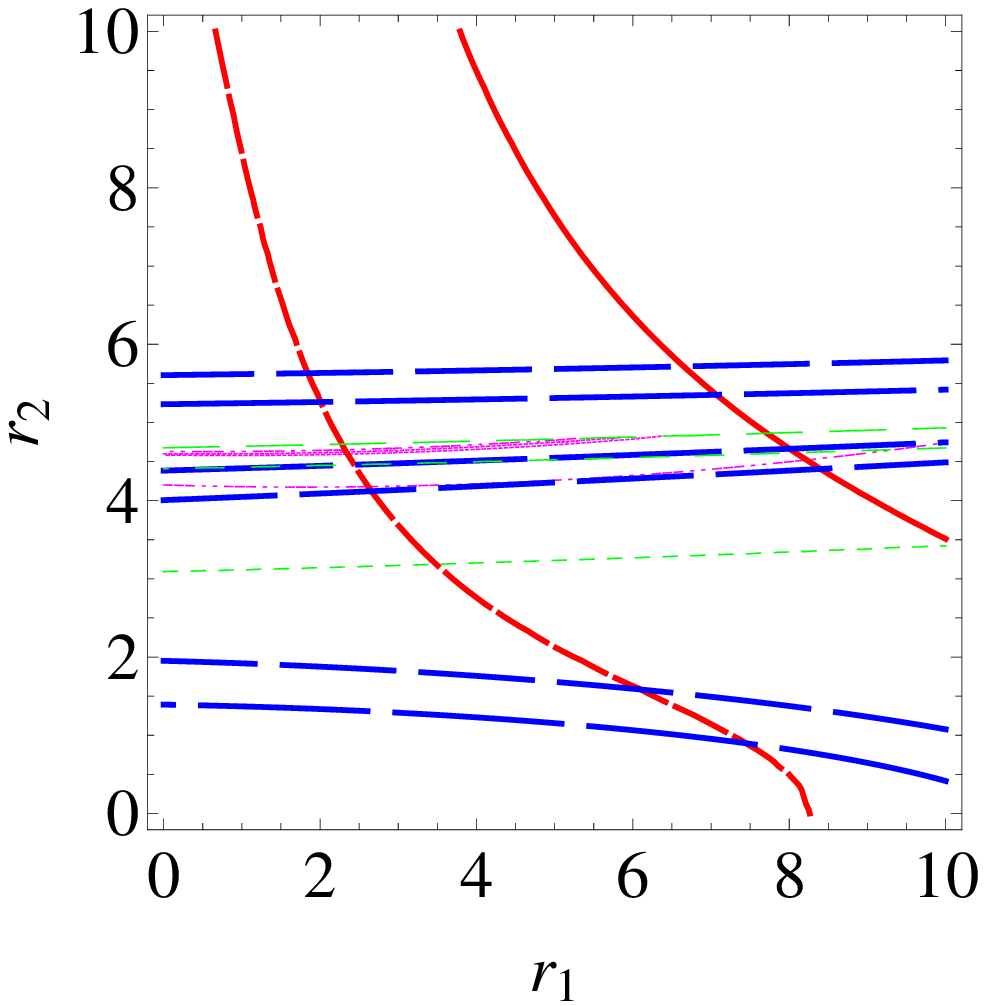}
\hfill
\includegraphics[width=0.35\linewidth]{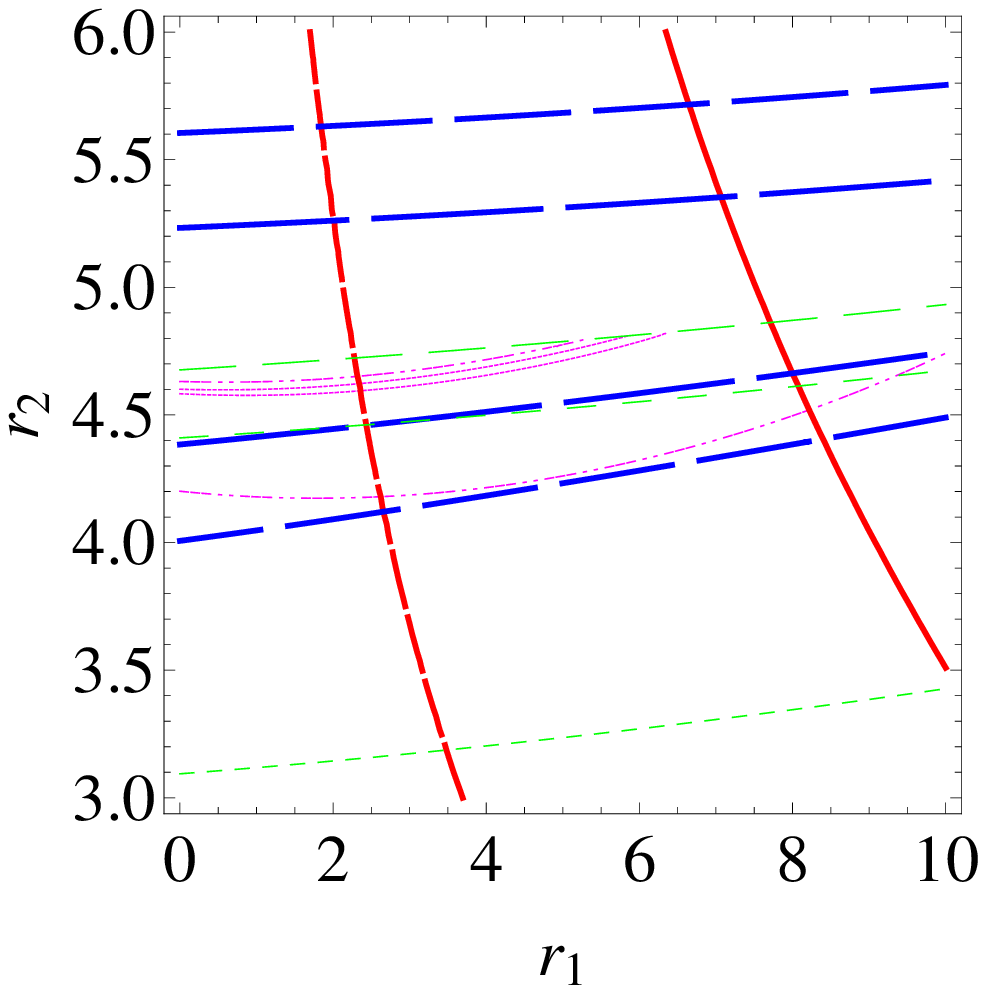}
\hfill
\includegraphics[width=0.25\linewidth]{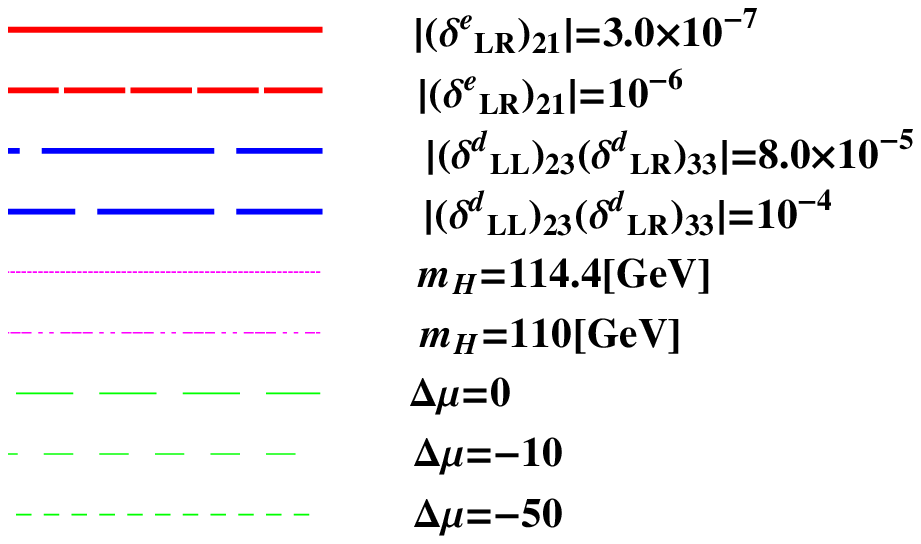}
\hfill
\caption{
Contours of 
$|(\delta_{LR}^e)_{21}|$, 
$|(\delta_{LL}^d)_{23}(\delta_{LR}^d)_{33}|$, 
$|\Delta_{\mu}|$ and 
$m_{H_0}$ 
as functions of the gaugino mass ratios 
$r_1=M_1/M_3$ and $r_2=M_2/M_3$ at $M_{\rm GUT}$. 
The region $3 \le r_2 \le 6$ 
is magnified in the right panel. 
The parameters are chosen as $M_{\rm SB}=100$~GeV,
$\alpha_1=1,\alpha_2=1/2$, $\alpha_3=1/(4\pi^2)$, and $\tan \beta =4$. 
The gluino mass~$M_3$ is set to 343~GeV at $M_{\rm GUT}$. 
}
\label{1t2}
\end{figure}
\begin{figure}[t]
\centering \leavevmode
\hfill
\includegraphics[width=0.35\linewidth]{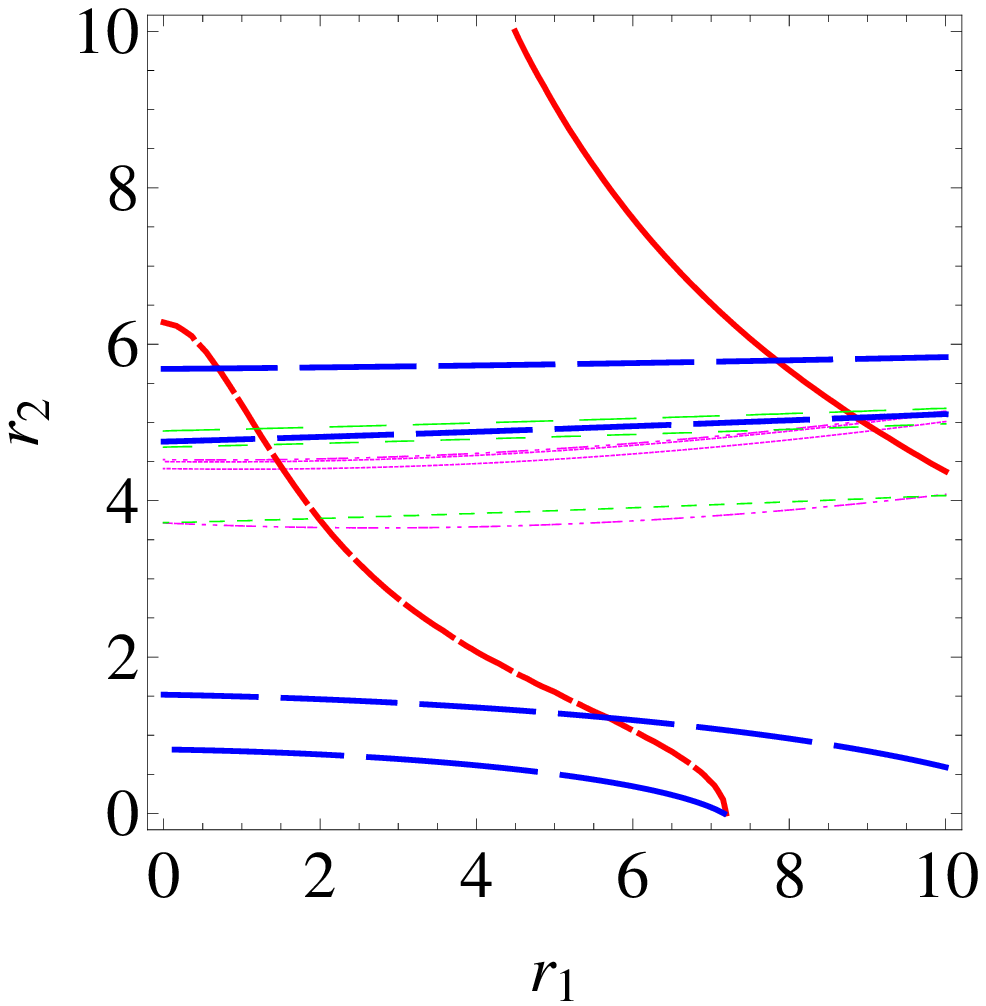}
\hfill 
\includegraphics[width=0.35\linewidth]{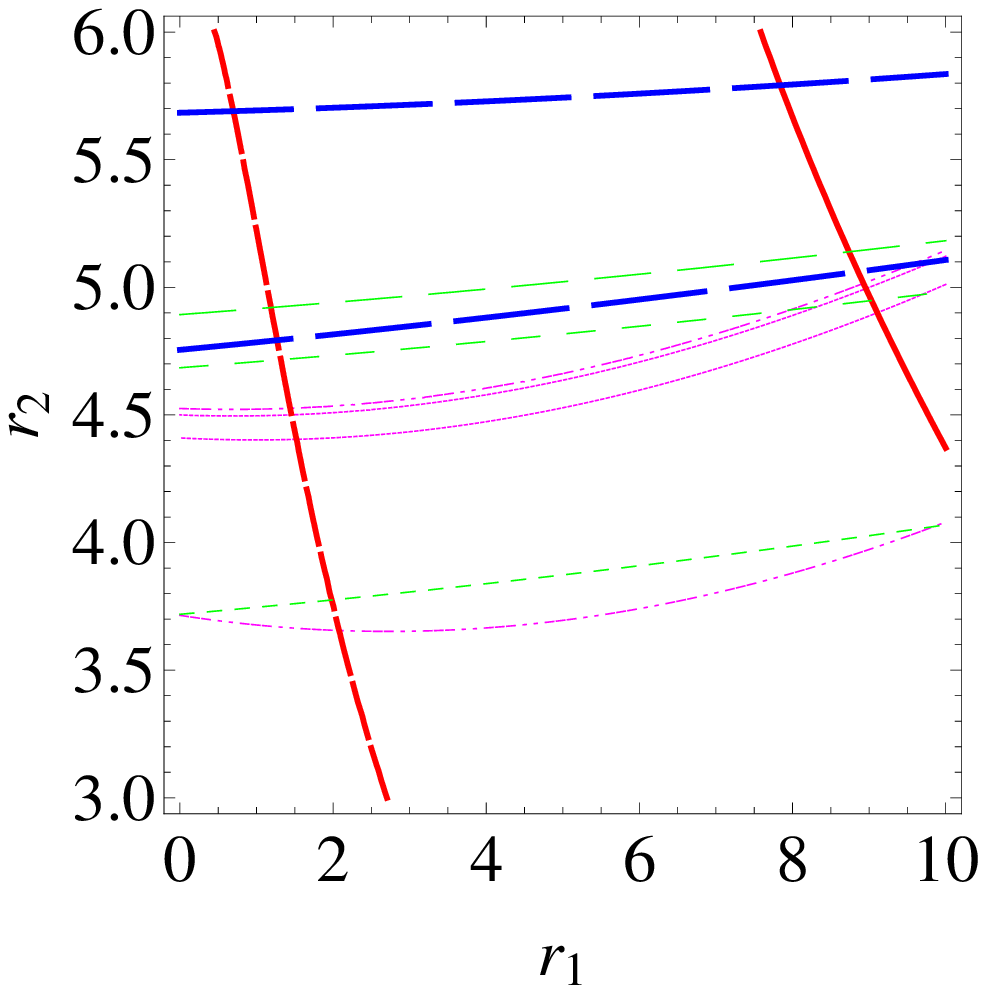}
\hfill 
\includegraphics[width=0.25\linewidth]{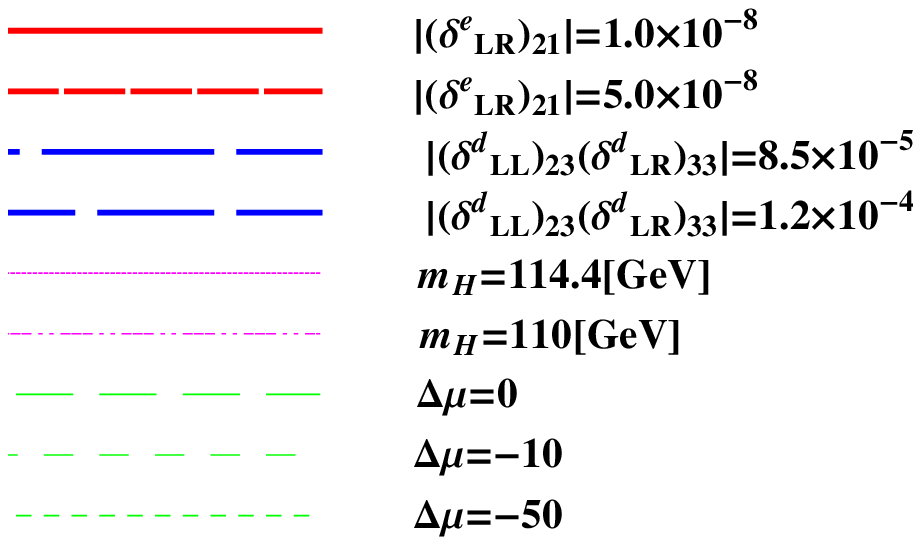}
\hfill
\caption{
Contours of 
$|(\delta_{LR}^e)_{21}|$, 
$|(\delta_{LL}^d)_{23}(\delta_{LR}^d)_{33}|$, 
$|\Delta_{\mu}|$ and 
$m_{H_0}$ 
as functions of the gaugino mass ratios 
$r_1=M_1/M_3$ and $r_2=M_2/M_3$ at $M_{\rm GUT}$. 
The region $3 \le r_2 \le 6$ 
is magnified in the right panel. 
The parameters are chosen as $M_{\rm SB}=200$~GeV,
$\alpha_1=1/2$, $\alpha_2=1/(8\pi^2)$, $\alpha_3=1/(4\pi^2)$,  
and $\tan \beta =4$. 
The gluino mass~$M_3$ is set to 383~GeV at $M_{\rm GUT}$. 
}
\label{pt2}
\end{figure}
\begin{figure}[t]
\centering \leavevmode
\hfill
\includegraphics[width=0.35\linewidth]{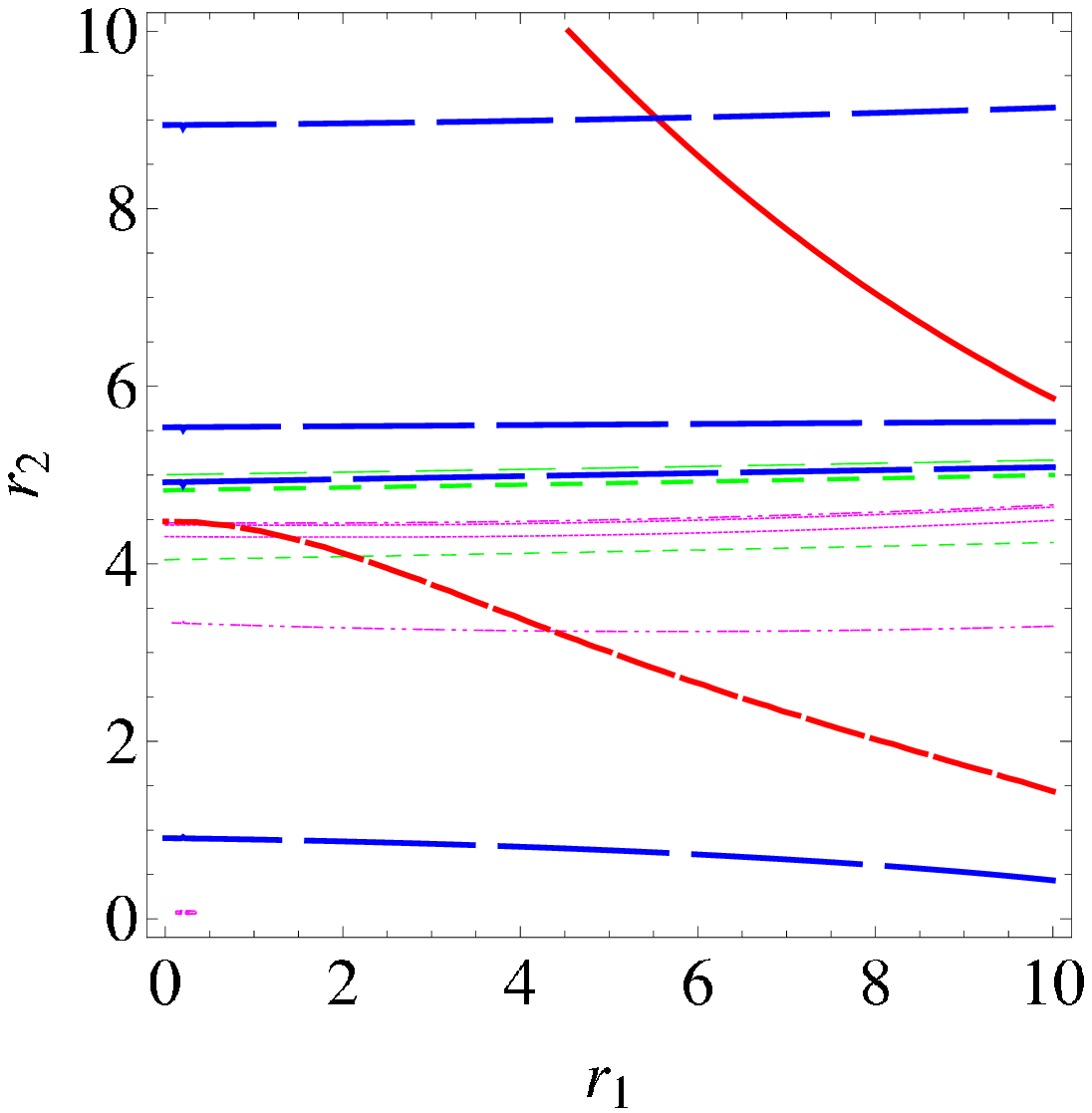}
\hfill 
\includegraphics[width=0.35\linewidth]{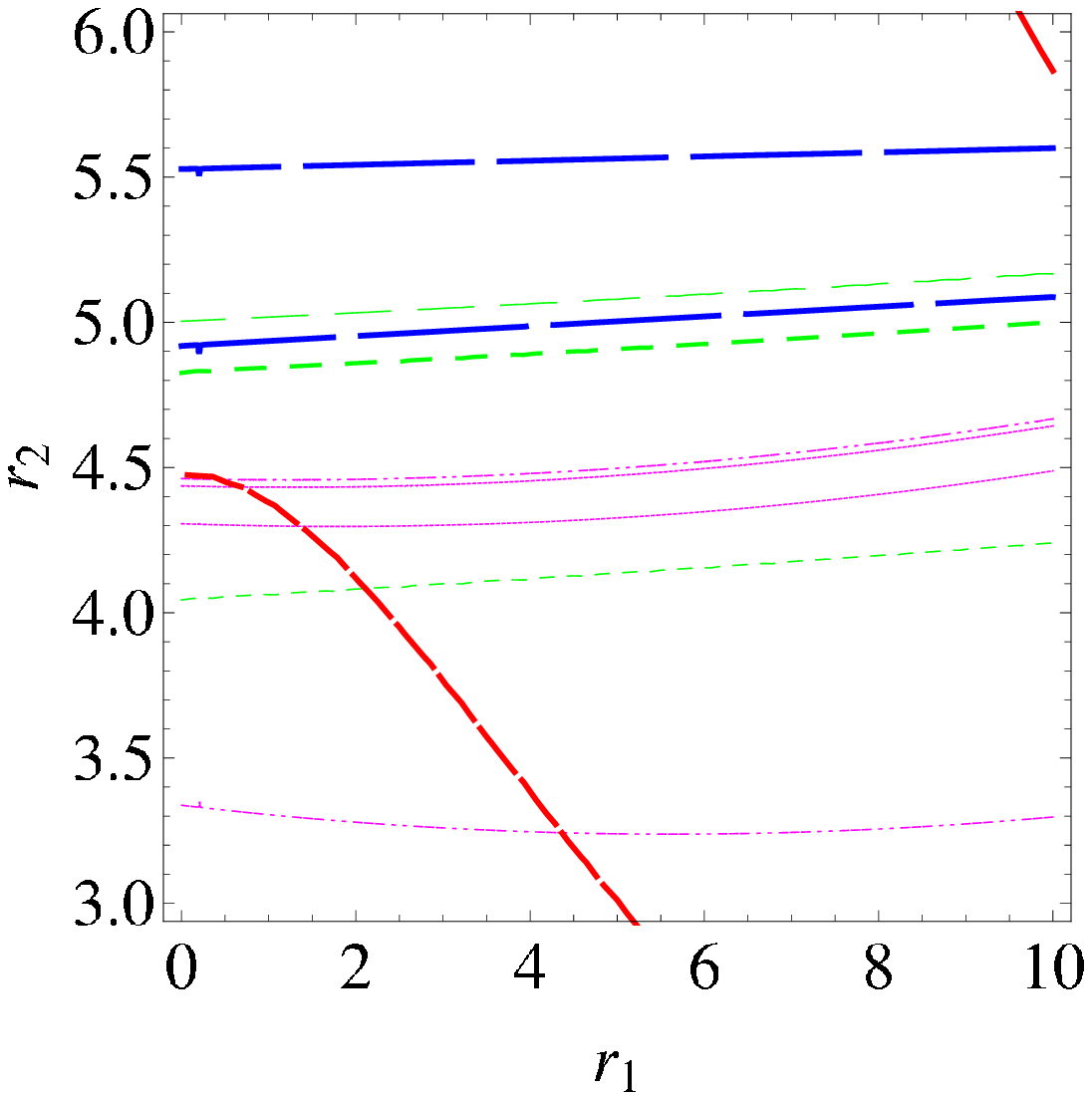}
\hfill 
\includegraphics[width=0.25\linewidth]{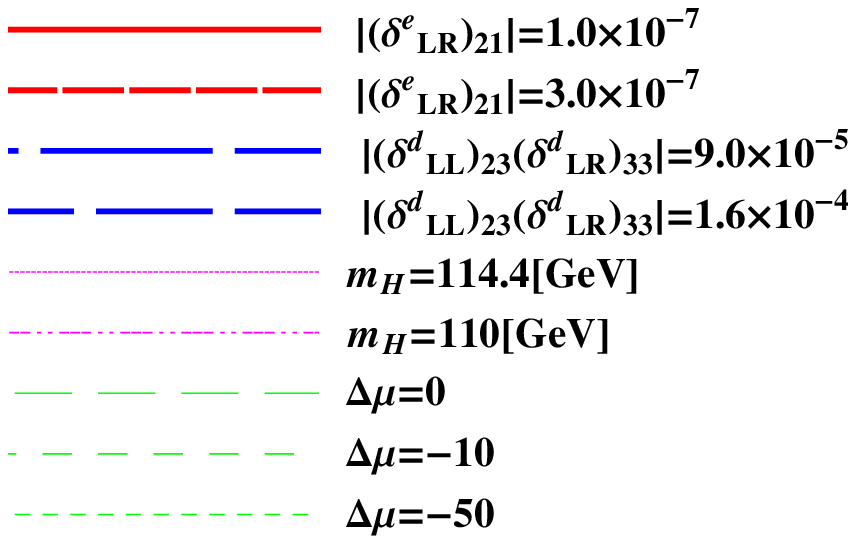}
\hfill 
\caption{
Contours of 
$|(\delta_{LR}^e)_{21}|$, 
$|(\delta_{LL}^d)_{23}(\delta_{LR}^d)_{33}|$, 
$|\Delta_{\mu}|$ and 
$m_{H_0}$ 
as functions of the gaugino mass ratios 
$r_1=M_1/M_3$ and $r_2=M_2/M_3$ at $M_{\rm GUT}$. 
The region $3 \le r_2 \le 6$ 
is magnified in the right panel. 
The parameters are chosen as $M_{\rm SB}=500$~GeV,
$\alpha_1=1/(4\pi^2),\alpha_2=2/(4 \pi^2)$, $\alpha_3=1/(4 \pi^2)$, 
and $\tan \beta =4$. 
The gluino mass~$M_3$ is set to 416~GeV at $M_{\rm GUT}$. 
}
\label{pti}
\end{figure}

Furthermore we introduce a quantity 
\be
 \Dlt_\mu \equiv \frac{\mu}{2M_Z^2}\frac{\der M_Z^2}{\der\mu}, 
\ee
which describes the sensitivity of the Z-boson mass $M_Z$ 
to the $\mu$ parameter~\cite{Barbieri:1987fn}. 

In Figs.~\ref{1t2}, \ref{pt2} and \ref{pti}, 
we plot contours of various quantities as functions of 
the gaugino mass ratios at $M_{\rm GUT}$, 
\be
r_1 \equiv M_1/M_3, \;\;\;\;\;
r_2 \equiv M_2/M_3. 
\ee
Contours of $|(\delta_{LR}^e)_{21}|$, 
$|(\delta_{LL}^d)_{23}(\delta_{LR}^d)_{33}|$, 
$|\Delta_{\mu}|$ and the lightest CP-even neutral Higgs mass~$m_{H_0}$ 
are shown in Fig.~\ref{1t2} for $M_{\rm SB}=100$~GeV, 
$\alpha_1=1$, $\alpha_2=1/2$ and $\alpha_3=1/(4\pi^2)$. 
Similar contours are plotted in Fig.~\ref{pt2} 
for$M_{\rm SB}=200$~GeV, $\alpha_1=1/2$, $\alpha_2=1/(8\pi^2)$ and $\alpha_3=1/(4\pi^2)$, and 
in Fig.~\ref{pti} for $M_{\rm SB}=500$ GeV, 
$\alpha_1=\alpha_3=1/(4 \pi^2)$ and 
$\alpha_2=2/(4 \pi^2)$. 
The gluino mass~$M_3$ at $M_{\rm GUT}$ is 343~GeV, 383~GeV and 
416~GeV in Figs.~\ref{1t2}, \ref{pt2} and \ref{pti}, respectively.  
Ratios of the gaugino masses at $M_{\rm EW}$ to those at $M_{\rm GUT}$ are 
given by 
\be
 \frac{M_1(M_{\rm EW})}{M_1(M_{\rm GUT})}=0.4, \;\;\;\;\;
 \frac{M_2(M_{\rm EW})}{M_2(M_{\rm GUT})}=0.8, \;\;\;\;\;
 \frac{M_3(M_{\rm EW})}{M_3(M_{\rm GUT})}=2.9. 
\ee
The gaugino masses at $M_{\rm EW}$ are read off from these relations. 

The curves with fixed values of $m_{H_0}$ represents 
the upper bound on the lightest CP-even Higgs mass 
in our model~\cite{Carena:1995wu}. 
The region surrounded by the curve representing 
$|\Delta_{\mu}| = 10$ is free from the little hierarchy problem of MSSM. 
There is no (less than 10\%) fine-tuning of $\mu$ 
in this region in order to realize the correct EW symmetry breaking. 

As for the SUSY flavor violations, 
the most stringent experimental constraints on the 
mass insertion parameters in our model are typically 
expressed as 
$|(\delta_{LR}^e)_{21}| \lesssim 10^{-7}$ and 
$|(\delta_{LL}^d)_{23}(\delta_{LR}^d)_{33}| \lesssim 10^{-4}$ 
which come from the upper bound on the branching ratios of 
$\mu \to e \gamma$ and $b \to s \gamma$, respectively~\cite{Abe:2004tq}. 
Recall that contribution of the direct mediation by $F^X$ 
has only weak flavor dependence for our charge assignment 
while that of the moduli mediation 
by $F^{T^{I'}}$ can disturb such flavor structures. 
Namely, the flavor dependence of the soft masses becomes weaker 
when $\alp_{I'}\ll 1$. 
We emphasize that the allowed region from $\mu \to e \gamma$ 
is much wider in Fig.~\ref{pt2} with $\alpha_{2}=1/(8 \pi^2)$, 
compared with the one in Fig.~\ref{1t2} with $\alpha_{2}=1/2$. 
This is because the flavor structure of the first and second 
generations in the lepton sector are governed by the $U(1)_2$ charges, 
and only $\alp_2$ affects the flavor structure. 
Note that the contours of 
$|(\delta_{LL}^d)_{23}(\delta_{LR}^d)_{33}|$ in Fig.~\ref{pti} 
is drawn differently compared with Figs.~\ref{1t2} and \ref{pt2}. 
This is because contributions of the direct mediation dominates 
over those of the moduli mediation 
since $\alp_{I'}\ll 1$ in Fig.~\ref{pti}, 
and then the scaling of the soft terms are changed. 

It is commonly said that models with the gravity-mediated SUSY breaking 
suffer from the SUSY flavor problem. 
However, 
we find that suitable charge assignments for $U(1)_{I'}$ do not 
cause the SUSY flavor problem while realize a viable flavor structure for quarks 
and leptons (without tachyonic squarks and sleptons). 
We should emphasize that this is due to the existence of 
multiple moduli, which induce additional contact terms~$\abs{Q_a}^2\abs{X}^2$ 
in 4D effective K\"ahler potential. 
This is in sharp contrast to models with a single modulus 
discussed in many papers. 

Finally we comment that the Higgs mass bound seems to be most stringent 
in Figs.~\ref{1t2}, \ref{pt2} and \ref{pti}. This is due to the fact that 
the visible sector is assumed to be MSSM with $\tan \beta =4$. 
It may become milder without affecting the flavor structure if we take 
a larger value of $\tan \beta$~\footnote{
For large values of $\tan\beta$, contributions of the bottom Yukawa coupling 
become important in the RG running, which we have neglected here. 
}  
and/or extend the Higgs sector, such as the next to minimal SUSY SM. 
Since we are focusing on the flavor structure here, 
we would leave analyses on such extensions for a separate paper. 
Even for MSSM with $\tan \beta =4$, 
we find a region where all the experimental constraints 
considered here are satisfied in Fig.~\ref{pt2}.

\section{Summary} \label{summary}
We have systematically studied the SUSY flavor structure of 
generic 5D SUGRA models, where all the hidden and visible 
sector fields are living in the whole 5D bulk spacetime, 
where $N=2$ SUSY exists. 
In order to realize the observed quark and lepton masses 
and mixings, the visible sector fields are quasi-localized 
in the extra dimension by a suitable charge assignment 
for $Z_2$-odd $U(1)_{I'}$ vector multiplets~$V^{I'}$. 
This type of models have been considered in many papers, 
but most of them assume that there is just a single 
$Z_2$-odd vector multiplet, \ie, the graviphoton multiplet. 
However, it has been shown in Ref.~\cite{Abe:2008ka} 
that induced squark and slepton masses 
become tachyonic in such a case. 
Besides, too large flavor violation generically occurs 
in the SUGRA models, \ie, the SUSY flavor problem.  
In our previous work~\cite{Abe:2008an}, we pointed out a new possibility 
of avoiding such problems by introducing 
an extra $Z_2$-odd vector multiplet 
other than the graviphoton multiplet. 
In such a case, additional contributions to 
4D effective K\"ahler potential~$K_{\rm eff}$  
appear after integrating out the $Z_2$-odd vector multiplets, 
and they affect the flavor structure of the soft SUSY-breaking mass matrices. 

In this paper, we have extended our previous work to more generic cases 
and specify conditions to solve the tachyonic sfermion 
problem~(\ref{cond:non-tachyonic}) and 
the SUSY flavor problem~(\ref{cond:d_a}). 
In fact, through a detailed phenomenological analysis, we have 
explicitly shown that the SUSY flavor problem can be avoided 
by introducing multiple vector multiplets $V^{I'}$ without 
encountering tachyonic sfermion problem mentioned above. 
Therefore we conclude that the SUSY flavor structure of 
gravity-mediated SUSY breaking scenario can be 
{\it controllable}~\cite{Conlon:2007dw} 
once it is concretely constructed, contrary to the general 
criticism that SUSY flavor violation is problematic for it. 

The additional contributions to $K_{\rm eff}$ by integrating out 
the bulk SUGRA fields 
have been discussed in the context of the string theory 
in Ref.~\cite{Anisimov:2001zz}. 
Because 5D SUGRA is the simplest set-up for the brane-world models, 
we can derive an explicit form of $K_{\rm eff}$ directly from 
the higher-dimensional theory. 
This enables us to perform detailed analyses on it, 
as we have done in this paper. 
Our systematic analyses also owe to the existence of 
the off-shell description of SUGRA~\cite{Kugo-Ohashi,Kugo:2002js}. 
It makes the derivation of the 4D effective theory transparent 
by utilizing the off-shell dimensional reduction~\cite{Abe:2007}. 


The results obtained in this paper are quite generic 
when the hierarchical flavor structure originates 
from the wave function localization in the extra dimension.  
Most of the results in Sec.~\ref{specific_model} do not much depend on 
the choice of the norm function~(\ref{nft1t2t3}) if we choose 
a suitable charge assignment for $V^{I'}$. 
Finally we emphasize that the multi moduli case discussed in this paper 
is naturally realized 
when we consider the low-energy effective theories of the string theory.

\section*{Acknowledgments}
This work was supported in part by the Waseda University
Grant for Special Research Projects No.2011B-177 (H. A.), 
and Grant-in-Aid for Young Scientists (B) No.22740187 
from Japan Society for the Promotion of Science (Y. S.). 
H. A. and Y. S. thank the Yukawa Institute for Theoretical 
Physics at Kyoto University. 
Discussions during the YITP workshop~"Summer Institute 2011" 
were useful to complete this work.

\appendix

\section{Derivation of effective action} \label{derive:L_eff}
In this section, we provide a systematic derivation of 
the 4D effective action. 
We basically follow the off-shell dimensional reduction 
developed in Refs.~\cite{Abe:2007,Abe:2008an}. 
This procedure keeps the $N=1$ off-shell structure. 

First, we neglect the kinetic terms for $Z_2$-odd $N=1$ multiplets 
because they do not have zero-modes that are dynamical below 
the compactification scale. 
Then the $Z_2$-odd multiplets play a role of Lagrange multipliers, 
and their equations of motion extract zero-modes 
from the $Z_2$-even multiplets. 
This is the basic strategy. 

\subsection{Gauge kinetic functions and superpotential}  \label{derive:fW}
The kinetic terms for $\Sgm^{I''}$ are included through 
the gauge-invariant quantities~$\cV^{I''}$ in the $d^4\tht$-integral  
of (\ref{5D_action2}). 
Thus, dropping the kinetic terms for $\Sgm^{I''}$ 
in a gauge-invariant way is equivalent to imposing the constraint, 
\be
 \cV^{I''}=0.  \label{constraint1}
\ee

Before dropping the kinetic terms for the other $Z_2$-odd multiplets, 
let us redefine the chiral multiplets~$\Phi^{\hat{a}}$ as 
\bea
 \Phi^1 \eql \phi^{\frac{3}{2}}\phi^c, \;\;\;\;\;
 \Phi^2 = \phi^{\frac{3}{2}}, \nonumber\\
 \Phi^{2a-1} \eql \phi^{\frac{3}{2}}Q^c_a, \;\;\;\;\;
 \Phi^{2a} = \phi^{\frac{3}{2}}Q_a. 
\eea
Then, $\phi$ has the Weyl weight 1 and plays a role of 
the chiral compensator multiplet, 
while $Q_a$ has the Weyl weight zero, just like the matter chiral multiplets
in 4D off-shell SUGRA. 

Now we drop the kinetic terms for $\phi^c$ and $Q^c$, and obtain 
\bea
 \cL \eql -2\sbk{\int\dr^2\tht\brc{
 \phi^{\frac{3}{2}}\phi^c\brkt{\der_y+3k\cdot\Sgm}\phi^{\frac{3}{2}}
 -\sum_a\phi^{\frac{3}{2}}Q^c_a
 \brkt{\der_y+2c_a\cdot\Sgm}\brkt{\phi^{\frac{3}{2}}Q_a}}+\hc} \nonumber\\
 &&+\cdots, 
\eea
where the ellipsis denotes terms independent of $\phi^c$ and $Q^c$. 
Thus their equations of motion are read off as 
\be
 \brkt{\der_y+3k\cdot\Sgm}\phi^{\frac{3}{2}} = 0, \;\;\;\;\;
 \brkt{\der_y+2c_a\cdot\Sgm}\brkt{\phi^{\frac{3}{2}}Q_a} = 0. 
\ee
By solving these, we find the $y$-dependence of $\phi$ and $Q_a$ as 
\bea
 \phi(y) \eql \exp\brc{-2k\cdot\int_0^y\dr y'\;\Sgm(y')}\phi(0), \nonumber\\
 Q_a(y) \eql \exp\brc{(3k-2c_a)\cdot\int_0^y\dr y'\;\Sgm(y')}Q_a(0), 
\eea
where the 4D superspace coordinates~$x^\mu$ and $\tht^\alp$ 
in the arguments are suppressed. 
Note that these $y$-dependent factors can be eliminated by 
5D gauge transformations, 
\bea
 \tl{\phi} \eql e^{-2k\cdot\Lmd}\phi, \;\;\;\;\;
 \tl{Q}_a = e^{(3k-2c_a)\cdot\Lmd}Q_a, \nonumber\\
 \tl{V}^I \eql V^I+\Lmd^I+\bar{\Lmd}^I, \;\;\;\;\;
 \tl{\Sgm}^I = \Sgm^I+\der_y\Lmd^I = 0,  \label{5D_gauge_trf}
\eea
with the transformation parameters, 
\be
  \Lmd^I(y) \equiv -\int_0^y\dr y'\;\Sgm^I(y'). 
\ee

Since $\Sgm^{I'}$ are $Z_2$-even, 
the transformation parameters~$\Lmd^{I'}$ are discontinuous at $y=L$. 
The discontinuities correspond to zero-modes for $\Sgm^{I'}$ as 
\be
 \Lmd^{I'}|_{y=L-\ep} = -\frac{1}{2}T^{I'}, \;\;\;\;\;
 \Lmd^{I'}|_{y=L} = 0,  \label{gap:Lmd}
\ee
where
\be
 T^{I'} \equiv 2\int_0^L\dr y\;\Sgm^{I'}(y).  \label{def:T^I}
\ee
The zero-modes~$T^{I'}$ are called the moduli multiplets 
in the following~\footnote{
In the single modulus case, it corresponds 
to the radion multiplet~$T_{\rm rad}$. }. 
As a result, $\tl{V}^{I'}$ also have gaps at $y=L$ as 
\be
 \tl{V}^{I'}|_{y=L-\ep} = -\Re T^{I'}, \;\;\;\;\;
 \tl{V}^{I'}|_{y=L} = 0.  \label{gap:tlV}
\ee

After the above gauge transformation, the expressions of $\cV^I$ 
reduce to 
\be
 \cV^I = -\der_y\tl{V}^I. 
\ee
Thus the constraint~(\ref{constraint1}) indicates that 
the gauge-transformed $Z_2$-even vector superfield~$\tl{V}^{I''}$ 
are independent of $y$, \ie, 4D vector superfields. 

Under the gauge transformation, $\cL_{\rm vec}$ defined in (\ref{def:L_CS}) 
is invariant up to a total derivative for $y$~\cite{Dudas:2004ni},\footnote{
We have dropped total derivatives for $x^\mu$ and $\tht^\alp$. }
\be
 \dlt\cL_{\rm vec} = 
 \int\dr^2\tht\;\der_y\brkt{
 -\frac{C_{IJK}}{2}\Lmd^I\tl{\cW}^J\tl{\cW}^K}+\hc. 
\ee
This becomes surface terms after the $y$-integration, and thus 
\bea
 \int\dr y\;\dlt\cL_{\rm vec} \eql 
 \int\dr^2\tht\;\sbk{-\frac{C_{I'J''K''}}{2}
 \Lmd^{I'}\tl{\cW}^{J''}\tl{\cW}^{K''}+\cdots}_0^{L-\ep}+\hc 
 \nonumber\\
 \eql \int\dr^2\tht\brc{\frac{C_{I'J''K''}}{4}T^{I'}\tl{\cW}^{J''}\tl{\cW}^{K''}
 +\cdots}+\hc,
\eea
where the ellipsis denotes terms involving 
a $Z_2$-odd vector superfield~$\tl{V}^{I'}$. 
We have used (\ref{gap:Lmd}) in the second equality. 

Finally we drop the kinetic terms for $\tl{V}^{I'}$, and 
perform the $y$-integration. 
By utilizing all the above results, we obtain 
\bea
 \cL_{\rm eff} \eql \sbk{\int\dr^2\tht\;
 \frac{C_{I'J''K''}}{4}T^{I'}\tl{\cW}^{J''}\tl{\cW}^{K''}+\hc} \nonumber\\
 &&-3\int\dr^4\tht\;\abs{\tl{\phi}}^2\brc{
 \int_0^{L-\ep}\dr y\;\hat{\cN}^{1/3}(-\der_y\tl{V})e^{2k\cdot\tl{V}}
 \brkt{1-\sum_{a=2}^{n_H+1}e^{(2c_a-3k)\cdot\tl{V}}\abs{\tl{Q}_a}^2}^{2/3}} 
 \nonumber\\
 &&+\sbk{\int\dr^2\tht\;\tl{\phi}^3\brc{W^{(0)}(\tl{Q}_a)
 +e^{-3k\cdot T}W^{(L)}\brkt{e^{\brkt{\frac{3}{2}k-c_a}\cdot T}\tl{Q}_a}}+\hc}, 
 \label{L_eff1}
\eea
where $\hat{\cN}$ is a truncated norm function defined by 
\be
 \hat{\cN}(X) \equiv C_{I'J'K'}X^{I'}X^{J'}X^{K'}. 
\ee
Here we have used the fact that the gauge-transformed $\tl{\phi}$ and 
$\tl{Q}_a$ are $y$-independent in the region~$0\leq y<L$ but 
has discontinuities at $y=L$ as 
\be
 \tl{\phi}|_{y\neq L} = e^{k\cdot T}\tl{\phi}|_{y=L}, \;\;\;\;\;
 \tl{Q}_a|_{y\neq L} = e^{\brkt{c_a-\frac{3}{2}k}\cdot T}\tl{Q}_a|_{y=L}. 
\ee
The chiral multiplets~$\tl{\phi}$ and $\tl{Q}_a$ in (\ref{L_eff1}) 
are the ones evaluated at $y\neq L$. 

From (\ref{L_eff1}), we can read off the gauge kinetic functions 
and the superpotential in the effective theory as (\ref{defining_fcns}). 
We omit the tilde in the expression in 
(\ref{L_eff}) and (\ref{defining_fcns}). 

Notice that the only $y$-dependent superfields in (\ref{L_eff1}) 
are $\tl{V}^{I'}$. 
Since we have dropped the kinetic terms for $\tl{V}^{I'}$, 
we can integrate them out by their equations of motion. 
We will perform this procedure and calculate 
the effective K\"{a}hler potential in the next subsection.

\subsection{K\"{a}hler potential} \label{derive:K_eff}
Now we integrate $\tl{V}^{I'}$ out 
in the expression~(\ref{L_eff1}), and derive the effective K\"{a}hler 
potential~$K_{\rm eff}$, or $\Omg_{\rm eff}\equiv -3e^{-K_{\rm eff}/3}$. 
It is written as 
\bea
 \Omg_{\rm eff} 
 \eql -3\int_0^{L-\ep}\dr y\;\hat{\cN}^{1/3}(-\der_y\tl{V})e^{2k\cdot\tl{V}}
 \brkt{1-\sum_a e^{2d_a\cdot\tl{V}}\abs{\tl{Q}_a}^2}^{2/3},  
 \label{Omg_eff1}
\eea
where $d_{aI'}\equiv c_{aI'}-\frac{3}{2}k_{I'}$. 
We will omit the prime of the indices~$I',J',\cdots$ in the following. 

At first sight, it seems to be possible to integrate $\tl{V}^I$ out by 
finding their function forms in terms of $y$ 
and performing the $y$-integral.  
However, this naive procedure fails because we cannot solve 
the equations of motion as functions of $y$. 
In fact, the equations of motion for $\tl{V}^I$ are rewritten as 
\be
 \brc{\der_y\brkt{\frac{\hat{\cN}_J}{\hat{\cN}^{2/3}}}
 +6k_J\hat{\cN}^{1/3}
 -\frac{4\sum_a d_{aJ}e^{2d_a\cdot\tl{V}}\abs{\tl{Q}_a}^2}
 {1-\sum_b e^{2d_b\cdot\tl{V}}\abs{\tl{Q}_b}^2}\hat{\cN}^{1/3}}
 \cP^J_{\;\;I} = 0,  \label{EOM:V^I}
\ee
where the argument~$(-\der_y\tl{V})$ 
of the norm function are suppressed, and 
the projection operator~$\cP^I_{\;\;J}$ is defined in (\ref{def:cP}). 
The presence of $\cP^I_{\;\;J}$ indicates that
the number of independent equations is less than 
that of $\cV^I$. 
Thus we cannot solve $\cV^I$ as functions of $y$. 
This stems from the fact that 5D vector multiplets cannot be expanded 
into KK modes keeping $N=1$ off-shell structure 
because component fields have different physical degrees of freedom 
within the multiplets, as mentioned in Sec.~\ref{sc_GF}. 
Hence we need another method to integrate $\tl{V}^I$ out.  

Let us define 
\be
 v^I \equiv \frac{\der_y\tl{V}^I}{\der_y U}, \;\;\;\;\;
 U \equiv k\cdot\tl{V}. 
\ee
Then $\cF_J\equiv\hat{\cN}_J/\hat{\cN}^{2/3}$ is a function of $v^I$, and 
\bea
 &&\hat{\cN}^{1/3}(-\der_y\tl{V}) = \hat{\cN}^{1/3}(v)\cdot \brkt{-\der_y U}, 
 \nonumber\\
 &&\der_y\cF_J = \der_y v^K\cF_{KJ} = -2\der_y v^K\hat{\cN}^{1/3}(v)
 \brkt{a\cdot\cP}_{KJ}(v), 
\eea
where $a_{IJ}$ is defined in (\ref{def:a_IJ}). 
Thus (\ref{EOM:V^I}) is rewritten as 
\be
 \brc{\der_y v^K a_{KJ}
 +\brkt{3k_J-\frac{2\sum_a d_{aJ}e^{2d_a\cdot\tl{V}}\abs{\tl{Q}_a}^2}
 {1-\sum_b e^{2d_b\cdot\tl{V}}\abs{\tl{Q}_b}^2}}\der_y U}
 \cP^J_{\;\;I} = 0.  \label{EOM:V^I2}
\ee
Here and henceforth, the arguments of the norm function 
and its derivatives are understood as $v^I$ unless specified. 
From (\ref{EOM:V^I2}) and the constraint~$k_I(dv^I/dU)=0$, we obtain   
\bea
 \frac{dv^I}{dU} \eql a^{IJ}\brkt{-3\tl{k}_J
 +\frac{2\sum_a\tl{d}_{aJ}e^{2d_a\cdot\tl{V}}\abs{\tl{Q}_a}^2}
 {1-\sum_b e^{2d_b\cdot\tl{V}}\abs{\tl{Q}_b}^2}} \nonumber\\
 \eql \cG^I(v)+2\sum_a a^{IJ}\tl{d}_{aJ}e^{2d_a\cdot\tl{V}}
 \abs{\tl{Q}_a}^2 \nonumber\\
 &&+2\sum_{a,b}a^{IJ}\tl{d}_{aJ}e^{2(d_a+d_b)\cdot\tl{V}}\abs{\tl{Q}_a}^2
 \abs{\tl{Q}_b}^2+\cO\brkt{\abs{\tl{Q}}^6},  \label{dv/dU}
\eea
where $a^{IJ}$ is an inverse matrix of $a_{IJ}$, 
$\cG^I(v)\equiv -3a^{IJ}\tl{k}_J$, and 
\bea
 \tl{k}_J \defa k_J-\frac{k_K a^{KL}k_L}{k_K a^{KL}\hat{\cN}_L}
 \hat{\cN}_J 
 = \brkt{k\cdot\cP}_J-\frac{k\cdot\cP a^{-1}\cdot k}
 {2\hat{\cN}}\hat{\cN}_J, \nonumber\\
 \tl{d}_J \defa d_{aJ}-\frac{k_K a^{KL}d_{aL}}
 {k_K a^{KL}\hat{\cN}_L}\hat{\cN}_J 
 = \brkt{d_a\cdot\cP}_J-\frac{k\cdot\cP a^{-1}\cdot d_a}{2\hat{\cN}}
 \hat{\cN}_J. 
\eea
Here we have used that 
\be
 k_I a^{IJ}\hat{\cN}_J = 2\hat{\cN}, 
\ee
which follows from the relation~$a^{IJ}\hat{\cN}_J=2\hat{\cN}v^I$. 

Since $\tl{V}^I$ can be regarded as a function of $U$ through 
\be
 \tl{V}^I = \int_0^y\dr y'\;\der_y\tl{V}^I 
 = \int_0^y\dr y'\;v^I\der_y U 
 = \int_0^U\dr U'\;v^I(U'),  \label{rel:V-v}
\ee
the $y$-integral in (\ref{Omg_eff1}) can be converted 
into the integral for $U$. 
Thus, if we find explicit function forms of $v^I$ in terms of $U$, 
we can integrate $\tl{V}^I$ out by using the boundary conditions, 
\be
 U|_{y=0} = 0, \;\;\;\;\;
 U|_{y=L-\ep} = -k\cdot\Re T. \label{limit:U}
\ee
This is actually possible (at least in principle) 
because the number of independent $v^I$ is equal to 
that of the equations of motion due to the constraint~$k_Iv^I=1$. 

We expand $v^I$ as 
\be
 v^I = \bar{v}^I(U)+\sum_a g_a^I(U)\abs{\tl{Q}_a}^2
 +\sum_{a,b}g_{ab}^I(U)\abs{\tl{Q}_a}^2\abs{\tl{Q}_b}^2
 +\cO\brkt{\abs{\tl{Q}}^6}.  \label{expand:v}
\ee
From (\ref{dv/dU}), 
the functions~$\bar{v}^I(U)$, $g_a^I(U)$ and $g_{ab}^I(U)$ satisfy 
\bea
 \frac{d\bar{v}^I}{dU} \eql \cG^I(\bar{v}), \nonumber\\
 \frac{dg_a^I}{dU} \eql \cG^I_J(\bar{v})g_a^J(U)
 +2\left.\brkt{a^{IJ}\cdot\tl{d}_{aJ}}\right|_{v=\bar{v}}
 e^{2\int_0^U\dr U'\;d_a\cdot\bar{v}}, \nonumber\\
 \frac{dg_{ab}^I}{dU} \eql \cG^I_J(\bar{v})g_{ab}^J
 +\frac{1}{2}\cG^I_{JK}(\bar{v})g_a^Jg_b^K
 +2e^{2\int_0^U\dr U'\;\brkt{d_a+d_b}\cdot\bar{v}}
 \left.\brkt{a^{IJ}\cdot\tl{d}_{aJ}}\right|_{v=\bar{v}}   \label{dg/dU} \\
 &&+2e^{2\int_0^U\dr U'\;d_a\cdot\bar{v}}
 \brc{\left.\der_K\brkt{a^{IJ}\cdot\tl{d}_{aJ}}\right|_{v=\bar{v}}g_b^K(U)
 +2\left.\brkt{a^{IJ}\cdot\tl{d}_{aJ}}\right|_{v=\bar{v}}
 \int_0^U\dr U'\;d_a\cdot g_b(U')}, \nonumber
\eea
where $\cG^I_J(v)\equiv\der\cG^I/\der v^J$ 
and $\cG^I_{JK}(v)\equiv\der^2\cG^I/\der v^J\der v^K$. 
From (\ref{rel:V-v}) and (\ref{limit:U}), it follows that 
\be
 -\Re T^I = \int_0^{-k\cdot\Re T}dU\;v^I(u), 
\ee
which leads to 
\be
 \int_0^{-k\cdot\Re T}dU\;\bar{v}^I(U) = -\Re T^I, \;\;\;\;\;
 \int_0^{-k\cdot\Re T}dU\;g_a^I(U) 
 = \int_0^{-k\cdot\Re T}dU\;g_{ab}^I(U) = 0.   \label{bdcd:bv}
\ee

Substituting (\ref{expand:v}) into (\ref{Omg_eff1}), we obtain 
\bea
 \Omg_{\rm eff} \eql 3\int_0^{-k\cdot\Re T}\dr U\;
 \hat{\cN}^{1/3}(v)e^{2U}\brkt{1-\sum_a e^{2\int_0^U\dr U'\;d_a\cdot v}
 \abs{\tl{Q}_a}^2}^{2/3} \nonumber\\
 \eql 3\int_0^{-k\cdot\Re T}\dr U\;e^{2U}\left[
 \hat{\cN}^{1/3}(\bar{v})
 +\sum_a\brc{\frac{\cF_I(\bar{v})}{3}g_a^I
 -\frac{2\hat{\cN}^{1/3}(\bar{v})}{3}e^{2d_a\cdot\bar{V}}}\abs{\tl{Q}_a}^2 
 \right. \nonumber\\
 &&\hspace{25mm}
 +\sum_{a,b}\left\{\frac{\cF_I(\bar{v})}{3}g_{ab}^I
 +\frac{\cF_{IJ}(\bar{v})}{6}g_a^Ig_b^I
 -\frac{2\cF_I(\bar{v})}{9}g_a^I e^{2d_b\cdot\bar{V}} \right. \nonumber\\
 &&\hspace{30mm}\left.\left. 
 -\hat{\cN}^{1/3}(\bar{v})\brkt{\frac{4}{3}e^{2d_a\cdot\bar{V}}
 \int_0^U\dr U'\;d_a\cdot g_b+\frac{e^{2(d_a+d_b)\cdot\bar{V}}}{9}}
 \right\}\abs{\tl{Q}_a}^2\abs{\tl{Q}_b}^2\right] \nonumber\\
 &&+\cO(\abs{\tl{Q}}^6),  \label{Omg_eff2} 
\eea
where 
\be
 \bar{V}^I(U) \equiv \int_0^U\dr U'\;\bar{v}^I(U'). 
\ee

Now we will find the function forms of $\bar{v}^I(U)$, $g_a^I(U)$ 
and $g_{ab}^I(U)$. 
Since it is difficult to find general solutions 
to the differential equations in (\ref{dg/dU}), 
we focus on a simple case where the condition, 
\be
 k_I\cP^I_{\;\;J}(\bar{v}) = 0, \label{mdl_cstrt1}
\ee
is satisfied. 
In this case, $\cG^I(\bar{v})=0$ and $\bar{v}^I$ reduces 
to a constant for $U$. 
By using (\ref{bdcd:bv}), it is determined as 
\be
 \bar{v}^I = \frac{\Re T^I}{k\cdot\Re T}. \label{expr:bv} 
\ee
Thus the condition~(\ref{mdl_cstrt1}) means that 
\be
 k_I\cP^I_{\;\;J}(\Re T) = 0. \label{mdl_cstrt2}
\ee
Substituting (\ref{expr:bv}) into the second equation in (\ref{dg/dU}), 
we obtain 
\be
 g_a^I(U) = \brc{e^{2d_a\cdot\bar{v}U}
 -Y(d_a\cdot T)Y\brkt{\frac{k\cdot T}{2}\bdm{\cG}}^{-1}
 e^{\bdm{\cG}U}}^I_{\;\;J}C_a^J,  \label{expr:g^I1}
\ee
where 
\be
 C_a^I \equiv \sbk{2\brkt{-\bdm{\cG}+2d_a\cdot\bar{v}}^{-1}
 \cdot a^{-1}\cdot\tl{d}_a}^I, 
 \;\;\;\;\;
 Y(z) \equiv \frac{1-e^{-2\Re z}}{2\Re z},
\ee
and $\bdm{\cG}$ is a matrix whose $(I,J)$-component is $\cG^I_J(\bar{v})$. 
Under the constraint~(\ref{mdl_cstrt1}), we find that 
$k\cdot a^{-1}\cdot\tl{d}_a=0$ and 
$\bdm{\cG}^I_{\;\;J}(\bar{v}) = -2\brkt{\dlt^I_{\;\;J}-k_J\bar{v}^I}$, 
which indicates that 
\be
 \brkt{\bdm{\cG}a^{-1}\tl{d}_a}^I = -2\brkt{a^{-1}\tl{d}_a}^I.  
\ee
Thus (\ref{expr:g^I1}) is rewritten as 
\be
 g_a^I(U) = \frac{1}{1+d_a\cdot\bar{v}}
 \brc{e^{2d_a\cdot\bar{v}U}
 -\frac{Y(d_a\cdot T)}{Y(-k\cdot T)}e^{-2U}}
 \brkt{a^{-1}\cdot\tl{d}_a}^I.  \label{expr:g^I2}
\ee

Substituting (\ref{expr:bv}) and (\ref{expr:g^I2}) into 
the third equation in (\ref{dg/dU}), we obtain 
\be
 g_{ab}^I(U) = \brkt{e^{\bdm{\cG}U}}^I_{\;\;J}
 \int_{U_0}^U\dr U'\;\brkt{e^{-\bdm{\cG}U'}}^J_{\;\;K}C_{ab}^K(U'), 
\ee
where $U_0$ is a constant, which is determined by (\ref{bdcd:bv}), and 
\bea
 C_{ab}^I(U) \defa \frac{1}{2}\cG^I_{JK}(\bar{v})g_a^Jg_b^K
 +2e^{2(d_a+d_b)\cdot\bar{v}U}
 \left.\brkt{a^{IJ}\cdot\tl{d}_{aJ}}\right|_{v=\bar{v}} \\
 &&+2e^{2d_a\cdot\bar{v}U}\brc{\left.\der_K\brkt{a^{IJ}\cdot\tl{d}_{aJ}}
 \right|_{v=\bar{v}}g_b^K(U)
 +2\left.\brkt{a^{IJ}\cdot\tl{d}_{aJ}}\right|_{v=\bar{v}}
 \int_0^U\dr U'\;d_a\cdot g_b(U')}.  \nonumber
\eea
Under the constraint~(\ref{mdl_cstrt1}), we can show that 
\be
 \hat{\cN}_I(\bar{v})C_{ab}^I(U) = 0, 
\ee
which leads to 
\be
 \cF_I(\bar{v})g_{ab}^I 
 = e^{-2U}\int_{U_0}^U\dr U'\;e^{2U'}\cF_I(\bar{v})C_{ab}^I(U') = 0. 
 \label{cFg}
\ee

Substituting (\ref{expr:bv}) and (\ref{expr:g^I2}) 
into (\ref{Omg_eff2}) and using (\ref{cFg}), we obtain the final result. 
\be
 \Omg_{\rm eff}
 = \Omg^{(0)}+\sum_a\Omg^{(2)}_a\abs{\tl{Q}_a}^2
 +\sum_{a,b}\Omg^{(4)}_{a,b}\abs{\tl{Q}_a}^2\abs{\tl{Q}_b}^2
 +\cO(\abs{\tl{Q}}^6), 
\ee
where 
\bea
 \Omg^{(0)} \eql -3\hat{\cN}^{1/3}(\Re T)Y(k\cdot T), \nonumber\\
 \Omg^{(2)}_a \eql 2\hat{\cN}^{1/3}(\Re T)Y((k+d_a)\cdot T), \nonumber\\
 \Omg^{(4)}_{a,b} \eql \hat{\cN}^{1/3}(\Re T)\left[
 -\frac{\brkt{d_a\cdot\cP a^{-1}\cdot d_b}\brc{Y((k+d_a+d_b)\cdot T)
 -\frac{Y(d_a\cdot T)Y(d_b\cdot T)}{Y(-k\cdot T)}}}
 {\brc{(k+d_a)\cdot\Re T}\brc{(k+d_b)\cdot\Re T}} \right. \nonumber\\
 &&\hspace{25mm}\left. 
 +\frac{Y\brkt{(k+d_a+d_b)\cdot T}}{3} \right]. 
\eea
We have used various relations stemming from 
the fact that $\hat{\cN}$ is a cubic polynomial. 
In (\ref{defining_fcns}) in the text, 
we omit the tilde of $\tl{Q}_a$.

\section{Comment on integration out of $\bdm{V^{I'}}$} \label{comment:4-fermi}
Here we provide a comment on the effect of integrating out 
the $Z_2$-odd vector multiplets~$V^{I'}$. 
As mentioned in Sec.~\ref{quartic_cp}, this procedure induces 
the quartic terms for $Q_a$ in the effective K\"ahler potential. 
Let us denote the components of $Q_a$ as 
$Q_a=q_a+\tht\chi_{Q_a}+\tht^2F_{Q_a}$. 
Then we have four-fermion interactions coming from the quartic terms 
in $\Omg_{\rm eff}$, 
\be
 \cL_{\mbox{\scriptsize 4-fermi}} = 
 \abs{\vev{\phi}}^2\frac{\hat{\cN}^{1/3}(\Re\vev{T})}{8}
 \sum_{a,b}\tl{\Omg}^{(4)}_{a,b}
 \abs{\chi_{Q_a}\chi_{Q_b}}^2.  \label{4-fermi}
\ee
To simplify the explanation, let us consider a case that 
$(k+d_a)\cdot\Re\vev{T}<0$ and $d_b\cdot\Re\vev{T}>0$. 
Then, 
$Y((k+d_a+d_b)\cdot\vev{T})\ll \frac{Y(d_a\cdot\vev{T})Y(d_b\cdot\vev{T})}
{Y(-k\cdot\vev{T})}$ and the above four-fermion terms are
proportional to $d_a\cdot\cP a^{-1}\cdot d_b$. 
However, since all the vector components in $V^{I'}$ 
including the graviphoton are physical, 
the contributions of the diagrams in Fig.~\ref{diagram} 
to $\cL_{\mbox{\scriptsize 4-fermi}}$ are expected to be proportional to 
$d_a\cdot a^{-1}\cdot d_b$, rather than $d_a\cdot\cP a^{-1}\cdot d_b$. 
Note that the coefficient of the first term in (\ref{tlOmg4}) 
is rewritten as 
\be
 \frac{d_a\cdot\cP a^{-1}\cdot d_b}{\brc{(k+d_a)\cdot\Re T}
 \brc{(k+d_b)\cdot\Re T}} 
 = \frac{d_a\cdot a^{-1}\cdot d_b}{\brc{(k+d_a)\cdot\Re T}
 \brc{(k+d_b)\cdot\Re T}}
 -\frac{2}{3},  \label{lack_contribution}
\ee
where we have used (\ref{moduli_align}). 
The second term in the left-hand side corresponds to 
the lack of the graviphoton contribution. 
In fact, it is restored when the promotion~(\ref{FDpromote}) 
is taken into account. 
For example, each derivative~$\der_\mu$ is promoted to 
the covariant derivative~$\cD_\mu$ that includes 
the $U(1)_A$ gauge field~$A_\mu$. 
($U(1)_A$ is the R-symmetry of the $N=1$ superconformal algebra.) 
This gauge field does not have a kinetic term in the action, 
and is an auxiliary field. 
By integrating out the auxiliary fields in the Weyl multiplet 
including $A_\mu$, we obtain an additional contribution to (\ref{4-fermi}),  
and cancel the second term in (\ref{lack_contribution}). 
Notice that this contribution does not originate from 
the $\abs{Q_a}^2\abs{Q_b}^2$ terms in $\Omg_{\rm eff}$, 
but is purely SUGRA interaction.  
Hence it is not accompanied with a term $\abs{q_a}^2\abs{F_{Q_b}}^2$, 
which contributes to the soft SUSY-breaking scalar masses. 

As a result, the dominant part of the four-fermion interaction 
is proportional to $d_a\cdot a^{-1}\cdot d_X$, while that of 
the soft scalar mass is to $d_a\cdot\cP a^{-1}\cdot d_X$, 
as shown in Sec.~\ref{quartic_cp}.

\section{Explicit expressions in the illustrative model} \label{explicit_forms}
Here we collect explicit expressions in the illustrative model 
discussed in Sec.~\ref{specific_model}. 

From the definition~$K_{\rm eff}=-3\ln(-\Omg_{\rm eff}/3)$ and 
(\ref{defining_fcns}), we obtain the K\"ahler metric as 
\bea
 K_{I'\bar{J'}} \defa \der_{T^{I'}}\der_{\bar{T}^{J'}}K_{\rm eff} 
 =\frac{1}{2}a_{I'J'}-\frac{3}{4}k_{I'}k_{J'}\cY(k\cdot\Re T), 
 \nonumber\\
 K_{Q_a\bar{Q}_b} \defa \der_{Q_a}\der_{\bar{Q}_b}K_{\rm eff} 
 = \frac{2Y((k+d_a)\cdot T)}{Y(k\cdot T)}\dlt_{ab},  \nonumber\\
 K_{X\bar{X}} \defa \der_X\der_{\bar{X}}K_{\rm eff} 
 = \frac{2Y((k+d_X)\cdot T)}{Y(k\cdot T)},  \label{K-metric}
\eea
where $\cY(x)$ is defined from $Y(x)$ ($x$: real) in (\ref{def:Y}) as 
\be
 \cY(x) \equiv \frac{Y\der_x^2 Y-(\der_x Y)^2}{Y^2}(x)
 = \frac{1+e^{4x}-2e^{2x}(1+2x^2)}{(1-e^{2x})^2x^2}.  \label{def:cY}
\ee
The function $\cY(x)$ is an even function and monotonically decreasing 
function of $\abs{x}$.
For example, $\cY(0)=1/3$ and $\cY(\pm\kp L)\simeq 0.032$ 
when $\kp L=5.6$. 

In Sec.~\ref{specific_model}, we have assumed that
only $F^{T^{I'}}$ ($I'=1,2,3$) and $F^X$ have nonvanishing VEVs. 
Thus, the soft scalar mass normalized by $M_{\rm SB}$ defined in (\ref{def:M_SB}) 
is calculated as 
\be
 \frac{m_a^2}{M_{\rm SB}^2} \simeq \frac{1}{K_{X\bar{X}}}\brc{
 -\frac{\tl{\Omg}^{(4)}_{a,X}(\Re\vev{T})}{Y(\tl{c}_a)}
 +\frac{F^{T^{I'}}\bar{F}^{T^{J'}}}{\left|F^X\right|^2}
 \brkt{\frac{a_{I'J'}}{6}-\frac{1}{4}(k+d_a)_{I'}(k+d_a)_{J'}\cY(\tl{c}_a)}}, 
 \label{explicit:ma/MSB}
\ee
where $\tl{c}_a\equiv\sum_{I'}\tl{c}_a^{I'}$. 

For the (truncated) norm function chosen as (\ref{nft1t2t3}), 
we can obtain explicit forms of various functions of 
$\cX^{I'}$, where $I'=1,2,3$ is the index of the $Z_2$-odd vector multiplets.  
The coefficient matrix of the vector boson kinetic term~$a_{I'J'}$ 
defined in (\ref{def:a_IJ}) is calculated as
\be
 a_{I'J'}(\cX) = \begin{pmatrix} \frac{1}{2(\cX^1)^2} & & \\
 & \frac{1}{2(\cX^2)^2} & \\ & & \frac{1}{2(\cX^3)^2} \end{pmatrix}, 
 \label{explicit:a_IJ}
\ee
and the projection operator~$\cP^I_{\;\;J}$ defined in (\ref{def:cP}) is 
\be
 \cP^I_{\;\;J}(\cX) = \frac{1}{3}
 \begin{pmatrix} 2 & -\frac{\cX^1}{\cX^2} & -\frac{\cX^1}{\cX^3} \\
 -\frac{\cX^2}{\cX^1} & 2 & -\frac{\cX^2}{\cX^3} \\
 -\frac{\cX^3}{\cX^1} & -\frac{\cX^3}{\cX^2} & 2 \end{pmatrix}. 
 \label{explicit:P^I_J}
\ee
Thus, the symmetric matrix~$\cP a^{-1}$ appearing in the expression of 
$\tl{\Omg}^{(4)}_{a,b}$ in (\ref{tlOmg4}) is obtained as 
\be
 \brc{\cP a^{-1}}(\cX) = \frac{2}{3}\begin{pmatrix}
 2(\cX^1)^2 & -\cX^1\cX^2 & -\cX^1\cX^3 \\ 
 -\cX^1\cX^2 & 2(\cX^2)^2 & -\cX^2\cX^3 \\
 -\cX^1\cX^3 & -\cX^2\cX^3 & 2(\cX^3)^2 \end{pmatrix}. \label{explicit:cPa}
\ee

\ignore{
The ratio of the $F$-terms~$\alp_{I'}$ defined in (\ref{def:M_SB}) are 
estimated as 
\be
 (\alp_1,\alp_2,\alp_3) \simeq \begin{cases}
 \frac{0.13}{\sqrt{Y(\tl{c}_X)}\left|F^X\right|}\brkt{
 \left|F^1\right|,2\left|F^2\right|,4\left|F^3\right|}, 
 & \brkt{\mbox{flat case ($\kp L=0$)}},  \\
 \frac{0.029}{\sqrt{Y(\tl{c}_X)}\left|F^X\right|}
 \brkt{\left|F^1\right|,2\left|F^2\right|,4\left|F^3\right|}, 
 & \brkt{\mbox{warped case ($\kp L=4.8$)}}, \end{cases}
\ee
where $\tl{c}_X$ is defined in (\ref{def:tlc}). 
Note that the second term of $K_{I'\bar{J}'}$ in (\ref{K-metric}) is
negligible in both cases. 
The typical SUSY-breaking scale~$M_{\rm SB}$ is estimated as 
\be
 M_{\rm SB} \simeq \begin{cases} 1.4\sqrt{Y(\tl{c}_X)}\left|F^X\right|, 
 & \brkt{\mbox{flat case ($\kp L=0$)}}, \\
 4.4\sqrt{Y(\tl{c}_X)}\left|F^X\right|. 
 & \brkt{\mbox{warped case ($\kp L=4.8$)}}. \end{cases}
\ee
}


\begin{thebibliography}{99}
\bibitem{Arkani-Schmaltz} N.~Arkani-hamed and M.~Schmaltz, 
 \PR{D61} (2000) 033005 [{\tt hep-ph/9903417}];
 D.E.~Kaplan and T.M.P.~Tait, \JH{0006} (2000) 020 
 [{\tt hep-ph/0004200}]. 

\bibitem{RS} L.~Randall and R.~Sundrum, \PRL{83} (1999) 
 3370 [{\tt hep-ph/9905221}]. 
 
\bibitem{Servant:2002aq}
  G.~Servant, T.~M.~P.~Tait,
  Nucl.\ Phys.\  {\bf B650 } (2003)  391-419.
  [hep-ph/0206071].
  
\bibitem{Kugo-Ohashi} T.~Kugo and K.~Ohashi, \PTP{105} (2001) 323 
 [{\tt hep-ph/0010288}]; T.~Fujita and K.~Ohashi, \PTP{106} (2001) 221 
 [{\tt hep-th/0104130}]; T.~Fujita, T.~Kugo and K.~Ohashi, \PTP{106} 
 (2001) 671 [{\tt hep-th/0106051}]. 

\bibitem{Kugo:2002js}
  T.~Kugo and K.~Ohashi,
  \PTP{108} (2002) 203 [{\tt hep-th/0203276}]. 
  
\bibitem{Abe:2007} H.~Abe and Y.~Sakamura, \NP{B796} (2008) 224 
 [{\tt arXiv:0709.3791}]. 

\bibitem{Paccetti Correia:2004ri}
  F.~Paccetti Correia, M.~G.~Schmidt and Z.~Tavartkiladze,
  \NP{B709} (2005) 141 [{\tt hep-th/0408138}]. 
  
\bibitem{Abe:2004}
  H.~Abe and Y.~Sakamura,
  \JH{0410} (2004) 013 [{\tt hep-th/0408224}]. 
  
\bibitem{Abe:2007zv}
  H.~Abe, Y.~Sakamura,
  Nucl.\ Phys.\  {\bf B796 } (2008)  224-245.
  [arXiv:0709.3791 [hep-th]]. 

\bibitem{SUSY_RS}
  R.~Altendorfer, J.~Bagger, D.~Nemeschansky,
  Phys.\ Rev.\  {\bf D63 } (2001)  125025.
  [hep-th/0003117]; 
  A.~Falkowski, Z.~Lalak, S.~Pokorski,
  Phys.\ Lett.\  {\bf B491 } (2000)  172-182.
  [hep-th/0004093];
  T.~Gherghetta, A.~Pomarol,
  Nucl.\ Phys.\  {\bf B586 } (2000)  141-162.
  [hep-ph/0003129].
  
\bibitem{5D_Mtheory} A.~Lukas, B.A.~Ovrut, K.S.~Stelle 
 and D.~Waldram, \PR{D59} (1999) 086001 [{\tt hep-th/9803235}];
 \NP{B552} (1999) 246 [{\tt hep-th/9806051}]. 
  
\ignore{
 \bibitem{Choi:2003di}
  K.~-w.~Choi, D.~Y.~Kim, I.~-W.~Kim, T.~Kobayashi,
  Eur.\ Phys.\ J.\  {\bf C35 } (2004)  267-275.
  [hep-ph/0305024].
}
  
\bibitem{Abe:2008ka}
  H.~Abe, T.~Higaki, T.~Kobayashi, Y.~Omura,
  JHEP {\bf 0804 } (2008)  072.
  [arXiv:0801.0998 [hep-th]]. 
  
\bibitem{Abe:2008an}
  H.~Abe, Y.~Sakamura,
  Phys.\ Rev.\  {\bf D79}, 045005 (2009).
  [arXiv:0807.3725 [hep-th]]. 
  
\bibitem{ArkaniHamed:2001tb}
  N.~Arkani-Hamed, T.~Gregoire, J.~G.~Wacker,
  JHEP {\bf 0203}, 055 (2002).
  [hep-th/0101233].

\bibitem{Correia:2006pj}
  F.~P.~Correia, M.~G.~Schmidt and Z.~Tavartkiladze,
  \NP{B751} (2006) 222 [{\tt hep-th/0602173}]. 
  
\bibitem{4Doffshell}
 M.~Kaku, P.K.~Townsend and P.~Van Nieuwenhuizen, \PRL{39} (1977) 1109; 
 \PL{B69} (1977) 304; \PR{D17} (1978) 3179; 
 T.~Kugo and S.~Uehara, \NP{B226} (1983) 49; \PTP{73} (1985) 235. 
 
\bibitem{Sakamura:2011df}
  Y.~Sakamura,
  [arXiv:1107.4247 [hep-th]].
 
\bibitem{Marti:2001iw}
  D.~Marti, A.~Pomarol,
  Phys.\ Rev.\  {\bf D64}, 105025 (2001).
  [hep-th/0106256]. 
  
\bibitem{Bergshoeff:2000zn}
  E.~Bergshoeff, R.~Kallosh, A.~Van Proeyen,
  JHEP {\bf 0010}, 033 (2000).
  [hep-th/0007044]. 
  
\bibitem{Wess:1992cp}
  J.~Wess and J.~Bagger,
{\it  Princeton, USA: Univ. Pr. (1992) 259 p}

\bibitem{Luty:2000ec}
  M.~A.~Luty, R.~Sundrum,
  Phys.\ Rev.\  {\bf D64}, 065012 (2001).
  [hep-th/0012158]. 

\bibitem{Choi:2004sx}
  K.~Choi, A.~Falkowski, H.~P.~Nilles, M.~Olechowski, S.~Pokorski,
  JHEP {\bf 0411}, 076 (2004).
  [hep-th/0411066].

\bibitem{Choi:2005ge}
  K.~Choi, A.~Falkowski, H.~P.~Nilles, M.~Olechowski,
  Nucl.\ Phys.\  {\bf B718}, 113-133 (2005).
  [hep-th/0503216].

\bibitem{Abe:2006xp}
  H.~Abe, T.~Higaki, T.~Kobayashi, Y.~Omura,
  Phys.\ Rev.\  {\bf D75}, 025019 (2007).
  [hep-th/0611024]; 
  H.~Abe, T.~Higaki, T.~Kobayashi,
  Phys.\ Rev.\  {\bf D76}, 105003 (2007).
  [arXiv:0707.2671 [hep-th]].

\bibitem{Dudas:2006gr}
  E.~Dudas, C.~Papineau, S.~Pokorski,
  JHEP {\bf 0702}, 028 (2007).
  [hep-th/0610297].

\bibitem{Nelson:1993nf}
  A.~E.~Nelson, N.~Seiberg,
  Nucl.\ Phys.\  {\bf B416}, 46-62 (1994).
  [hep-ph/9309299].

\bibitem{Kaplunovsky:1993rd}
  V.~S.~Kaplunovsky and J.~Louis,
  Phys.\ Lett.\  B {\bf 306} (1993) 269
  [arXiv:hep-th/9303040].
  
\bibitem{Davoudiasl:1999tf}
  H.~Davoudiasl, J.~L.~Hewett, T.~G.~Rizzo,
  Phys.\ Lett.\  {\bf B473 } (2000)  43-49.
  [hep-ph/9911262].

\bibitem{Nakamura:2010zzi}
  K.~Nakamura {\it et al.} [ Particle Data Group Collaboration ],
  J.\ Phys.\ G {\bf G37}, 075021 (2010).

\bibitem{Misiak:1997ei}
  M.~Misiak, S.~Pokorski, J.~Rosiek,
  Adv.\ Ser.\ Direct.\ High Energy Phys.\  {\bf 15}, 795-828 (1998).
  [hep-ph/9703442].

\bibitem{Barbieri:1987fn}
  R.~Barbieri, G.~F.~Giudice,
  Nucl.\ Phys.\  {\bf B306}, 63 (1988).

\bibitem{Carena:1995wu}
  M.~S.~Carena, M.~Quiros, C.~E.~M.~Wagner,
  Nucl.\ Phys.\  {\bf B461}, 407-436 (1996).
  [hep-ph/9508343]; 
  M.~S.~Carena, H.~E.~Haber, S.~Heinemeyer, W.~Hollik, C.~E.~M.~Wagner, G.~Weiglein,
  Nucl.\ Phys.\  {\bf B580}, 29-57 (2000).
  [hep-ph/0001002].

\bibitem{Abe:2004tq}
  H.~Abe, K.~Choi, K.~-S.~Jeong, K.~-i.~Okumura,
  JHEP {\bf 0409 } (2004)  015.
  [hep-ph/0407005].

\bibitem{Conlon:2007dw}
  J.~P.~Conlon,
  JHEP {\bf 0803}, 025 (2008).
  [arXiv:0710.0873 [hep-th]].

\bibitem{Anisimov:2001zz}
  A.~Anisimov, M.~Dine, M.~Graesser, S.~D.~Thomas,
  Phys.\ Rev.\  {\bf D65}, 105011 (2002).
  [hep-th/0111235].

\bibitem{Dudas:2004ni}
  E.~Dudas, T.~Gherghetta and S.~Groot Nibbelink,
  Phys.\ Rev.\  D {\bf 70} (2004) 086012
  [arXiv:hep-th/0404094].

\end{thebibliography}
\end{document}